\documentclass[preprint,amsmath,amssymb,aps,prd,nofootinbib]{revtex4}
\usepackage{epsfig,natbib,graphicx,color,appendix,fge}
\begin{document}
\def\CP{{\it CP}~}
\def\cp{{\it CP}}
\title{\mbox{}\\[15pt]
 Simple Modular invariant model\\for Quark, Lepton, and flavored-QCD axion}

\author{Y. H. Ahn$^{1}$\footnote{Email: axionahn@naver.com},
Sin Kyu Kang$^{1,2}$\footnote{Email: skkang@seoultech.ac.kr}
}
\affiliation{$^{1}$Institute for Convergence Fundamental Study, Seoul National University of Science and Technology, \\
       Seoul 139-743, Korea}
\affiliation{$^2$School of Liberal Arts, Seoul National University of Science and Technology,
       Seoul 139-743, Korea}



\begin{abstract}
We propose a minimal extension of the Standard Model by incorporating sterile neutrinos and a QCD axion to account for
the mass and mixing hierarchies of quarks and leptons and to solve the strong CP problem, and by introducing $G_{\rm SM}\times \Gamma_N\times U(1)_X$ symmetry. 
We demonstrate that the K{\"a}hler transformation corrects the weight of modular forms in the superpotential and show that the model is consistent with the modular and $U(1)_X$ anomaly-free conditions. This enables a simple construction of a modular-independent superpotential for scalar potential. Using minimal supermultiplets, we demonstrate a level 3 modular form-induced superpotential. Sterile neutrinos explain small active neutrino masses via the seesaw mechanism and provide a well-motivated $U(1)_X$ breaking scale, whereas gauge singlet scalar fields play crucial roles in generating the QCD axion, heavy neutrino mass, and fermion mass hierarchy. The model predicts a range for the $U(1)_X$ breaking scale from $10^{13}$ GeV to $10^{15}$ GeV for $1\,\mbox{TeV}< m_{3/2}<10^6\,\mbox{TeV}$. In the supersymmetric limit, all Yukawa coefficients in the superpotential are given by complex numbers with an absolute value of unity, implying a democratic distribution. 
Performing numerical analysis, we study how model parameters are constrained by current experimental results.
In particular, the model predicts that the value of the quark Dirac CP phase  falls between $38^\circ$ to $87^\circ$, which is consistent with experimental data, and the favored value of  the neutrino Dirac CP phase is around $250^\circ$. Furthermore, the model can be tested by ongoing and future experiments on axion searches, neutrino oscillations, and $0\nu\beta\beta$-decay.

\end{abstract}

\maketitle 
\section{Introduction}
Despite being theoretically self-consistent and successfully demonstrating experimental results in low-energy experiments so far, the Standard Model (SM) of particle physics leaves unanswered questions in theoretical and cosmological issues and fails to explain some physical phenomena such as neutrino oscillations, muon $g-2$, etc.
Various attempts have been made to extend the SM in order to address these questions and account for experimental results that cannot be explained within the SM. For instance, the canonical seesaw mechanism \cite{Minkowski:1977sc}  has been proposed to explain the tiny masses of neutrinos by introducing new heavy neutral fermions alongside the SM particles. Additionally, the Peccei-Quinn (PQ) mechanism \cite{Peccei-Quinn}  has been suggested to solve the strong CP problem in quantum chromodynamics (QCD) by extending the SM to include an anomalous $U(1)_X$ symmetry.

Recently, Feruglio\,\cite{Feruglio:2017spp} proposed  a new idea regarding the origin of the structure of lepton mixing.
He applied modular invariance\,\footnote{Modular invariance was analyzed for supersymmetric field theories in Ref.\cite{Ferrara:1989bc}.} under the modular group to determine the flavor structure of leptons without introducing a number of scalar fields. This approach requires the Yukawa couplings among twisted states to be modular forms. 
It is a string-derived mechanism that naturally restricts the possible variations in the flavor structure of quarks and leptons, which are unconstrained by the SM gauge invariance. However, explaining the hierarchies of the masses and mixing in the quark and lepton sectors remains a challenge. 
As studied in most references\,\cite{modular},  the Yukawa coefficients are assumed to be free parameters \footnote{In Ref.\cite{modular2}, the Yukawa coefficients are assumed to be of order unity.}  which can be determined by matching them with experimental data on fermion mass and mixing hierarchies.
This approach is not significantly different from that in the SM, except for the introduction of modular forms. Alternatively, it is also possible to take the Yukawa coefficient to be of order unity, accommodating the hierarchies of the fermion masses and mixing.
Recently, Ref.\cite{Feruglio:2023uof} has demonstrated that the vanishing QCD angle, a large CKM phase, and the reproduction of quark and lepton masses and mixings can be achieved by using coefficients up to order one, see also Ref.\cite{Kobayashi:2020oji}.

To incorporate sterile neutrinos and a QCD axion into the SM and provide a natural explanation for the mass and mixing hierarchies of quarks and leptons, we propose an extension of a modular invariant model based on the  four dimensional (4D) effective action  derived from superstring theory with $G_{\rm SM}\times \Gamma_N\times U(1)_X$ symmetry.
The non-Abelian discrete symmetry $\Gamma_N$ with $N=2,3,4,5$ plays a role of modular invariance, and  may originate from superstring theory in compactified extra dimensions, where it acts as a finite subgroup of the modular group\,\cite{deAdelhartToorop:2011re}.
To ensure the validity of a modular invariant model with $G_{\rm SM}\times \Gamma_N\times U(1)_X$,
we take the followings into account:
\begin{description}
\item[(i)] T-duality relates one type of superstring theory to another, and it also appears in the 4D low-energy effective field theory derived from superstring theory (for a review, see Ref.\cite{string_book}). In particular, 4D low-energy effective field theory of type IIA string theory with a certain compactification is invariant under the modular transformation of the modulus $\tau$,
{\begin{eqnarray}
 \tau\rightarrow\gamma\tau=\frac{a\tau+b}{c\tau+d}\,,\quad(a, b, c, d\in{Z},~~ ad-bc=1)\,.
  \label{mt1}
\end{eqnarray}}
So, the 4D action we conisder is required to be invariant 
under the modular transformation and gauged $U(1)_X$ symmetry, as well as the K{\"a}hler transformation (refer to Eq.(\ref{tr1})). This is necessary to cancel out the modular anomaly (see Ref.\cite{Derendinger:1991hq}) associated with the modular transformation Eq.(\ref{mt1}) under the non-local modular group $\Gamma_N$ and the gauged $U(1)_X$ anomaly, at the quantum level.
\item[(ii)] While type II string theory allows for low axion decay constant models via D-branes, leading to the gauged $U(1)_X$  that becomes a global PQ symmetry when  the $U(1)_X$ gauge boson is decoupled\,\cite{Dine:1987xk}, heterotic string theory typically has a $U(1)_X$ breaking scale with a decay constant close to the string scale.
The broken $U(1)_X$ gauge symmetry leaves behind a protected global $U(1)_X$ that is immune to quantum-gravitational effects, achieved via the Green-Schwarz (GS) mechanism\,\cite{Green:1984sg}.
The PQ breaking scale, or the low axion decay constant, can be determined by taking into account both supersymmetry (SUSY)-breaking effects\,\cite{Burgess:2003ic} and supersymmetric next-leading-order Planck-suppressed terms\,\cite{Nanopoulos:1983sp, Ahn:2016hbn, Ahn:2017dpf}.
\end{description}
The model features a minimal set of fields that transform based on representations of $G_{\rm SM}\times \Gamma_N\times U(1)_X$, and includes modular forms of level $N$. These modular forms act as Yukawa couplings and transform under the modular group $\Gamma_N$. It should be noted that the K{\"a}hler transformation (refer to Eq.(\ref{tr1})) corrects the weight of modular forms in the superpotential due to the modular invariance of both the superpotential and K{\"a}hler potential, see Eq.(\ref{req}). This enables a simple construction of a $\tau$-independent superpotential for scalar potential.
The so-called flavored-PQ symmetry $U(1)_X$ guarantees the absence of bare mass terms\,\cite{Ahn:2014gva}. 
We minimally extend the model by incorporating three right-handed neutrinos $N^c$ and SM gauge singlet scalar fields $\chi(\tilde{\chi})$. The scalar fields with a modular weight of zero and charged by $+(-)$ under $U(1)_X$ play a crucial role in generating the QCD axion, heavy neutrino mass, and fermion mass hierarchy.
Then the complex scalar field ${\cal F}=\chi(\tilde{\chi})$ with modular weight zero acts on dimension-four(three) operators well-sewed by $G_{\rm SM}\times \Gamma_N\times U(1)_X$ and modular invariance with different orders, which generate the effective interactions for the SM and the right-handed neutrinos as follows,
\begin{eqnarray}
 \tilde{c}_{1}\,{\cal O}_{3}\,({\cal F})^{1}+{\cal O}_{4}\sum_{n=0}^{\rm finite} c_{n}\,\left(\frac{{\cal F}}{\Lambda}\right)^n+...
  \label{AFN}
\end{eqnarray}
Here, $\Lambda$ is the scale of flavor dynamics above which unknown physics exists as a UV cutoff,
and Yukawa coefficients $c_n(\tilde{c}_1)$ are all complex numbers assumed to have a unit absolute value ($|\tilde{c}_{1}|,|c_n|=1$). 
 The dimension-$4(3)$ operators ${\cal O}_{4(3)}$ are determined by $G_{\rm SM}\times \Gamma_N\times U(1)_X$ and modular invariance in the supersymmetric limit. These operators include modular forms of level $N$, which transform according to the representation of $\Gamma_N$\,\cite{Feruglio:2017spp}. We will demonstrate that any additive finite correction terms, which could potentially be generated by higher weight modular forms, are prohibited due to the modular weight of the $\chi(\tilde{\chi})$ fields being zero.
Note that there exist the infinite series of higher-dimensional operators induced solely by the combination of $\chi\tilde{\chi}$ in the supersymmetric limit. These operators, represented by dots in Eq.(\ref{AFN}), can be absorbed into the finite leading order terms and effectively modify the coefficients $\tilde{c}_{1}$ and $c_{n}$ at the leading order.
Furthermore, to avoid the breaking effects of the axionic shift symmetry caused by gravity that spoil the axion solution to the strong CP problem\,\cite{Kamionkowski:1992mf}, we imposed a $U(1)_X$-mixed gravitational anomaly-free condition\,\cite{Ahn:2016hbn, Ahn:2018cau, Ahn:2021ndu}.

The rest of this paper is organized as follows: The next section discusses modular and $U(1)_X$ anomaly-free conditions under $G_{\rm SM}\times \Gamma_N\times U(1)_X$ symmetry, along with the modular forms of superpotential corrected by K{\"a}hler transformation. Sec.\ref{exam} presents an example of a superpotential induced by level 3 modular forms. We introduce minimal supermultiplets to address the challenges of tiny neutrino masses, the strong CP problem, and the hierarchies of SM fermion mass and mixing. For our purpose, we show how to derive Yukawa superpotentials and  a modular-independent superpotential for the scalar potential and determining the relevant $U(1)_X$ PQ symmetry breaking scale (or seesaw scale).
Additionally, we provide comments on the modular invariant model.
In Sec.\ref{visu}, we visually demonstrate the interconnections between quarks, leptons, and flavored-QCD axion.
In Sec.\ref{num1}, we present numerical values of physical parameters that satisfy the current experimental data on flavor mixing and mass for quarks and leptons while also favoring the assumption in Eq.(\ref{AFN}). The study predicts the Dirac CP phases of quarks and leptons, as well as the mass of the flavored-QCD axion and its coupling to photons and electrons.
The final section provides a summary of our work.

\section{Modular and U(1) anomaly-free}
\label{int}
T-duality relates different types of superstring theory and is also present in the 4D low-energy effective field theory derived from superstring theory (see Ref.\cite{string_book} for a review). In particular, type IIA intersecting D-brane models are related to magnetized D-brane models through T-duality\,\cite{string_book}.
The group $\Gamma(N)$ acts on the complex variable $\tau$, varying in the upper-half complex plane ${\rm Im}(\tau)>0$, as the modular transformation Eq.(\ref{mt1}).
Then the low-energy effective field theory of type IIA intersecting D-brane models must have the symmetry under the modular transformation Eq.(\ref{mt1}).
First, we shortly review the modular symmetry. The infinite groups $\Gamma(N)$, called principal congruence subgroups of level $N=1,2,3,...$, are defined by
{\begin{eqnarray}
 \Gamma(N)=\Big\{\begin{pmatrix} a & b  \\ c & d  \end{pmatrix}\in SL(2, Z), \begin{pmatrix} a & b  \\ c & d  \end{pmatrix}=\begin{pmatrix} 1 & 0  \\ 0 & 1  \end{pmatrix}\quad(\text{mod}~N)\Big\}\,,
\end{eqnarray}}
which are normal subgroups of homogeneous modular group $\Gamma\equiv\Gamma(1)\simeq SL(2,Z)$, where $SL(2, Z)$ is the group of $2\times2$ matrices with integer entries and determinant equal to one. The projective principal congruence subgroups are defined as $\bar{\Gamma}(N)=\Gamma(N)/\{\pm {I}\}$ for $N=1,2$. For $N\geq3$ we have $\bar{\Gamma}(N)=\Gamma(N)$ because the elements $-I$ does not belong to $\Gamma(N)$.  
The modular group $\bar{\Gamma}\equiv \Gamma/\{\pm {I}\}$ is generated by two elements $S$ and $T$,
{\begin{eqnarray}
 S:\tau\rightarrow-\frac{1}{\tau}\,,\qquad\qquad T:\tau\rightarrow\tau+1\,,
\end{eqnarray}}
satisfying
{\begin{eqnarray}
S^2=(ST)^3=(TS)^3={\bf 1}\,.
\label{st1}
\end{eqnarray}}
They can be represented by the $PSL(2, Z)$ matrices
{\begin{eqnarray}
 S=\begin{pmatrix} 0 & 1  \\ -1 & 0  \end{pmatrix}\,,\qquad T=\begin{pmatrix} 1 & 1  \\ 0 & 1  \end{pmatrix}\,.
\end{eqnarray}}
The groups $\Gamma_N$ are finite modular groups obtained by imposing the condition $T^N={\bf 1}$ in addition to Eq.(\ref{st1}), where $\Gamma_N\equiv\bar{\Gamma}/\bar{\Gamma}(N)$. The groups $\Gamma_N$ are isomorphic to the permutation groups $S_3$, $A_4$, $S_4$, and $A_5$ for $N = 2, 3, 4, 5$, respectively\,\cite{deAdelhartToorop:2011re}.

We work in the 4D ${\cal N}=1$ string-derived supergravity framework defined by a general K{\"a}hler potential $G(\Phi,\bar{\Phi})$ of the chiral superfields $\Phi$ and their conjugates,
{\begin{eqnarray}
G(\Phi,\bar{\Phi})=\frac{K(\Phi,\bar{\Phi})}{M^2_P}+\ln\frac{|W(\Phi)|^2}{M^6_P}\,,
\label{ge_k}
\end{eqnarray}}
and by an analytic gauge kinetic function $f(\Phi)$ of the chiral superfields $\Phi$, where $M_P=(8\pi G_N)^{-1/2}=2.436\times10^{18}$ GeV is the reduced Planck mass with Newton's gravitational constant $G_N$, $K(\Phi,\bar{\Phi})$ is a real gauge-invariant function of $\Phi$ and $\bar{\Phi}$, and $W(\Phi)$ is a holomorphic gauge-invariant function of $\Phi$.
Based on the 4D effective field theory derived from type IIA intersecting D-brane models, we build a modular invariant model with minimal chiral superfields transforming according to representations of $G_{\rm SM}\times \Gamma_N\times U(1)_X$. Here we assume that the non-Abelian discrete symmetry $\Gamma_N$ as a finite subgroup of the modular group\,\cite{Feruglio:2017spp}, and the anomalous gauged $U(1)_X$ including the SM gauge symmetry $G_{\rm SM}$ may arise from several stacks on D-brane models\,\cite{string_book}. 
 In the 4D global supersymmetry the most general form of the action can be written as
{\begin{eqnarray}
 {\cal S}=\int d^4x d^2\theta d^2\bar{\theta}K(\Phi, \bar{\Phi}e^{2A})+\Big\{\int d^4x d^2\theta\Big(W(\Phi)+\frac{f_{ab}(\Phi)}{4}{\cal W}^{\alpha a}{\cal W}^b_{\alpha}\Big)+{\rm h.c.}\Big\}\,,
 \label{ac1}
\end{eqnarray}}
where $A\equiv A^aT^a$ is the gauge multiplet containing Yang-Mills multiplet and $T^a$ are the gauge group generators, and ${\cal W}_\alpha$ is a gauge-invariant chiral spinor superfield containing the Yang-Mills field strength.  
The chiral superfields $\Phi$ denote all chiral supermultiplets with K{\"a}hler moduli, complex structure moduli, axio-dilaton, and matter superfields, transforming under $G_{\rm SM}\times\Gamma_N\times U(1)_X$. We assume that the low-energy K{\"a}hler potential $K$, superpotential $W$, and gauge kinetic function $f$ for moduli and matter superfields are given at a scale where K{\"a}hler moduli and complex structure moduli are stabilized through fluxes (see Ref.\cite{Gukov:1999ya, Giddings:2001yu, Dasgupta:1999ss}) leading to a consistent low-energy SM gauge theory.

Under the modular transformation Eq.(\ref{mt1}) and the gauged $U(1)_X$ symmetry, the action (\ref{ac1}) should be invariant with the transformations\,\footnote{The upper two shifts in Eq.(\ref{tr1}) of the K{\"a}hler potential and superpotential are known as the `K{\"a}hler transformation' with reference to Eq.(\ref{ge_k}).}
 {\begin{eqnarray}
&K(\Phi, \bar{\Phi}e^{2A})\rightarrow K(\Phi, \bar{\Phi}e^{2A})+\big(g(\Phi)+g(\bar{\Phi})\big)M^2_P\,,\nonumber\\
&W(\Phi)\rightarrow W(\Phi)e^{-g(\Phi)}\,,\nonumber\\
&f(\Phi){\cal W}^\alpha{\cal W}_\alpha\rightarrow f(\Phi){\cal W}^\alpha{\cal W}_\alpha\,,
\label{tr1}
\end{eqnarray}}
where $g(\Phi)$ is a function of modulus $\tau$. 
Then the given symmetry $G_{\rm SM}\times\Gamma_N\times U(1)_X$ can be violated at the quantum level by (i) an anomalous triangle graph associated with modular transformation Eq.(\ref{mt1}) under the non-local modular group $\Gamma_N$, and (ii) anomalous triangle graphs with external states $A^a_{\nu}A^b_\rho V_{X\mu}$ where $A^a_\nu$, $A^b_\rho$ are gauge bosons of the SM gauge group $G_{\rm SM}$ and $V^\mu_X$ is the connection associated with the gauged $U(1)_X$. These anomalies can be cancelled by the GS mechanism\,\cite{Green:1984sg}.

\subsection{Modular anomaly-free and modular forms of level $N$}
To demonstrate the invariance of $K(\Phi, \bar{\Phi}e^{2A})$ and $f(\Phi){\cal W}^\alpha{\cal W}_\alpha$ of Eq.(\ref{ac1}) under the finite modular group $\Gamma_N$ and the gauged $U(1)_X$, we consider a low-energy K{\"a}hler potential\,\footnote{It is similar to the one-loop K{\"a}hler potential presented in Ref.\cite{Derendinger:1991hq}.}:
{\begin{eqnarray}
K&=&-M^2_P\ln\Big\{\Big(S+\bar{S}-3\tilde{c}\ln(-i\tau+i\bar{\tau})\Big)\Big(U_X+\bar{U}_X-\frac{\delta^{\rm GS}_X}{16\pi^2}V_X\Big)\prod^2_{i=1}({\cal U}_i+\bar{\cal U}_i)\Big\}
\nonumber\\
&-&M^2_P\ln\big(-i\tau+i\bar{\tau}\big)^3+(-i\tau+i\bar{\tau})^{-k}|\varphi|^2+Z_X\varphi^\dag_X e^{-XV_X}\varphi_X+...\,,
\label{kalher1}
\end{eqnarray}}
where $-k$ is the modular weight, $Z_X$ is the normalization factor, 
$S$ denotes the axio-dilaton field, $\tau$ represents the overall Kähler modulus, and $U_X$ and ${\cal U}_i$ correspond to the complex structure moduli.
The dots in Eq.(\ref{kalher1}) denote the contributions of non-renormalizable terms scaled by an UV cutoff $M_P$ invariant under $G_{\rm SM}\times\Gamma_N\times U(1)_X$.
We note that the matter fields $\varphi_X$ with $U(1)_X$ charge, complex structure modulus $U_X$ and the vector superfield $V_X$ of the gauged $U(1)_X$ including the gauge field $A^\mu_X$ participate in the 4D GS mechanism. 
We take the holomorphic gauge kinetic function to be linear in the complex structure moduli $U_X$ and  ${\cal U}_i$, $f_{ab}(\Phi)\supset \delta_{ab}(S+U_X + {\cal U}_i) $.
The complex structure moduli  are associated with the SM gauge theory, which we will not be focusing on.
The GS parameter $\delta^{\rm GS}_X$ characterizes the coupling of the anomalous gauge boson to the axion $\theta_X$.
The matter superfields in $K$ consist of all scalar fields that are not moduli and do not have Planck-sized vacuum expectation values (VEVs). The scalar components of $\varphi$ and $\varphi_X$ are neutral under the $U(1)_X$ symmetry and the modular group $\Gamma_N$, respectively. 

Calculating $K_{IJ}=\partial_I\partial_J$ from the K{\"a}hler potential Eq.(\ref{kalher1}), we obtain the kinetic terms for the scalar components of the supermultiplets which are approximated well for $M_P\gg\langle\varphi\rangle, \langle\varphi_X\rangle$ and $V_X=0$ as follows,
\begin{eqnarray}
{\cal L}_{\rm kinetic}&\simeq&\frac{3M^2_P}{\langle-i\tau+i\bar{\tau}\rangle^2}\partial_\mu\bar{\tau}\partial^\mu\tau+\frac{M^2_P}{\langle U_X+\bar{U}_X\rangle^2}\partial_\mu\bar{U}_X\partial^{\mu}U_X\nonumber\\
&+&\frac{M^2_P}{\langle S+\bar{S}-3\tilde{c}\ln(-i\tau+i\bar{\tau})\rangle^2}\partial_\mu\bar{S}\partial^{\mu}S+K_{\varphi\bar{\varphi}}\partial_\mu\bar{\varphi}\partial^{\mu}\varphi+K_{\varphi_X\bar{\varphi}_X}\partial_\mu\bar{\varphi}_X\partial^{\mu}\varphi_X,
\label{kne}
\end{eqnarray}
where $K_{\varphi\bar{\varphi}}=K_{\varphi_X\bar{\varphi}_X}=1$ for canonically normalized scalar fields achieved by rescaling the fields $\varphi$ and $\varphi_X$ for given values of the VEVs of $\tau$ and $U_X$.
The $U(1)_X$ charged modulus $U_X$ and scalar field $\varphi_X$ can be decomposed as:
\begin{eqnarray}
U_X=\frac{\rho_X}{2}+i\theta_X,\qquad \varphi_X\big|_{\theta=\bar{\theta}=0}=\frac{1}{\sqrt{2}}e^{i\frac{A_X}{v_X}}(v_X+h_X),
\end{eqnarray}
where $Re[U_X]=\rho_X/2=1/g^2_X$ with $g_X$ being 4D $U(1)_X$ gauge coupling,  $A_X$, $v_X$, and $h_X$ are the axion, VEV, and Higgs boson of scalar components, respectively.
Due to the axionic shift symmetry, the kinetic terms of Eq.(\ref{kne}) for the axionic and size part of $U_X$ do not mix in perturbation theory, where any non-perturbative violations are small enough to be irrelevant, and the same goes for the axion and Higgs boson of the scalar components of $\varphi_X$ for $v_X\rightarrow\infty$.

Since the matter superfields $\varphi$ and axio-dilaton $S$ transform as,
 {\begin{eqnarray}
\varphi\rightarrow(c\tau+d)^{-k}\rho(\gamma)\varphi\,,\qquad S\rightarrow S-3\tilde{c}\ln(c\tau+d)\,,
\label{mt2}
\end{eqnarray}}
where $\rho(\gamma)$ is the unitary representation of the modular group $\Gamma_N$ and $\tilde{c}$ is a constant,
the transformation of the  K{\"a}hler potential $K$ given in Eq.(\ref{tr1}) leads us to
 {\begin{eqnarray}
g(\tau)=\ln(c\tau+d)^3\,.
\label{redu}
\end{eqnarray}}
Generically, the transformation of $K$ in Eq.(\ref{tr1}) incorporating Eq.(\ref{redu}) gives rise to a modular anomaly arising from $\delta S=-\tilde{c}\frac{1}{4}\int d^4xd^2\theta{\cal W}^{\alpha}{\cal W}_{\alpha}\,g(\tau)+{\rm h.c.}$\,\cite{Derendinger:1991hq},
 {\begin{eqnarray}
-\frac{1}{8}\tilde{c}\Big\{\big(g(\tau)+g(\bar{\tau})\big)Q^{\mu\nu}Q_{\mu\nu}+i\big(g(\tau)-g(\bar{\tau})\big)Q^{\mu\nu}\tilde{Q}_{\mu\nu}\Big\}
\label{mo_a}
\end{eqnarray}}
where $\tilde{Q}_{\mu\nu}=\frac{1}{2}\varepsilon_{\mu\nu\rho\sigma}Q^{\rho\sigma}$ with associated gauge field strengths $Q$, and the first term in the bracket represents the kinetic term for gauge bosons and the second term is the CP-odd term.
After receiving a  correction due to the modular transformation of $S$ in Eq.(\ref{mt2}), 
The gauge kinetic function $f_{ab}$  is given at leading order  by,
\begin{eqnarray}
f^{\rm 1-loop}(\Phi)=\delta_{ab}(S+U_X)-\tilde{c}\ln(c\tau+d)^3\, .
\end{eqnarray}
where the second term in the right hand side is the correction.
It is worthwhile to notice that this correction cancels the modular anomaly Eq.(\ref{mo_a}) generated by $g(\tau), g(\bar{\tau})$.

The modular invariance $W(\Phi)$ under the modular group $\Gamma_N$ ($N\geq2$) provides a strong restriction on the flavor structure\,\cite{Feruglio:2017spp}. The superpotential $W(\Phi)$ can be expanded in power series of the multiplets $\varphi$ which are separated into brane sectors $\varphi_{(I)}$
 {\begin{eqnarray}
W(\Phi)=\sum_n Y_{I_1...I_n}(\tau)\varphi_{(I_1)}\cdot\cdot\cdot\varphi_{(I_n)}\,,
\label{ms1}
\end{eqnarray}}
where the functions $Y_{I_1...I_n}(\tau)$ are generically\,\footnote{In Type II string orientifold compactifications, the Yukawa couplings have modular properties\,\cite{Cremades:2003qj}.} $\tau$-dependent in type IIA intersecting D-brane models\,\cite{string_book, Cremades:2004wa}.
The superpotential $W(\Phi)$ must have modular invariance under the transformation $W(\Phi) \rightarrow W(\Phi)e^{-g(\tau)}$, where $g(\tau)$ is given by Eq.(\ref{redu}). To ensure this, we need to satisfy two conditions:
 (i) the matter superfields $\varphi_{I_i}$ of the brane sector $I_i$ should transform
 {\begin{eqnarray}
\varphi_{(I_i)}\rightarrow (c\tau+d)^{-k_{I_i}}\rho_{(I_i)}(\gamma)\varphi_{(I_i)}
\label{mt3}
\end{eqnarray}}
in a representation $\rho_{(I_i)}(\gamma)$ of the modular group $\Gamma_N$, where $-k_{I_i}$ is the modular weight of sector $I_i$, and (ii) the functions $Y_{I_1...I_n}(\tau)$ should be modular forms of weight $k_Y(n)$ transforming in the representation $\rho(\gamma)$ of $\Gamma_N$,
{\begin{eqnarray}
Y_{I_1...I_n}(\gamma\tau)=(c\tau+d)^{k_Y(n)}\rho(\gamma)Y_{I_1...I_n}(\tau)\,,
\label{mof}
\end{eqnarray}}
with the requirements
{\begin{eqnarray}
&&k_Y(n)-3=k_{I_1}+...+k_{I_n}\,,\nonumber\\
&&\rho(\gamma)\otimes\rho_{(I_1)}\otimes\cdot\cdot\cdot\otimes\rho_{(I_n)}\ni{\bf 1}\,.
\label{req}
\end{eqnarray}}
The weight of modular forms in the superpotential is corrected by the K{\"a}hler transformation in Eq.(\ref{tr1}) due to the modular invariance of both the superpotential and K{\"a}hler potential. 
For example, a $\tau$-independent superpotential for scalar potential  can be simply constructed by the matter supermultiplets that belong to the untwisted sector in the orbifold compactification of type II string theory (see Eq.(\ref{super_d})).
We will show an explicit example of the superpotential induced by the modular forms of level 3 in the section\,\ref{exam}.

\subsection{Gauged U(1) anomaly-free}
The 4D action given by Eq.(\ref{ac1}) should also be $U(1)_X$ gauge invariant.
Under the $U(1)_X$ gauge transformation $V_X\rightarrow V_X+i(\Lambda_X-\bar{\Lambda}_X)$,
the matter superfields $\Phi_X$ and complex structure modulus $U_X$ transform
as,
{\begin{eqnarray}
 \Phi_X   \rightarrow  e^{iX\Lambda_X}\Phi_X\,,\qquad U_X\rightarrow U_X+i\frac{\delta^{\rm GS}_X}{16\pi^2}\Lambda_X\,,
\end{eqnarray}}
where $\Lambda_X(\bar{\Lambda}_X)$ are (anti-) chiral superfields parameterizing $U(1)_X$ transformation on the superspace. 
So the axionic modulus $\theta_X$ and axion $a_X$ have shift symmetries
{\begin{eqnarray}
\theta_X\rightarrow\theta_X-\frac{\delta^{\rm GS}_X}{16\pi^2}\xi_X\,,\qquad a_X\rightarrow a_X+\frac{\delta^{\rm GS}_X}{\delta^Q_X}f_X\xi_X\,,
\label{gat1}
\end{eqnarray}
where $\xi_X=-{\rm Re}\Lambda_X\big|_{\theta=\bar{\theta}=0}$,  $f_X=Xv_X$ is the axion decay constant, and $\delta^Q_X$ are anomaly coefficients defined in Eq.(\ref{anoc}).
Then, the $U(1)_X$ gauge field $A^\mu_X$ transforms as
{\begin{eqnarray}
A^\mu_X\rightarrow A^\mu_X-\partial^\mu\xi_X\,.
\label{gat2}
\end{eqnarray}}}
Since the gauged $U(1)_X$ is anomalous, the axion $a_X$ and axionic modulus $\theta_X$ couple to the (non)-Abelian Chern-Pontryagin densities for the SM gauge group in the compactification.
In type II string vacuum, the $U(1)_X$ anomalies should be cancelled by appropriate shifts of Ramond-Ramond axions in the bulk\,\cite{Sagnotti:1992qw, Ibanez:1998qp, Poppitz:1998dj, Lalak:1999bk}. 
The 4D effective action of the axions, $\theta_X$ and $a_X$, and its corresponding gauge field $A^\mu_X$ contains the followings\,\cite{Ahn:2017dpf, Ahn:2016typ}:
{\begin{eqnarray}
 &&K_{U_X\bar{U}_X}\Big(\partial^\mu\theta_X-\frac{\delta^{\rm GS}_X}{16\pi^2}A^\mu_X\Big)^2-\frac{1}{4g^2_X}F^{\mu\nu}_XF_{X\mu\nu}-g_X\xi^{\rm FI}_XD_X+D_Xg_XX|\varphi_X|^2\nonumber\\
 &&+|D_\mu\varphi_X|^2+\theta_X{\rm Tr}(Q^{\mu\nu}\tilde{Q}_{\mu\nu})+\frac{a_X}{f_X}\frac{\delta^Q_X}{16\pi^2}{\rm Tr}(Q^{\mu\nu}\tilde{Q}_{\mu\nu})\,,
 \label{act1}
\end{eqnarray}}
where the gauge field strengths $Q=G, W, Y$ for $SU(3)_C$, $SU(2)_L$, and $U(1)_Y$, respectively, and their gauge couplings are absorbed into their corresponding gauge field strengths. $F^{\mu\nu}_X$ is the $U(1)_X$ gauge field strength defined by $F^{\mu\nu}_X=\partial^\mu A^\nu_X-\partial^\nu A^\mu_X$. In $|D_\mu\varphi_X|^2$ the scalar components of $\varphi_X$ couple to the $U(1)_X$ gauge boson, where the gauge coupling $g_X$ is absorbed into the gauge boson $A^\mu_X$ in the $U(1)_X$ gauge covariant derivative $D^\mu=\partial^\mu-iXA^\mu_X$.
 The coefficients of  the mixed $U(1)_X\times[SU(3)_C]^2$, $U(1)_X\times[SU(2)_L]^2$, and $U(1)_X\times[U(1)_Y]^2$ anomalies are given respectively by
{\begin{eqnarray}
\delta^{G}_X=2{\rm Tr}[XT^2_{SU(3)}]\,,\quad \delta^{W}_X=2{\rm Tr}[XT^2_{SU(2)}]\,,\quad\delta^{Y}_X=2{\rm Tr}[XY^2]\,.
\label{anoc}
\end{eqnarray}}
 Here $U(n)$ generators ($n\geq2$) are normalized according to ${\rm Tr}[T^aT^b]=\delta_{ab}/2$, and for convenience, $\delta^Y_X=2{\rm Tr}[XY^2]$  is defined for hypercharge.
The FI term ${\cal L}^{\rm FI}_X=-\xi^{\rm FI}_{X}\int d^2\theta V_X=-\xi^{\rm FI}_{X}g_XD_X$ with $D_X=g_X(\xi^{\rm FI}_{X}-X|\varphi_X|^2)$ leads to D-term potential for the anomalous $U(1)_X$,
{\begin{eqnarray}
V_D=\frac{1}{U_X+\bar{U}_X}\big(-\xi^{\rm FI}_{X}+X|\varphi_X|^2\big)^2\,,
\label{dv}
\end{eqnarray}}
where  $\xi^{\rm FI}_X$ is FI factor produced by expanding the K{\"a}hler potential Eq.(\ref{kalher1}) in components linear in $V_X$ and depends on the closed string modulus ${\rm Re}[U_X]=\rho_X/2$.
Since the FI term is controlled by the string coupling it can not be zero. The re-stabilization of VEVs by $\varphi_X$ necessarily implies spontaneous breaking of the anomalous $U(1)_X$, which will be shown later.

The first, third, fourth, and fifth terms in Eq.(\ref{act1}) result from expanding the K{\"a}hler potential of Eq.(\ref{kalher1}). The first and sixth terms together, and the fifth and seventh terms in Eq.(\ref{act1}), are gauge invariant under the anomalous $U(1)_X$ gauge transformations of Eqs.(\ref{gat1}, \ref{gat2}).
The gauge invariant  interaction  Lagrangian is given by
{\begin{eqnarray}
{\cal L}^{\rm int}_{A\theta}=-A^\mu_X J^\theta_\mu+\theta_X{\rm Tr}(Q^{\mu\nu}\tilde{Q}_{\mu\nu})-A^\mu_X J^X_\mu+\frac{a_X}{f_X}\frac{\delta^Q_X}{16\pi^2}{\rm Tr}(Q^{\mu\nu}\tilde{Q}_{\mu\nu})
\label{intL}
\end{eqnarray}}
where the anomalous currents $J^X_\mu$ and $J^\theta_\mu$ coupling to the gauge boson $A^\mu_X$ (that is, $\partial_\mu J^\mu_X=\frac{\delta^{\rm GS}_X}{16\pi^2}{\rm Tr}(Q^{\mu\nu}\tilde{Q}_{\mu\nu})=-\partial_\mu J^\mu_\theta$ with $\delta^{\rm GS}_X=\alpha^Q_X\delta^Q_X$ ) are represented  by
$J^\theta_\mu=K_{U_X\bar{U}_X}\frac{\delta^{\rm GS}_X}{8\pi^2}\partial_\mu\theta_X$ and  $J^X_\mu=-iX\varphi_X^\dag\overleftrightarrow{\partial_\mu}\varphi_X$.

Expanding Eq.(\ref{act1}) and setting $\theta_X=a_\theta/8\pi^2 f_\theta$ with $f_\theta=\sqrt{\frac{2K_{U_X\bar{U}_X}}{(8\pi^2)^2}}$ to canonically normalize,  ${\cal L}^{\rm int}_{A\theta}$ becomes
{\begin{eqnarray}
 &&\frac{1}{2}(\partial^\mu a_\theta)^2+\frac{a_\theta}{8\pi^2 f_\theta}{\rm Tr}(Q^{\mu\nu}\tilde{Q}_{\mu\nu})+\frac{1}{2}(\partial^\mu A_X)^2+\frac{A_X}{f_X}\frac{\delta^Q_X}{16\pi^2}{\rm Tr}(Q^{\mu\nu}\tilde{Q}_{\mu\nu})\nonumber\\
 &&-A^\mu_X(J^X_\mu+J^\theta_\mu)+\frac{1}{2g^2_X}m^2_XA^\mu_XA_{X\mu}-\frac{1}{4g^2_X}F^{\mu\nu}_XF_{X\mu\nu}-\frac{g^2_X}{2}(\xi^{\rm FI}_X-X|\varphi_X|^2)^2\,,
 \label{act2}
\end{eqnarray}}
where the gauge boson mass $m_X$ obtained by the super-Higgs mechanism is given by $m_X=\sqrt{2K_{U_X\bar{U}_X}(\delta^{\rm GS}_X/16\pi^2)^2+2f^2_X}$. Then the open string axion $a_X$ (decay constant $f_X$) is mixed linearly with the closed string $a_\theta$ (decay constant $f_\theta$):
{\begin{eqnarray}
\tilde{A}=\frac{a_X\frac{\delta^{\rm GS}_X}{2}f_\theta-a_\theta f_X}{\sqrt{f^2_X+(\frac{\delta^{\rm GS}_X}{2}f_\theta)^2}}\approx a_X\,,\qquad G=\frac{a_\theta\frac{\delta^{\rm GS}_X}{2}f_\theta+A_X f_X}{\sqrt{f^2_X+(\frac{\delta^{\rm GS}_X}{2}f_\theta)^2}}\approx a_\theta\,.
 \label{ax1}
\end{eqnarray}}
where the approximations are valid under the assumption that $f_\theta$ is much larger than $f_X$. The gauged $U(1)_X$ absorbs one linear combination of $a_X$ and $a_\theta$, denoted $G$, giving it a string scale mass through the $U(1)_X$ gauge boson, while the other combination, $\tilde{A}\approx a_X$, remains at low energies and contributes to the QCD axion. At energies below the scale $m_X$, the gauge boson decouples, leaving behind an anomalous global $U(1)_X$ symmetry.

\section{Minimal model set-up}
\label{exam}
For our purpose, we take into account $\Gamma(3)$ modular symmetry, which gives the modular forms of level 3.
The group $\Gamma_3$ is isomorphic to $A_4$ which is the symmetry group of the tetrahedron and the finite groups of the even permutation of four objects having four irreducible representations. Its irreducible representations are three singlets ${\bf 1}, {\bf 1}'$, and ${\bf 1}''$ and one triplet ${\bf 3}$ with the multiplication rules ${\mathbf3}\otimes{\mathbf3}={\mathbf3}_{s}\oplus{\mathbf3}_{a}\oplus{\mathbf1}\oplus{\mathbf1}'\oplus{\mathbf1}''$ and ${\mathbf1}'\oplus{\mathbf1}'={\mathbf1}''$, where the subscripts $s$ and $a$ denote symmetric and antisymmetric combinations respectively. Let $(a_1, a_2, a_3)$ and $(b_1, b_2, b_3)$ denote the basis vectors for two ${\mathbf3}$'s. Then we have 
{\begin{eqnarray}
 (a\otimes b)_{{\mathbf 3}_s}&=&\frac{1}{\sqrt{3}}\big(2a_1b_1-a_2b_3-a_3b_2, ~2a_3b_3-a_1b_2-a_2b_1, ~2a_2b_2-a_3b_1-a_1b_3\big)\,,\nonumber\\
 (a\otimes b)_{{\mathbf 3}_a}&=&\big(a_2b_3-a_3b_2, ~a_1b_2-a_2b_1, ~a_3b_1-a_1b_3\big)\,,\nonumber\\
 (a\otimes b)_{\mathbf 1}&=& a_1b_1+a_2b_3+a_3b_2\,,\nonumber\\
 (a\otimes b)_{{\mathbf 1}'}&=& a_3b_3+a_1b_2+a_2b_1\,,\nonumber\\
  (a\otimes b)_{{\mathbf 1}''}&=& a_2b_2+a_3b_1+a_1b_3\,.
  \label{A4x}
\end{eqnarray}} 
The details of the $A_4$ group are shown in Appendix\,\ref{A4_i}.
The modular forms $f(\tau)$ of level $3$ and weight $k$, such as Eq.(\ref{mof}), are holomorphic functions of the complex variable $\tau$ with well-defined transformation properties 
{\begin{eqnarray}
 f(\gamma\tau)=(c\tau+d)^{k}f(\tau)\quad \gamma=\begin{pmatrix} a & b  \\ c & d  \end{pmatrix}\in\Gamma_3
 \end{eqnarray}}
with an integer $k\geq0$, under the group $\Gamma_3$. The three linearly independent weight 2 and level-3 modular forms are given by\,\cite{Feruglio:2017spp}
{\begin{eqnarray}
&&Y_1(\tau)=\frac{i}{2\pi}\Big[\frac{\eta'(\frac{\tau}{3})}{\eta(\frac{\tau}{3})}+\frac{\eta'(\frac{\tau+1}{3})}{\eta(\frac{\tau+1}{3})}+\frac{\eta'(\frac{\tau+2}{3})}{\eta(\frac{\tau+2}{3})}-\frac{27\eta'(3\tau)}{\eta(3\tau)}\Big]\,,\nonumber\\
&&Y_2(\tau)=\frac{-i}{\pi}\Big[\frac{\eta'(\frac{\tau}{3})}{\eta(\frac{\tau}{3})}+\omega^2\frac{\eta'(\frac{\tau+1}{3})}{\eta(\frac{\tau+1}{3})}+\omega\frac{\eta'(\frac{\tau+2}{3})}{\eta(\frac{\tau+2}{3})}\Big]\,,\nonumber\\
&&Y_3(\tau)=\frac{-i}{\pi}\Big[\frac{\eta'(\frac{\tau}{3})}{\eta(\frac{\tau}{3})}+\omega\frac{\eta'(\frac{\tau+1}{3})}{\eta(\frac{\tau+1}{3})}+\omega^2\frac{\eta'(\frac{\tau+2}{3})}{\eta(\frac{\tau+2}{3})}\Big]\,,
\end{eqnarray}}
where $\omega=-1/2+i\sqrt{3}/2$ and $\eta(\tau)$ is the Dedekind eta-function defined by
{\begin{eqnarray}
 \eta(\tau)=q^{1/24}\prod^{\infty}_{n=1}(1-q^n)\quad\text{with}~q\equiv e^{i2\pi\tau}~\text{and}~{\rm Im}(\tau)>0\,.
\end{eqnarray}}
The Dedekind eta-function satisfies
{\begin{eqnarray}
 \eta(-1/\tau)=\sqrt{-i\tau}\,\eta(\tau)\,,\qquad\eta(\tau+1)=e^{i\pi/12}\,\eta(\tau)\,.
\end{eqnarray}}
The three linear independent modular functions transform as a triplet of $A_4$, {\it i.e.} $Y^{(2)}_{\bf 3}=(Y_1, Y_2, Y_3)$. The $q$-expansion of $Y_i(\tau)$ reads
{\begin{eqnarray}
 &&Y_1(\tau)=1+12q+36q^2+12q^3+...\nonumber\\
 &&Y_2(\tau)=-6q^{1/3}(17q+8q^2+...)\nonumber\\
 &&Y_3(\tau)=-18q^{2/3}(1+2q+5q^2+...)\,.
\end{eqnarray}}
$Y^{(2)}_{\bf 3}$ is constrained by the relation,
{\begin{eqnarray}
 (Y^{(2)}_{\bf 3}Y^{(2)}_{\bf 3})_{{\bf 1}''}=Y^2_2+2Y_1Y_3=0\,.
\end{eqnarray}}

\subsection{Modular invariant supersymmetric potential and a Nambu-Goldstone mode}
\label{exam01}
Using Eqs.(\ref{ms1}-\ref{req}),  we construct unique supersymmetric and modular invariant scalar potential
by introducing minimal supermultiplets.
Those include SM singlet fields $\chi_0$ \,\footnote{The field $\chi_0$ can act as an inflaton\,\cite{Ahn:2017dpf}.} with modular weight three and $\chi(\tilde{\chi})$ with modular weight zero. Additionally, we have the usual two Higgs doublets $H_{u,d}$ with modular weight zero, which are responsible for electroweak (EW) symmetry breaking. The fields $\chi$ and $\tilde{\chi}$ are charged by $+1$ and $-1$, respectively, and are ensured by the extended $U(1)_X$ symmetry due to the holomorphy of the superpotential. (If the seesaw mechanism\,\cite{Minkowski:1977sc} is implemented, the field $\chi$ or $\tilde{\chi}$ may be responsible for the heavy neutrino mass term).

Under $k_I\times A_4\times U(1)_X$ with the modular weights $k_I$ according to Eq.(\ref{req}), we assign the two Higgs doublets $H_{u,d}$ to be $(0, {\mathbf 1}, 0)$ and three SM gauge singlets $\chi$, $\tilde{\chi}$, $\chi_0$ to be $(0, {\bf 1}, +1)$, $(0, {\bf 1}, -1)$, $(3, {\bf 1}, 0)$, respectively\,\footnote{As a consequence of $k_I\times A_4\times U(1)_X$ the other superpotential term $\kappa_{\alpha}L_{\alpha}H_{u}$ and the terms violating the lepton and baryon number symmetries are not allowed. Besides, dimension 6 supersymmetric operators like $Q_{i}Q_{j}Q_{k}L_{l}$ ($i,j,k$ must not all be the same) are not allowed either, and stabilizing proton.}. The $A_4$-singlet $\chi_0$ field with modular weight three ensures that the functions $Y_{I_1...I_n}(\tau)$ are independent of $\tau$.
Then the supersymmetric scalar potential
invariant under $G_{\rm SM}\times U(1)_{X}\times A_4$ is given at leading order by
\begin{eqnarray}
 W_v&=&g_{\chi_0}\chi_0\,H_uH_d+\chi_0(g_\chi\chi\tilde{\chi}-\mu^2_\chi)\,\,,
  \label{super_d}
\end{eqnarray}
where dimensionless coupling constants $g_{\chi_0}$ and $g_{\chi}$ are assumed to be equal to one, but are modified to Eq.(\ref{AFN1}) by considering all higher order terms induced by $\chi\tilde{\chi}$ combinations.
Note that the PQ breaking parameter $\mu_\chi$ corresponds to the scale of the spontaneous symmetry breaking.

 In the global SUSY limit, {\it i.e.} $M_P\rightarrow\infty$, the scalar potential obtained by the $F$- and $D$-term of all fields is required to vanish.  Then the relevant $F$-term from Eq.(\ref{super_d}) and $D$-term of the scalar potential given by Eq.(\ref{dv})  reads
\begin{eqnarray}
 V_F^{\rm global}&=&\big|g_\chi\chi\tilde{\chi}-\mu^2_\chi\big|^2\,,\qquad  V_D^{\rm global}=\frac{|X|^2g^2_X}{2}\Big(-\frac{\xi^{\rm FI}_X}{|X|}+|\chi|^2-|\tilde{\chi}|^2\Big)^2\,\,.
  \label{super_1}
\end{eqnarray}
The scalar fields $\chi$ and $\tilde{\chi}$ have $X$-charges $+1$ and $-1$, respectively, {\it i.e.}, 
\begin{eqnarray}
 \chi\rightarrow e^{+i\xi}\chi\,,\qquad \tilde{\chi}\rightarrow e^{-i\xi}\tilde{\chi}\,,
  \label{super_2}
\end{eqnarray}
with a constant $\xi$. So the potential $V_{\rm SUSY}$ has $U(1)_X$ symmetry. Since SUSY is preserved after the spontaneous breaking of $U(1)_X$, the scalar potential in the limit of $M_P\rightarrow\infty$ vanishes at its ground states, {\it i.e.} $\langle V_F^{\rm global}\rangle=0$ and $\langle V_D^{\rm global}\rangle=0$ vanishing $F$-term must have also vanishing $D$-term. From the minimization of the $F$-term scalar potential we obtain
 \begin{eqnarray}
 \langle\chi\rangle=\langle\tilde{\chi}\rangle=\frac{v_{\chi}}{\sqrt{2}}\qquad\text{with}~\mu_\chi=v_\chi\sqrt{\frac{g_{\chi}}{2}}\,.
 \label{vev}
 \end{eqnarray}
 where we have assumed $\langle\chi\rangle, \langle\tilde{\chi}\rangle\gg\langle H_{u,d}\rangle$. 
 The above supersymmetric solution is taken by the $D$-flatness condition for\,\cite{Ahn:2016hbn,Ahn:2017dpf}
 \begin{eqnarray}
\xi^{\rm FI}_X=0\,,\qquad \langle\chi\rangle=\langle\tilde{\chi}\rangle\,.
 \label{vevd}
 \end{eqnarray}
 
The tension between $\langle\chi\rangle=\langle\tilde{\chi}\rangle$ and $\xi^{\rm FI}_X\neq0$ arises because the FI term cannot be cancelled, unless the VEV of flux in the FI term is below the string scale\,\cite{Achucarro:2006zf, Burgess:2003ic}. The FI term acts as an uplifting potential, 
 \begin{eqnarray}
\xi^{\rm FI}_X=M^2_P\frac{\delta^{\rm GS}_X}{16\pi^2}\frac{\Delta\rho}{\rho_0}\,,
 \label{fi1}
 \end{eqnarray}
where $\Delta\rho=\rho_X-\rho_0$, which raises the Anti-de Sitter minimum to the de Sitter minimum\,\cite{Burgess:2003ic}. To achieve this, the $F$-term must necessarily break SUSY for the $D$-term to act as an uplifting potential.
The PQ scale $\mu_\chi$ can be determined by taking into account both the SUSY-breaking effect, which lifts up the flat direction, and supersymmetric next-leading-order Planck-suppressed terms\,\cite{Nanopoulos:1983sp, Ahn:2016hbn,Ahn:2017dpf}.
The supersymmetric next-to-leading order term invariant under $A_4\times U(1)_X$ satisfying Eq.(\ref{req}) are given by
 \begin{eqnarray}
\Delta W_v\simeq\frac{\alpha}{M^2_P}\chi_0(\chi\tilde{\chi})^2\,,
 \label{deW}
 \end{eqnarray}
where $\alpha$ is assumed to be a real-valued constant being of unity.
 Since soft SUSY-breaking terms are already present at the scale relevant to flavor dynamics, the scalar potential for $\chi$, $\tilde{\chi}$ at leading order read
 \begin{eqnarray}
V(\chi,\tilde{\chi})\simeq-\alpha_1m^2_{3/2}|\chi|^2-\alpha_2m^2_{3/2}|\tilde{\chi}|^2+\alpha^2\frac{|\chi|^4|\tilde{\chi}|^4}{M^4_P}\,,
 \label{sof1}
 \end{eqnarray}
where $m_{3/2}$ represents soft SUSY-breaking mass, and $\alpha_1, \alpha_2$ are real-valued constants. It leads to the PQ breaking scale (equivalently, the seesaw scale),
 \begin{eqnarray}
\mu_\chi\simeq \big(\frac{g^6_\chi\alpha_1\alpha_2}{16\alpha^4}\big)^{\frac{1}{12}} \big(m_{3/2}M^2_P\big)^{\frac{1}{3}}\,,
 \label{sof2}
 \end{eqnarray} 
indicating that $\mu_\chi$ lies within the range of approximately $1.2\times10^{13}$ to $1.7\times10^{14}$ GeV (or $2.6\times10^{13}$ to $1.2\times10^{15}$ GeV) for $m_{3/2}$ values ranging from 1 TeV to $10^3$ TeV (or from 10 TeV to $10^6$ TeV) for $\alpha_1$ and $\alpha_2$ of order unity.

The model includes the SM gauge singlet scalar fields $\chi$ and $\tilde{\chi}$ charged under $U(1)_X$, which have interactions invariant under $G_{\rm SM}\times U(1)_X\times A_4$ with the transformations Eq.(\ref{tr1}). These interactions result in a chiral symmetry, which is reflected in the form of the kinetic and Yukawa terms, as well as the scalar potential $V_{\rm SUSY}$ in the SUSY limit:
{\begin{eqnarray}
 {\cal L} &\supset& \partial_\mu\chi^\ast\partial^\mu\chi+ \partial_\mu\tilde{\chi}^\ast\partial^\mu\tilde{\chi}+{\cal L}_Y-V_{\rm SUSY}+{\cal L}_{\vartheta}+\overline{\psi}\,i\! \! \not\!\partial \psi+\frac{1}{2}\overline{N}\,i\! \! \not\!\partial N +\frac{1}{2}\overline{\nu}\,i\! \! \not\!\partial\nu \,,
\label{lag0}
 \end{eqnarray}}
where $\psi$ denotes Dirac fermions, and $V_{\rm SUSY}$ is replaced by $V_{\rm total}$ when SUSY breaking effects are considered. The above kinetic terms for $\chi(\tilde{\chi})$ are canonically normalized from the K{\"a}hler potential Eq.(\ref{kalher1}). Here four component Majorana spinors ($N^c=N$ and $\nu^c=\nu$) are used.
The global $U(1)_X$ PQ symmetry guarantees the absence of bare mass term in the Yukawa Lagrangian ${\cal L}_Y$ in Eq.(\ref{lag0}).
The QCD Lagrangian has a CP-violating term 
{\begin{eqnarray}
{\cal L}_{\vartheta}&=&\vartheta_{\rm QCD}\,\frac{g^{2}_{s}}{32\pi^{2}}\,G^{a\mu\nu}\tilde{G}^{a}_{\mu\nu}
 \end{eqnarray}}
where $g_s$ stands for the gauge coupling constant of $SU(3)_C$, and $G^{a\mu\nu}$ is the color field strength tensor and its dual $\tilde{G}^{a}_{\mu\nu}=\frac{1}{2}\varepsilon_{\mu\nu\rho\sigma}G^{a\mu\nu}$ (here $a$ is an $SU(3)$-adjoint index), coming from the strong interaction. 
After obtaining VEV $\langle\chi\rangle\neq0$, which generates the heavy neutrino masses given by Eq.(\ref{lagrangian_l}), the PQ $U(1)_X$ symmetry breaks spontaneously at a much higher scale than EW scale. This is manifested through the existence of the Nambu-Goldstone (NG) mode $A_X$, which interacts with ordinary quarks and leptons via Yukawa interactions, see Eqs.(\ref{fl_in}, \ref{fla_1}, \ref{neut1}). 
To extract the associated boson resulting from spontaneous breaking of $U(1)_X$, we set the decomposition of complex scalar fields\,\cite{Ahn:2014gva, Ahn:2016hbn, Ahn:2018cau} as follows
 \begin{eqnarray}
\chi=\frac{v_{\chi}}{\sqrt{2}}e^{i\frac{A_X}{u_{\chi}}}\left(1+\frac{h_{\chi}}{u_{\chi}}\right)\,,\quad\,\tilde{\chi}=\frac{v_{\tilde{\chi}}}{\sqrt{2}}e^{-i\frac{A_X}{u_{\chi}}}\left(1+\frac{h_{\tilde{\chi}}}{u_{\chi}}\right)\qquad\text{with}~u_{\chi}=\sqrt{v^2_{\chi}+v^2_{\tilde{\chi}}}\,,
  \label{NGboson}
 \end{eqnarray}
in which $A_X$ is the NG mode and we set $v_\chi=v_{\tilde{\chi}}$ and $h_{\chi}=h_{\tilde{\chi}}$ in the supersymmetric limit. The derivative coupling of NG boson $A_X$ arises from the kinetic term 
{\begin{eqnarray}
\partial_\mu\chi^\ast\partial^\mu\chi+\partial_\mu\tilde{\chi}^\ast\partial^\mu\tilde{\chi}=\frac{1}{2}(\partial_\mu A_X)^2\Big(1+\frac{h_\chi}{u_\chi}\Big)^2+\frac{1}{2}(\partial_\mu h_\chi)^2\,.
 \end{eqnarray}}
Performing $u_\chi\rightarrow\infty$, the NG mode $A_X$, whose interaction is determined by symmetry, is distinguished from the radial mode $h_\chi$, which is invariant under the symmetry $U(1)_X$.
\subsection{Modular invariant Yukawa superpotentials and Anomaly coefficients}
By introducing just two $A_4$-singlet fields, $\chi$ and $\tilde{\chi}$, with modular weight zero and charged under $U(1)_X$ by $+1$ and $-1$, respectively, and using economic weight modular forms, we construct Yukawa superpotentials that are invariant under $G_{\rm SM}\times U(1)_{X}\times A_4$ satisfying Eq.(\ref{req}). This approach can explain the observed hierarchy of fermion masses and mixing given by the Cabibbo-Kobayashi-Maskawa (CKM) matrix for quarks as well as by Pontecorvo-Maki-Nakagawa-Sakata (PMNS) matrix for leptons. Furthermore, the approach provides a solution to the strong CP problem by breaking down the $U(1)_X$ flavor symmetry.
Since the modular weights of the fields $\chi(\tilde{\chi})$ are zero, any additive correction terms induced by higher weight modular forms are forbidden in the superpotential (see Eqs.(\ref{super_d}, \ref{lagrangian_q}, \ref{lagrangian_l})). However, higher-order corrections arising from the combination $\chi\tilde{\chi}$ are allowed, but they do not modify the leading-order flavor structure.

Now let us assign $A_4\times U(1)_X$ representations and quantum numbers as well as modular weights $k_I$ to the SM quarks and leptons including SM gauge singlet Majorana neutrinos as presented in Table-\ref{reps_q}\,\footnote{All fields appearing in Table-\ref{reps_q} are left-handed particles/antiparticles.}. 
Here,  three quark $SU(2)_L$ doublets and three up-type quark singlets are denoted as $Q_{i(=1,2,3)}$ and $(u^{c}, c^{c}, t^{c})$, respectively. ${\cal D}^c=\{d^{c}, s^{c}, b^c\}$ represents the down-type quark singlets.
\begin{table}[h]
\caption{\label{reps_q} Representations and quantum numbers of the quark fields under $G_{\rm SM}\times A_4\times U(1)_{X}$ and modular weight $k_I$ according to Eq.(\ref{req}). In $({\cal Q}_1, {\cal Q}_2)_ Y$ of $G_{\rm SM}$, ${\cal Q}_1$ and ${\cal Q}_2$ are the representations under $SU(3)_C$ and $SU(2)_L$ respectively, and the script $Y$ denotes the $U(1)$ hypercharge.}
\begin{ruledtabular}
\begin{tabular}{cccccccc}
Field &$Q_{1}$& $Q_{2}$& $Q_{3}$&${\cal D}^c$ & $u^c$&$c^c$ & $t^c$\\
\hline
$G_{\rm SM}$&$(3,2)_{1/6}$&$(3,2)_{1/6}$&$(3,2)_{1/6}$&$(3,1)_{1/3}$&$(3,1)_{-2/3}$&$(3,1)_{-2/3}$&$(3,1)_{-2/3}$\\
$A_4$&$\mathbf{1}$ &$\mathbf{1}''$ & $\mathbf{1^{\prime}}$ & $\mathbf{3}$&$\mathbf{1}$&$\mathbf{1}'$&$\mathbf{1^{\prime\prime}}$\\
$k_I$&$0$&$0$&$0$ &$-3$&$-3$&$-3$&$3$\\
$U(1)_{X}$&$f_b-f_d$&$f_b-f_s$&$0$ &$-f_b$&$f_d-f_b-f_u$&$f_s-f_b-f_c$&$0$
\end{tabular}
\end{ruledtabular}
\end{table}
Then the quark Yukawa superpotential invariant under $G_{\rm SM}\times A_4\times U(1)_X$ with modular forms is sewed with ${\cal F}=\{\chi$ or $\tilde{\chi}\}$ through Eq.(\ref{AFN}) as
{\begin{eqnarray}
 W_q &=&
  \alpha_{t}^{(0)}\,t^cQ_{3}H_u
  +\alpha_{c}^{(0)}\Big(\frac{\cal F}{\Lambda}\Big)^{|f_c|}Y^{(6)}_{\bf1}c^cQ_2H_u+\alpha_{u}^{(0)}\Big(\frac{\cal F}{\Lambda}\Big)^{|f_u|}Y^{(6)}_{\bf1}u^cQ_1H_u\nonumber\\
  &+& \alpha_{b}^{(0)}\Big(\frac{\cal F}{\Lambda}\Big)^{|f_b|}(Y^{(6)}_{\bf3}{\cal D}^c)_{{\bf 1}''}Q_{3}H_d+\alpha_{s}^{(0)}\Big(\frac{\cal F}{\Lambda}\Big)^{|f_s|}(Y^{(6)}_{\bf3}{\cal D}^c)_{{\bf 1}'}Q_{2}H_d\nonumber\\
  &+&\alpha_{d}^{(0)}\Big(\frac{\cal F}{\Lambda}\Big)^{|f_d|}(Y^{(6)}_{\bf3}{\cal D}^c)_{{\bf 1}}Q_{1}H_d+W^{(h)}_q
\,,
 \label{lagrangian_q}
 \end{eqnarray}}
where $\alpha_i^{(0)}$ denotes coefficient at leading order and $W^{(h)}_q$ stand for higher order contributions, which are simply contructed by the leadging order operators in Eq.(\ref{lagrangian_q}) multiplied by
$\sum^{\infty}_{n=1}\Big(\frac{\chi\tilde{\chi}}{\Lambda^{2}}\Big)^n$.
 Note that all Yukawa coefficients in the above suprpotential, $\alpha_i^{(0)}$, are assumed to be complex numbers with an absolute value of unity.
Since it is hard to reproduce the experimental data of fermion masses and mixing with Yukawa terms constructed with modular forms of weight 4 in quark and charged-lepton sectors in this model, we  take into account Yukawa terms with modular forms of weight 6 which are decomposed
as ${\bf1}\oplus{\bf 3}\oplus{\bf 3}$ under $A_4$ given explicitly by \cite{Feruglio:2017spp}
{\begin{eqnarray}
  &&Y^{(6)}_{\bf 1}=Y^3_1+Y^3_2+Y^3_3-3Y_1Y_2Y_3\,\nonumber\\
  &&Y^{(6)}_{{\bf 3},1}=(Y^3_1+2Y_1Y_2Y_3, Y^2_1Y_2+2Y^2_2Y_3, Y^2_1Y_3+2Y^2_3Y_2)\nonumber\\
  &&Y^{(6)}_{{\bf 3},2}=(Y^3_3+2Y_1Y_2Y_3, Y^2_2Y_1+2Y^2_1Y_2, Y^2_3Y_2+2Y^2_2Y_1)\,.
\label{modu6}
 \end{eqnarray}}
 In the above superpotential only the top quark operator is renormalizable and does not contain a modular form, leading to the top quark mass as the pole mass,
while the other quark operators driven by $\chi$ (or $\tilde{\chi})$ are dependent on modular forms. Using modular forms of weight 6, $Y^{(6)}_{\bf 1}$ and $Y^{(6)}_{\bf 3}$, with the quark fields charged under $A_4\times U(1)_X$, which does not allow mixing among up-type quarks, the off-diagonal entries in the up-type quark mass matrix are forbidden, as indicated in Eq.(\ref{Ch2}).
 From the above superpotential the effective Yukawa couplings of quarks can be visualized as functions of the SM gauge-singlet fields $\chi(\tilde{\chi})$ and modular forms $Y^{(6)}_{\bf 1(3)}$, except for the top Yukawa coupling (see the details given in Sec.\,\ref{visu}).
 
 According to the quantum numbers of the quark sectors as in Table-\ref{reps_q}, the color anomaly coefficient of $U(1)_{X}\times[SU(3)_C]^2$ defined as $N_C\equiv 2{\rm Tr}[X_\psi T^2_{SU(3)_C}]$ reads
 \begin{eqnarray}
  N_C=-(f_u+f_c+f_d+f_s+f_b)\,.
 \label{dGi}
 \end{eqnarray}
  Note that $U(n)$ generators ($n\geq2$) are normalized according to ${\rm Tr}[T^aT^b]=\delta^{ab}/2$. The $U(1)_X$ is broken down to its discrete subgroup $Z_{N_{\rm DW}}$ in the backgrounds of QCD instanton, and the quantity $N_C$ (non-zero integer) is given by the axionic domain-wall number $N_{\rm DW}$.
 At the QCD phase transition, each axionic string becomes the edge to $N_{\rm DW}$ domain-walls, and the process
of axion radiation stops. To avoid the domain-wall problem one should consider $N_{\rm DW}=1$ or the PQ phase transition occurred during (or before) inflation for $N_{\rm DW}>1$.

 Next, we turn to the lepton sector, where the fields are charged under $G_{\rm SM}\times A_4\times U(1)_X$ with modular weight $k_I$. 
 Remark that the sterile neutrinos $N^c$ (which interact with gravity) are introduced {\bf(i)} to solve the anomaly-free condition of $U(1)\times[gravity]^2$, {\bf (ii)} to explain the small active neutrino masses via the seesaw mechanism, and {\bf (iii)} to provide a theoretically well-motivated PQ symmetry breaking scale. 
\begin{table}[h]
\caption{\label{reps_l1} Representations and quantum numbers of the lepton fields under $G_{\rm SM}\times A_4\times U(1)_{X}$ and modular weight $k_I$ determined according to Eq.(\ref{req}).}
\begin{ruledtabular}
\begin{tabular}{cccccccc}
Field &$L_e$&$L_\mu$&$L_\tau$&$e^c$&$\mu^c$ & $\tau^c$&$N^c$\\
\hline
$G_{\rm SM}$&$(1,2)_{-1/2}$&$(1,2)_{-1/2}$&$(1,2)_{-1/2}$&$(1,1)_1$&$(1,1)_1$&$(1,1)_1$&$(1,1)_0$\\
$A_4$&$\mathbf{1}$&$\mathbf{1}'$&$\mathbf{1}''$&$\mathbf{1}$ & $\mathbf{1^{\prime\prime}}$ & $\mathbf{1^\prime}$&$\mathbf{3}$\\
$k_I$& $\frac{5}{2}$ & $\frac{5}{2}$ & $\frac{5}{2}$ & $-\frac{11}{2}$ & $-\frac{11}{2}$ & $-\frac{11}{2}$ & $-\frac{3}{2}$ \\
$U(1)_{X}$& $\frac{1}{2}-g_e$ & $\frac{1}{2}-g_\mu$  & $\frac{1}{2}-g_\tau$  & $g_e-\frac{1}{2}-f_e$  & $g_\mu-\frac{1}{2}-f_\mu$ & $g_\tau-\frac{1}{2}-f_\tau$ & $-\frac{1}{2}$
\end{tabular}
\end{ruledtabular}
\end{table}
In Table \ref{reps_l1}, the representations and quantum numbers of the lepton fields as well as modular weight
$k_f$ determined along with Eq.(\ref{req}) are presented.
Here, $L_e$, $L_\mu$, and $L_\tau$ denote $SU(2)_L$ lepton doublets and $e^c$, $\mu^c$, and $\tau^c$
are three charged lepton singlets. The field $N^c$ represents the right-handed $SU(2)_L$ singlet neutrino, which
is introduced to generate active neutrino masses via canonical seesaw mechanism \cite{Minkowski:1977sc}.

We note that mixing between different charged-leptons does not occur when
lepton Yukawa superpotential is economically constructed with modular forms of weight 6 prevents, which leads to 
the diagonal form of the charged lepton mass matrix as can be seen in
 Eq.(\ref{ChL1})).
In contrast, modular forms $Y^{(2)}_{\bf 3}$, $Y^{(6)}_{\bf 1}$, and $Y^{(6)}_{\bf 3}$ are used
to construct neutrino mass matrices 
\footnote{By selecting appropriate modular weight of particle contents, lower weight modular forms can be used, such as (i) $Y^{(2)}_{\bf 3}$ in the Dirac neutrino sector, and $Y^{(4)}_{\bf 1(1', 1'')}$ and $Y^{(4)}_{\bf 3}$ in the Majorana neutrino sector. However, this leads to additional interactions, including $\frac{1}{2}Y^{(4)}_{\bf 1}(N^cN^c)_{\bf 1}\chi$, $\frac{1}{2}Y^{(4)}_{{\bf 1}'}(N^cN^c)_{{\bf 1}''}\chi$, $\frac{1}{2}Y^{(4)}_{{\bf 1}''}(N^cN^c)_{{\bf 1}'}\chi$, and $\frac{1}{2}Y^{(4)}_{\bf 3}(N^cN^c)_{\bf 3}\chi$. (ii) Another option is to use $Y^{(2)}_{\bf 3}$ in the Dirac neutrino sector and no modular form in the Majorana neutrino sector, which results in only $\frac{1}{2}(N^cN^c)_{\bf 1}\chi$ and degenerate heavy Majorana neutrino mass states at the seesaw scale. However, we have found that this approach is difficult to reconcile with experimental neutrino data.} 
%
Then the Yukawa superpotential for lepton invariant under $G_{\rm SM}\times A_4\times U(1)_X$ with economic modular forms are sewed with ${\cal F}=\{\chi$ or $\tilde{\chi}\}$ through Eq.(\ref{AFN}), respectively, as
\begin{eqnarray}
 W_{\ell\nu} &=&
\alpha_\tau^{(0)}\Big(\frac{\cal F}{\Lambda}\Big)^{|f_\tau|}Y^{(6)}_{\bf 1}\tau^c L_\tau H_d+\alpha_\mu^{(0)}\Big(\frac{\cal F}{\Lambda}\Big)^{|f_\mu|}Y^{(6)}_{\bf 1} \mu^c L_\mu H_d+\alpha_e^{(0)}\Big(\frac{\cal F}{\Lambda}\Big)^{|f_e|}Y^{(6)}_{\bf 1}e^c L_e H_d\nonumber\\
 &+& \beta_1^{(0)}\Big(\frac{\cal F}{\Lambda}\Big)^{|g_e|}(Y^{(2)}_{\bf 3}N^c)_{{\bf 1}}L_eH_{u}+ \beta_2^{(0)}\Big(\frac{\cal F}{\Lambda}\Big)^{|g_\mu|}(Y^{(2)}_{\bf 3}N^c)_{{\bf 1}''}L_\mu H_{u}\nonumber\\
 &+& \beta_3^{(0)}\Big(\frac{\cal F}{\Lambda}\Big)^{|g_\tau|}(Y^{(2)}_{\bf 3}N^c)_{{\bf 1}'}L_\tau H_{u}
 +\gamma_{1}^{(0)}\frac{1}{2}Y^{(6)}_{\bf 1}(N^cN^c)_{{\bf 1}}\chi+\gamma_{2}^{(0)}\frac{1}{2}Y^{(6)}_{\bf 3}(N^cN^c)_{{\bf 3}}\chi+W^{(h)}_{l\nu},
 \label{lagrangian_l}
 \end{eqnarray}
where $\alpha_i^{(0)},\beta_i^{(0)}, \gamma_i^{(0)}$ denote coefficients at leading order and $W^{(h)}_{l\nu}$ stands for higher order contributions triggered by the combination $\chi\tilde{\chi}$. Like in the quark sector, the Yukawa coefficients in the above superpotential, such as $\alpha_i^{(0)},\beta_i^{(0)}, \gamma_i^{(0)}$, are assumed to be complex numbers with an absolute value of unity.
In the above superpotential, the charged-lepton and Dirac neutrino parts have three distinct Yukawa terms each, with their common modular forms being $Y^{(6)}_{\bf 1}$ and $Y^{(2)}_{\bf 3}$, respectively. Each term involves ${\cal F}/\Lambda$ to the power of an appropriate $U(1)_X$ quantum number. The flavored $U(1)_X$ PQ symmetry allows for two renormalizable terms for the right-handed neutrino $N^c$, which implement the seesaw mechanism\,\cite{Minkowski:1977sc} by making the VEV $\langle\chi\rangle$ large.
 The details on how the active neutrino masses and mixing are predicted will be presented in Sec.\,\ref{nu_sec}.
 
Nonperturbative quantum gravitational anomaly effects\,\cite{Kamionkowski:1992mf} violate the conservation of the corresponding current, $\partial_\mu J^{\mu}_X\propto R\tilde{R}$, where $R$ is the Riemann tensor and $\tilde{R}$ is its dual, and make the axion solution to the strong CP problem problematic. To consistently couple gravity to matter charged under $U(1)_X$, the mixed-gravitational anomaly $U(1)_X\times[{\rm gravity}]^2$ (related to the color anomaly $U(1)_X\times[SU(3)_C]^2$) must be cancelled, as shown in Refs.\,\cite{Ahn:2016hbn, Ahn:2018cau, Ahn:2021ndu}, which leads to the relation,
 \begin{eqnarray}
3N_C=f_e+f_\mu+f_\tau+g_e+g_\mu+g_\tau\,.
 \label{uyxf}
 \end{eqnarray}
Thus the choice of $U(1)_X$ charge for ordinary quarks and leptons is strictly restricted.
 
Below the $U(1)_X$ symmetry breaking scale (here, equivalent to the seesaw scale) the effective interactions of QCD axion with the weak and hypercharge gauge bosons and with the photon are expressed through the chiral rotation of Eq.(\ref{X-tr}), respectively, as
\begin{eqnarray}
{\cal L}^{WY}_{A}&=&\frac{A_X}{f_A}\frac{1}{32\pi^2}\Big\{g^2_W\,N_W\,W^{\mu\nu}\tilde{W}_{\mu\nu}+g^2_Y\,N_Y\,Y^{\mu\nu}\tilde{Y}_{\mu\nu}\Big\}\,,\label{}\\
{\cal L}^{\gamma}_{A}&=&\frac{A_X}{f_A}\frac{e^2}{32\pi^2}\,E\,F^{\mu\nu}\tilde{F}_{\mu\nu}\,,
\label{ema}
\end{eqnarray}
where $g_{W}$, $g_Y$, and $e$ stand for the gauge coupling constant of $SU(2)_L$, $U(1)_Y$, and $U(1)_{EM}$, respectively; their corresponding gauge field strengths $W^{\mu\nu}, Y^{\mu\nu}$, and $F^{\mu\nu}$ with their dual forms $\tilde{W}_{\mu\nu}, \tilde{Y}_{\mu\nu}$, and $\tilde{F}_{\mu\nu}$, respectively. Here $N_W\equiv2{\rm Tr}[X_{\psi_f}T^2_{SU(2)}]$ and $N_Y\equiv2{\rm Tr}[X_{\psi_f}(Q_f^Y)^2]$ are the anomaly coefficients of $U(1)_{X}\times[SU(2)_L]^2$ and $U(1)_{X}\times[U(1)_Y]^2$, respectively. And the electromagnetic anomaly coefficient $E$ of $U(1)_{X}\times[U(1)_{EM}]^2$ defined by $E=2\sum_{\psi_f} X_{\psi_f}(Q^{\rm em}_{\psi_f})^2$ with $Q^{\rm em}_{\psi_f}$ being the $U(1)_{\rm EM}$ charge of field $\psi_f$ is expressed as 
  \begin{eqnarray}
  E&=&N_W+N_Y=-2(f_e+f_\mu+f_\tau)-\frac{2}{3}(4f_u+4f_c+f_d+f_s+f_b)\,.
 \label{eano}
 \end{eqnarray}
The physical quantities of QCD axion, such as axion mass $m_a$ and axion-photon coupling $g_{a\gamma\gamma}$, are dependent on the ratio of electromagnetic anomaly coefficient $E$ to color one $N_C$. The value of $E/N_C$ is determined in terms of the $X$-charges for quarks and leptons by  the relation,
  \begin{eqnarray}
  \frac{E}{N_C}&=&\frac{2(f_e+f_\mu+f_\tau)+\frac{2}{3}(4f_u+4f_c+f_d+f_s+f_b)}{f_u+f_c+f_d+f_s+f_b}\,\nonumber\\
  &=&\frac{6(f_e+f_\mu+f_\tau)+2(4f_u+4f_c+f_d+f_s+f_b)}{-f_e-f_\mu-f_\tau-g_e-g_\mu-g_\tau}\,,
 \label{eano}
 \end{eqnarray}
where the first and second equality follow from Eqs.(\ref{dGi}) and (\ref{uyxf}), respectively. Our model with a specific value of $E/N_C$ can be tested by ongoing experiments such as  KLASH\,\cite{Alesini:2019nzq} and FLASH\,\cite{Gatti:2021cel}( see Eq.(\ref{gagg}) and Fig.\ref{Fig1} and \ref{Fig2})  by considering the scale of $U(1)_X$ breakdown induced by Eq.(\ref{sof2}).

 Compared to conventional $A_4$ symmetry models resulting in tribimaximal\,\cite{Ma:2001dn} or nearly tribimaximal\,\cite{Ahn:2012cg} mixing in the neutrino sector, the modular invariant model 
leads to neutrino mixing without the need for special breaking patterns and the introduction of multiple scalar fields. Our model can be {\it uniquely} realized for quark sector by assigning  $A_4\times U(1)_X$ quantum numbers to matter fields with appropriate modular forms based on Eq.(\ref{AFN}). Some comments are worth noting.
(i) By selecting the appropriate modular weight for the right-handed down-type quark fields, it is possible to construct
down-type quark Yukawa superpotential with  lower modular weight forms $Y^{(2)}_{\bf 3}$ or $Y^{(4)}_{\bf 3}$  while keeping the same up-type quark Yukawa superpotential given in Eq.(\ref{lagrangian_q}). However, it is hard to reproduce the experimental data for quark masses and mixing hierarchies in this way due to the limited number of parameters.
(ii) Unlike the case in Table-\ref{reps_q}, the quark $SU(2)_L$ doublets and singlets can be assigned to $A_4$ triplets and singlets by choosing appropriate modular weight forms and $U(1)_X$ quantum numbers, respectively.
 In this case, the quark mass hierarchies can be realized in the limit of $\langle\tau\rangle=i\infty$ ({\it i.e.} $Y_1\rightarrow1$, $Y_2\rightarrow0$, and $Y_3\rightarrow0$), whereas it is hard to reproduce the CKM mixing angles since additive correction terms induced by higher weight modular forms are forbidden by the modular weight zero of $\chi(\tilde{\chi})$ fields.
(iii) In the opposite scenario where the quark $SU(2)_L$ doublets and singlets are assigned to $A_4$ singlets and triplets, respectively, it is not possible to account for the observed quark mass hierarchy due to the charge assignment of $U(1)_X$.
(iv) For leptons,  unlike the case in Table-\ref{reps_l1}, the left-handed charged lepton $SU(2)_{L}$ doublets $L$ can be assigned to the $A_4$ triplet and their $U(1)_X$ quantum numbers are taken to be $ \frac{1}{2}-g_l$, whereas $SU(2)_L$ singlets ($e^c$, $\mu^c$, $\tau^c$) are assigned to the $A_4$ singlets ({\bf 1},{\bf 1$^{''}$},{\bf 1$^{'}$})  and $U(1)_X$ quantum numbers are taken to be $(g_l-f_e-\frac{1}{2}, g_l-f_\mu-\frac{1}{2}, g_l-f_\tau-\frac{1}{2})$.
To generate neutrino mass through the seesaw mechanism, $N^c$ is assigned to the $A_4$ triplet, and
$U(1)_X$ quantum number is taken to be $ -\frac{1}{2}$. 
In this case, we have the freedom to select the weights. For instance, we can choose the following weights: $k_L=\frac{5}{2}$, $k_{e^c}=k_{\mu^c}=k_{\tau^c}=-\frac{3}{2}$, and $k_{N^c}=\frac{1}{2}$.
Then the lepton Yukawa superpotential reads
\begin{eqnarray}
 W_{\ell\nu} &=&
\Big[\alpha_\tau^{(0)}\Big(\frac{\cal F}{\Lambda}\Big)^{|f_\tau|}(Y^{(2)}_{\bf 3} L)_{{\bf 1}''}\tau^c+\alpha_\mu^{(0)}\Big(\frac{\cal F}{\Lambda}\Big)^{|f_\mu|}(Y^{(2)}_{\bf 3} L)_{{\bf 1}'} \mu^c+\alpha_e^{(0)}\Big(\frac{\cal F}{\Lambda}\Big)^{|f_e|}(Y^{(2)}_{\bf 3} L)_{\bf 1}e^c\Big] H_d\nonumber\\
 &+& \beta^{(0)}\Big(\frac{\cal F}{\Lambda}\Big)^{|g_l|}(N^c L)_{{\bf 1}}H_{u}
 +\gamma^{(0)}\frac{1}{2}(Y^{(2)}_{\bf 3}N^cN^c)_{{\bf 1}}\chi+...\,,
 \label{lagrangian_l1}
 \end{eqnarray}
where dots stand for higher order contributions triggered by the combination $\chi\tilde{\chi}$. It is worth noting that the above superpotential enables mixing between different charged leptons, analogous to the down-type quark sector. Additionally, the Dirac neutrino Yukawa matrix, denoted as $m_D$, exhibits a proportional relationship to $m_D^\dag m_D \propto (1,1,1)$, and the heavy Majorana neutrino mass term follows a similar form, as found in Ref.\cite{Feruglio:2017spp}.
By selecting another specific weights, namely $k_L=\frac{9}{2}$, $k_{e^c}=k_{\mu^c}=k_{\tau^c}=-\frac{7}{2}$, and $k_{N^c}=-\frac{3}{2}$, a notable change occurs in the modular form of the Majorana neutrino operator. Specifically, the term $Y^{(2)}_{\bf 3}$ transforms into $Y^{(6)}_{\bf 1(3)}$, resulting in an expression that aligns with the form presented in Eq.(\ref{lagrangian_l}).
While these cases show potential for reproducing lepton mass and mixing, further investigation is necessary to confirm its viability.
(v) However, the assignment of the right-handed charged lepton $E^c=(e^c,\mu^c,\tau^c)$ to the $A_4$ triplet, the left-handed charged lepton denoted as $L_{e,\mu,\tau}$ to $A_4$ singlets ({\bf 1},{\bf 1$^{''}$},{\bf 1$^{'}$}), and $N^c$ to the $A_4$ triplet may not provide an explanation for the observed charged-lepton mass hierarchy. This difficulty arises from the charge assignment of $U(1)_X$.

\section{Quark and Lepton interactions with QCD axion}
\label{visu}
Let us discuss how quark and lepton masses and mixings are derived from Yukawa interactions within a framework based on $A_4\times U(1)_X$ symmetries with modular invariance. 
Non-zero VEVs of scalar fields $\chi(\tilde{\chi})$ spontaneously break  the flavor symmetry $U(1)_{X}$ \,\footnote{If the symmetry $U(1)_{X}$ is broken spontaneously, the massless mode $A_X$ of the scalar $\chi$ appear as a phase.} at high energies above EW scale and create a heavy Majorana neutrino mass term.
Then, the effective Yukawa structures in the low-energy limit depend on a small dimensionless parameter $\langle {\cal F}\rangle/\Lambda\equiv \Delta_{\cal F}$.
The higher order contributions of 
superpotentials $W^{(h)}_{q (l\nu)}$ become
 $\sum^{\infty}_{n=1}\tilde{c}_i \,\Delta^{2n}_\chi\cdot${\it (leading order operators)} 
with $\tilde{c}_i=e^{i\tilde{\theta}_i}$, which make
the Yukawa coefficients of the leading order terms in the superpotentials given in Eqs.(\ref{super_d}, \ref{lagrangian_q}, \ref{lagrangian_l}) shifted.
Denoting the effective Yukawa coefficients shifted by higher order contributions as $\alpha_i,\beta_i,\gamma_i$, we see that they are constrained as 
\begin{eqnarray}
1-\frac{\Delta^2_\chi}{1-\Delta^2_\chi}\leq|\alpha_i|,|\beta_|,|\gamma_i|\leq1+\frac{\Delta^2_\chi}{1-\Delta^2_\chi}\qquad\text{with}~\Delta_\chi\equiv \frac{v_\chi}{\sqrt{2}\,\Lambda},
  \label{AFN1}
\end{eqnarray}
where the lower (upper) limit corresponds to the sum of higher order terms
with $\tilde{\theta}_i=\pi (0)$.
When $H_{u(d)}$ acquire non-zero VEVs, all quarks and leptons obtain masses.
The relevant quark and lepton interactions with their chiral fermions are given by 
 \begin{eqnarray}
  -{\cal L} &\supset&
  \overline{q^{u}_{R}}\,\mathcal{M}_{u}\,q^{u}_{L}+\overline{q^{d}_{R}}\,\mathcal{M}_{d}\,q^{d}_{L}+\frac{g}{\sqrt{2}}W^+_\mu\overline{q^u_L}\gamma^\mu\,q^d_L \nonumber\\
  &+&\overline{\ell_{R}}\,{\cal M}_{\ell}\,\ell_{L}+ \frac{1}{2} \begin{pmatrix} \overline{\nu^c_L} & \overline{N_R} \end{pmatrix} \begin{pmatrix} 0 & m^T_{D}  \\ m_{D} & ~~~ e^{i\frac{A_X}{u_{\chi}}}\,M_R  \end{pmatrix} \begin{pmatrix} \nu_L \\ N^c_R  \end{pmatrix}+\frac{g}{\sqrt{2}}W^-_\mu\overline{\ell_L}\gamma^\mu\,\nu_L+\text{h.c.}\,,
  \label{AxionLag2}
 \end{eqnarray}
where $g$ is the $SU(2)_L$ coupling constant, $q^{u}=(u,c,t)$, $q^{d}=(d,s,b)$, $\ell=(e, \mu, \tau)$, $\nu=(\nu_e,\nu_\mu,\nu_\tau)$, and $N=(N_1,N_2, N_3)$. $M_R$ contains a VEV of $\chi $ presented by
Eq.(\ref{NGboson}). The explicit forms of $\mathcal{M}_{u,d,l}$ will be given
later.
The above Lagrangian of the fermions, including their kinetic terms of Eq.(\ref{lag0}), should be invaruant under $U(1)_X$:
 \begin{eqnarray}
  \psi_f\rightarrow e^{if_{\psi_f}\frac{\gamma_5}{2}\alpha}\psi_f\,,\quad t=\text{invariant}\,,\quad N\rightarrow e^{i\frac{\gamma_5}{2}\alpha}N
 \label{X-tr}
 \end{eqnarray}
where $\psi_f=\{u,c,d,s,b,e,\mu,\tau\}$ and $\alpha$ is a transformation constant parameter.

\subsection{Quark and flavored-QCD axion}
As axion models, the axion-Yukawa coupling matrices and quark mass matrices in our model can be aresimultaneously diagonalized.
The quark mass matrices are diagonalized
through bi-unitary transformations: $V^{\psi}_R{\cal M}_{\psi}V^{\psi\dag}_L=\hat{{\cal M}}_\psi$ (diagonal form) and the mass eigenstates are $\psi'_R=V^\psi_R\,\psi_R$ and $\psi'_L=V^\psi_L\,\psi_L$. These transformation include, in particular, the chiral transformation of Eq.(\ref{X-tr}) that  necessarily  makes ${\cal M}_{u,d,\ell}$ real and positive. 
This induces a contribution to the QCD vacuum angle in Eq.(\ref{lag0}), {\it i.e.}, 
 \begin{eqnarray}
  \vartheta_{\rm QCD}\rightarrow\vartheta_{\rm eff}=\vartheta_{\rm QCD}+\arg\{\det({\cal M}_u)\det({\cal M}_d)\}
 \label{}
 \end{eqnarray}
with $-\pi\leq\vartheta_{\rm eff}\leq\pi$. Then one obtains the vanishing QCD anomaly term
 \begin{eqnarray}
 {\cal L}_\vartheta=\Big(\vartheta_{\rm eff}+\frac{A_X}{F_a}\Big)\frac{\alpha_s^{\prime}}{8\pi}G^{a\mu\nu}\tilde{G}^a_{\mu\nu}\qquad\text{with}~F_a=\frac{f_a}{N_C}\,,
 \label{}
 \end{eqnarray}
where $\alpha_s^{\prime}=g^2_s/4\pi$ and the axion decay constant $F_a$ with $f_a=u_\chi$ of Eq.(\ref{NGboson}). At low energies $A_X$ will get a VEV, $\langle A_X\rangle=-F_a\vartheta_{\rm eff}$, eliminating the constant $\vartheta_{\rm eff}$ term. The QCD axion then is the excitation of the $A_X$ field, $a=A_X-\langle A_X\rangle$.

Substituting the VEV of  Eq.(\ref{vev}) into the superpotential Eq.(\ref{lagrangian_q}), the mass matrices ${\cal M}_{u}$ and ${\cal M}_{d}$ for up- and down-type quarks given in the Lagrangian\,(\ref{AxionLag2}) are derived as,
 \begin{eqnarray}
  &{\cal M}_{u}={\left(\begin{array}{ccc}
 \alpha_u\Delta^{|f_u|}_\chi\,Y^{(6)}_{\bf 1}e^{if_u\frac{A_X}{u_\chi}} &  0 &  0 \\
 0  &  \alpha_c\Delta^{|f_c|}_\chi\,Y^{(6)}_{\bf 1}e^{if_c\frac{A_X}{u_\chi}}  &  0   \\
 0  &  0  & \alpha_t
 \end{array}\right)}\,v_u\,,
 \label{Ch2}\\
 &{\cal M}_{d}=\Big[{\left(\begin{array}{ccc}
 \alpha_d\Delta^{|f_d|}_\chi &  \alpha_s\,x\Delta^{|f_s|}_\chi &  \alpha_b\,y\Delta^{|f_b|}_\chi \\
 \alpha_d\,y\Delta^{|f_d|}_\chi &  \alpha_s\Delta^{|f_s|}_\chi &  \alpha_b\,x\Delta^{|f_b|}_\chi   \\
 \alpha_d\,x\Delta^{|f_d|}_\chi &  \alpha_s\,y\Delta^{|f_s|}_\chi  &  \alpha_b\Delta^{|f_b|}_\chi
 \end{array}\right)}(1+2xy)\nonumber\\
 &\qquad\qquad\qquad\qquad+{\left(\begin{array}{ccc}
 \tilde{\alpha}_d\,y\Delta^{|f_d|}_\chi &  \tilde{\alpha}_s\Delta^{|f_s|}_\chi &  \tilde{\alpha}_b\,x\Delta^{|f_b|}_\chi \\
 \tilde{\alpha}_d\,x\Delta^{|f_d|}_\chi &  \tilde{\alpha}_s\,y\Delta^{|f_s|}_\chi &  \tilde{\alpha}_b\Delta^{|f_b|}_\chi   \\
 \tilde{\alpha}_d\Delta^{|f_d|}_\chi &  \tilde{\alpha}_s\,x\Delta^{|f_s|}_\chi  &  \tilde{\alpha}_b\,y\Delta^{|f_b|}_\chi
 \end{array}\right)}(y^2+2x)\Big]\tilde{C}\,Y^3_1\,v_d\,,  
 \label{Ch1}
 \end{eqnarray}
where $v_{d}\equiv\langle H_{d}\rangle=v\cos\beta/\sqrt{2}$, $v_{u}\equiv\langle H_{u}\rangle =v\sin\beta/\sqrt{2}$ with $v\simeq246$ GeV, and 
 \begin{eqnarray}
 \tilde{C}={\rm diag.}\big(e^{if_d\frac{A_X}{u_{\chi}}} , e^{if_s\frac{A_X}{u_{\chi}}},
e^{if_b\frac{A_X}{u_{\chi}}}\big)\,, \quad x=\frac{Y_2}{Y_1}\,,\quad y=\frac{Y_3}{Y_1}\,.
 \label{c_num}
 \end{eqnarray}
The terms with $\alpha_{d, s, b}$ in Eq.(\ref{Ch1}) generate  by taking the modular form $Y^{(6)}_{{\bf 3},1}$ given in Eq.(\ref{modu6}), whereas the terms with $\tilde{\alpha}_{d, s, b}$ in Eq.(\ref{Ch1}) generate by taking $Y^{(6)}_{{\bf 3},2}$.

The quark mass matrices ${\cal M}_{u}$ in Eq.(\ref{Ch2}) and ${\cal M}_{d}$ in Eq.(\ref{Ch1}) generate the up- and down-type quark masses:
 \begin{eqnarray}
 \hat{\mathcal{M}}_{u}=V^{u}_{R}\,{\cal M}_{u}\,V^{u\dag}_{L}
 ={\rm diag}(m_{u},m_{c},m_{t})\,,\quad \hat{\mathcal{M}}_{d}=V^{d}_{R}\,{\cal M}_{d}\,V^{d\dag}_{L}
 ={\rm diag}(m_{d},m_{s},m_{b})\,.
 \label{Quark21}
 \end{eqnarray}
Diagonalizing the matrices ${\cal M}_f^\dagger {\cal M}_f$ and ${\cal M}_f {\cal M}_f^\dagger$ ($f=u,d$) determine the mixing matrices $V_L^f$ and $V_R^f$, respectively \cite{Ahn:2011yj}. The left-handed quark mixing matrices $V_L^u$ and $V_L^d$ are components of the CKM matrix $V_{\rm CKM}=V_L^u V_L^{d\dagger}$, which is generated from the down-type quark matrix in Eq.(\ref{Ch1}) due to the diagonal form of the up-type quark mass matrix in Eq.(\ref{Ch2}). The CKM matrix is parameterized by the Wolfenstein parametrization\,\cite{Wolfenstein:1983yz}, see Eq.(\ref{ckm0}), and has been determined with high precision\,\cite{Ahn:2011fg}. The current best-fit values of the CKM mixing angles in the standard parameterization\,\cite{Chau:1984fp} read in the $3\sigma$ range\,\cite{ckm}
 \begin{eqnarray}
  \theta^q_{23}[^\circ]=2.376^{+0.054}_{-0.070}\,,\quad\theta^q_{13}[^\circ]=0.210^{+0.016}_{-0.010}\,,\quad\theta^q_{12}[^\circ]=13.003^{+0.048}_{-0.036}\,,\quad\delta^q_{CP}[^\circ]=65.5^{+3.1}_{-4.9}\,.
 \label{ckmmixing}
 \end{eqnarray} 
The physical structure of the up- and down-type quark Lagrangian should match up with the empirical results calculated from the Particle Data Group (PDG)\,\cite{PDG}:
 \begin{eqnarray}
 m_d&=&4.67^{+0.48}_{-0.17}\,{\rm MeV}\,,\qquad~m_s=93^{+11}_{-5}\,{\rm MeV}\,,\qquad\qquad~~m_b=4.18^{+0.03}_{-0.02}\,{\rm GeV}\,,  \nonumber\\
 m_u&=&2.16^{+0.49}_{-0.29}\,{\rm MeV}\,,\qquad~ m_c=1.27\pm0.02\,{\rm GeV}\,,\qquad~ m_t=173.1\pm0.9\,{\rm GeV}\,,
\label{qumas}
 \end{eqnarray}
where $t$-quark mass is the pole mass, $c$- and $b$-quark masses are the running masses in the $\overline{\rm MS}$ scheme, and the light $u$-, $d$-, $s$-quark masses are the current quark masses in the $\overline{\rm MS}$ scheme at the momentum scale $\mu\approx2$ GeV. Below the scale of spontaneous $SU(2)_L\times U(1)_Y$ gauge symmetry breaking, the running masses of $c$- and $b$-quark receive corrections from QCD and QED loops\,\cite{PDG}. The top quark mass at scales below the pole mass is unphysical since the $t$-quark decouples at its scale, and its mass is determined more directly by experiments\,\cite{PDG}.

After diagonalizing the mass matrices of Eqs.(\ref{Ch2}, \ref{Ch1}), the flavored-QCD axion to quark interactions are written at leading order as 
\begin{eqnarray}
-{\cal L}^{aq}&\simeq&\frac{\partial_\mu a}{2u_{\chi}}\Big\{f_u\,\bar{u}\gamma^\mu\gamma_5 u+f_c\,\bar{c}\gamma^\mu\gamma_5c+f_d\,\bar{d}\gamma^\mu\gamma_5 d+f_s\,\bar{s}\gamma^\mu\gamma_5s+f_b\,\bar{b}\gamma^\mu\gamma_5b\Big\}\nonumber\\
&+&\frac{\partial_\mu a}{2u_{\chi}}\Big\{(f_d-f_s)\lambda\Big(1-\frac{\lambda^2}{2}\Big)\bar{d}\gamma^\mu\gamma_5 s+(f_s-f_b)A_d\lambda^2\,\bar{s}\gamma^\mu\gamma_5 b\nonumber\\
&&\qquad\qquad+A_d\lambda^3\Big(f_d(\rho+i\eta)-f_s+f_b(1-\rho-i\eta)\Big)\bar{b}\gamma^\mu\gamma_5 d+{\rm h.c.}\Big\}\nonumber\\
&+&i\frac{a}{2u_{\chi}}\Big\{(f_d-f_s)\lambda\Big(1-\frac{\lambda^2}{2}\Big)(m_d-m_s)\,\bar{d}s+(f_s-f_b)A_d\lambda^2(m_s-m_b)\,\bar{s}b\nonumber\\
&&\qquad\qquad+A_d\lambda^3\Big(f_d(\rho+i\eta)-f_s+f_b(1-\rho-i\eta)\Big)(m_b-m_d)\,\bar{b}d+{\rm h.c.}\Big\}\nonumber\\
&+&m_u\,\bar{u}u+m_c\,\bar{c}c+m_t\,\bar{t}t+m_d\,\bar{d}d+m_s\,\bar{s}s+m_b\,\bar{b}b-\bar{q}i\! \! \not\! D\, q\,,
\label{fl_in}
\end{eqnarray}
where $V^{d\dag}_L= V_{\rm CKM}$ of Eq.(\ref{ckm0}) is used by rotating the phases in ${\cal M}_u$ away, which is the result of a direct interaction of the SM gauge singlet scalar field $\chi$
with the SM quarks  charged under $U(1)_X$.
The flavored-QCD axion $a$ is produced by flavor-changing neutral Yukawa interactions in Eq.(\ref{fl_in}), which leads to induced rare flavor-changing processes. The strongest bound on the QCD axion decay constant is from the flavor-changing process $K^+\rightarrow\pi^++a$\,\cite{Wilczek:1982rv, Bolton:1988af, Artamonov:2008qb, raredecay, Berezhiani:1989fp}, induced by the flavored-QCD axion $a$.
From Eq.(\ref{fl_in}) the flavored-QCD axion interactions with the flavor violating coupling to the $s$- and $d$-quark are given by
\begin{eqnarray}
-{\cal L}^{asd}_Y\simeq \frac{i}{2}(f_d-f_s)\frac{a}{N_CF_a}\bar{s}d\,(m_d-m_s)\lambda\Big(1-\frac{\lambda^2}{2}\Big)\,.
  \label{}
\end{eqnarray}
 Then the decay width of $K^+\rightarrow\pi^++a$ is given by
 \begin{eqnarray}
   \Gamma(K^+\rightarrow\pi^++a)=\frac{m^3_K}{16\pi}\Big(1-\frac{m^2_{\pi}}{m^2_{K}}\Big)^3\Big|\frac{f_d-f_s}{2\,F_{a}N_C}\lambda\Big(1-\frac{\lambda^2}{2}\Big)\Big|^2\,,
  \label{Gkp}
 \end{eqnarray}
 where $m_{K^{\pm}}=493.677\pm0.013$ MeV, $m_{\pi^{\pm}}=139.57061\pm0.00024$ MeV\,\cite{PDG}.
From the present experimental upper bound ${\rm Br}(K^+\rightarrow\pi^+a)<(3-6)\times10^{-11} (1\times 10^{-11})$ for $m_{a}=0-110$ (160-260) MeV at $90\%$ CL with ${\rm Br}(K^+\rightarrow\pi^+\nu\bar{\nu})=(10.6^{+4.0}_{-3.4}|_{\rm stat}\pm0.9_{\rm syst})\times10^{-11}$ at $68\%$CL\,\cite{NA62:2021zjw},
we obtain the lower limit on the QCD axion decay constant,
 \begin{eqnarray}
  F_{a}\frac{2|N_C|}{|f_d-f_s|}\gtrsim(0.86-1.90)\times10^{11}\,{\rm GeV}\,.
 \label{cons_1}
 \end{eqnarray}

The QCD axion mass $m_a$ in terms of the pion mass and pion decay constant reads\,\cite{Ahn:2014gva, Ahn:2016hbn}
 \begin{eqnarray}
 m^{2}_{a}F^{2}_{a}=m^{2}_{\pi^0}f^{2}_{\pi}F(z,w)\,,
\label{axiMass2}
 \end{eqnarray}
where $f_\pi\simeq92.1$ MeV\,\cite{PDG} and
 $F(z,w)=z/(1+z)(1+z+w)$ with $\omega=0.315\,z$.
Here the Weinberg value lies in $z\equiv m^{\overline{\rm MS}}_u(2\,{\rm GeV})/m^{\overline{\rm MS}}_d(2\,{\rm GeV})=0.47^{+0.06}_{-0.07}$\,\cite{PDG}.
After integrating out the heavy $\pi^{0}$ and $\eta$ at low energies, there is an effective low energy Lagrangian with an axion-photon coupling $g_{a\gamma\gamma}$:
${\cal L}_{a\gamma\gamma}= -g_{a\gamma\gamma}\,a\,\vec{E}\cdot\vec{B}$
where $\vec{E}$ and $\vec{B}$ are the electromagnetic field components. The axion-photon coupling is expressed in terms of the QCD axion mass, pion mass, pion decay constant, $z$ and $w$,
 \begin{eqnarray}
 g_{a\gamma\gamma}=\frac{\alpha_{\rm em}}{2\pi}\frac{m_a}{f_{\pi}m_{\pi^0}}\frac{1}{\sqrt{F(z,w)}}\left(\frac{E}{N_C}-\frac{2}{3}\,\frac{4+z+w}{1+z+w}\right)\,.
 \label{gagg}
 \end{eqnarray}
The upper bound on the axion-photon coupling, derived from the recent analysis of the horizontal branch stars in galactic globular clusters\,\cite{Ayala:2014pea}, can be translated to 
 \begin{eqnarray}
 |g_{a\gamma\gamma}|<6.6\times10^{-11}\,{\rm GeV}^{-1}\,(95\%\,{\rm CL})\Leftrightarrow F_a\gtrsim2.525\times10^{7}\Big|\frac{E}{N_C}-1.903\Big|\,{\rm GeV}\,,
 \label{gagg_1}
 \end{eqnarray}
where $z=0.47$ is used.

\subsection{Charged-lepton and flavored-QCD axion}
Substituting the VEV ofEq.(\ref{vev}) into the superpotential (\ref{lagrangian_l}),  the charged-lepton mass matrix given in the Lagrangian (\ref{AxionLag2}) is derived as,
 \begin{eqnarray}
 {\cal M}_{\ell}&=& {\left(\begin{array}{ccc}
 \alpha_{e}\Delta^{|f_e|}_\chi\,e^{if_e\frac{A_X}{u_\chi}} & 0 &  0 \\
 0 &  \alpha_{\mu}\,\Delta^{|f_\mu|}_\chi\,e^{if_\mu\frac{A_X}{u_\chi}} & 0 \\
 0 & 0 &  \alpha_{\tau}\,\Delta^{|f_\tau|}_\chi\,e^{if_\tau\frac{A_X}{u_\chi}}
 \end{array}\right)}Y^{(6)}_{\bf 1}\,v_d\,.
 \label{ChL1}
 \end{eqnarray}
Recall that the coefficients $\alpha_i$ are complex numbers with an effective absolute value satisfying Eq.(\ref{AFN1}).
Then, the corresponding charged-lepton masses are given by
 \begin{eqnarray}
 m_{e}&=&\alpha_{e}\,Y^{(6)}_{\bf 1}\Delta^{|f_e|}_{\chi}\,v_d\,,\quad m_{\mu}=\alpha_{\mu}\,Y^{(6)}_{\bf 1}\Delta^{|f_\mu|}_{\chi}\,v_d\,,\quad
 m_{\tau}=\alpha_{\tau}\,Y^{(6)}_{\bf 1}\Delta^{|f_\tau|}_{\chi}\,v_d\,,
 \label{cLep1}
 \end{eqnarray}
 where $Y^{(6)}_{\bf 1}$ is given in Eq.(\ref{modu6}) and the phases in each term can be absorbed into $(l_i)_R$. They
are matched with the empirical values from the PDG\,\cite{PDG} given by
  \begin{eqnarray}
 m_{e}&=&0.511\,{\rm MeV}\,,\qquad m_{\mu}=105.658\,{\rm MeV}\,,\qquad
 m_{\tau}=1776.86\pm0.12\,{\rm MeV}\,.
 \label{clepmass}
 \end{eqnarray}
 Flavored axions typically interact with charged leptons (electrons, muons, taus)\,\cite{Ahn:2014gva, Ahn:2016hbn, Ahn:2018cau, Ahn:2021ndu} and can be emitted through atomic axio-recombination, axio-deexcitation, axio-bremsstrahlung in electron-ion or electron-electron collisions, and Compton scatterings\,\cite{Redondo:2013wwa}.
 Then the flavored-QCD axion to charged-lepton interactions read 
 \begin{eqnarray}
  -{\cal L}^{a\ell} &\simeq& \frac{\partial_\mu a}{2u_{\chi}}\Big(f_e\,\bar{e}\gamma^\mu\gamma_5 e+f_\mu\,\bar{\mu}\gamma^\mu\gamma_5\mu+f_\tau\,\bar{\tau}\gamma^\mu\gamma_5\tau\Big)+\sum_{\ell=e,\mu,\tau}\big(m_\ell\bar{\ell}\ell-\bar{\ell}i\! \! \not\! \partial\, \ell\big)\,.
  \label{fla_1}
 \end{eqnarray}
Like rare neutral flavor-changing decays in particle physics, the interaction of the flavored-QCD axion $a$ with leptons makes it possible to search for the QCD axion in astroparticle physics through stellar evolution.
The flavored-QCD axion coupling to electrons reads
\begin{eqnarray}
g_{aee}&=&|f_e|\,\frac{m_e}{u_\chi}\,.
\label{gaTe}
\end{eqnarray}
Stars in the red giant branch (RGB) of color-magnitude diagrams in globular clusters provide a strict constraint on axion-electron couplings which leads to a lower bound on the axion decay constant. This constraint is expressed as\,\cite{Viaux:2013lha}
\begin{eqnarray}
 |g_{aee}|<4.3\times10^{-13}\quad(95\%\,{\rm CL})~\Leftrightarrow N_CF_a\gtrsim1.19|f_e|\times10^9\,{\rm GeV}\,.
   \label{alee0}
\end{eqnarray}
Bremsstrahlung off electrons $e+Ze\rightarrow Ze+e+a$ in white dwarfs is an effective process for detecting axions as the Primakoff and Compton processes are suppressed due to the large plasma frequency. Comparing the theoretical and observed WD luminosity functions (WDLFs) provides a way to place limits\,\footnote{Note that Refs.\,\cite{WD01,Bertolami:2014wua} have pointed out features in some WDLFs\,\cite{DeGennaro:2007yw, Rowell:2011wp} that could imply axion-electron couplings in the range $7.2\times10^{-14}\lesssim|g_{aee}|\lesssim2.2\times10^{-13}$.}  on $|g_{aee}|$\,\cite{Raffelt:1985nj}. Recent analyses of WDLFs, using detailed WD cooling treatment and new data on the WDLF of the Galactic disk, suggest electron couplings $|g_{aee}|\lesssim2.8\times10^{-13}$\,\cite{Bertolami:2014wua}. However, these results come with large theoretical and observational uncertainties. 

We note that the entries of the quark and charged lepton mass matrices given in Eqs.(\ref{Ch2}), (\ref{Ch1}), and (\ref{ChL1}) except for the 
entry corresponding to the top quark are expressed as a combination of  $\Delta_\chi^{|f_\alpha|}$ and modular forms for each component. Accurate determination of the values of $\Delta_\chi$, its power, and the value of $\tau$ is crucial to reproduce the observed CKM mixing angles given in Eq.(\ref{ckmmixing}) and quark masses in Eq.(\ref{qumas}). The values of those parameters are also closely linked to those in the lepton sector, and they should necessarily be determined in order to reproduce the observed values of charged lepton masses and to predict light active neutrinos derived from Eqs.(\ref{MR1}) and (\ref{Ynu1}). 
The $U(1)_X$ PQ scale, which corresponds to the seesaw scale (as shown n Eq.(\ref{MR5})), can be estimated as $\mu_\chi\sim6\times10^{13}$ GeV from Eq.(\ref{sof2}) for $m_{3/2}\sim100$ TeV. 

\subsection{Neutrino}
\label{nu_sec}
Similar to the case of charged lepton mass matrix, the heavy Majorana mass matrix given in the Lagrangian\,(\ref{AxionLag2})
is derived from the superpotential (\ref{lagrangian_l}) as
\begin{eqnarray}
 M_{R}&=&M {\left(\begin{array}{ccc}
1+y^3-5xy & 0 &  0 \\
0 & 0 & 1+y^3-5xy \\
0 & 1+y^3-5xy & 0
 \end{array}\right)}\nonumber\\
 &+&M{\left(\begin{array}{ccc}
\frac{2}{\sqrt{3}}(\gamma\,p+\gamma'\,yr) & -\frac{1}{\sqrt{3}}(\gamma\,yp+\gamma'\,xr) &  -\frac{1}{\sqrt{3}}(\gamma\,xp+\gamma'\,r) \\
-\frac{1}{\sqrt{3}}(\gamma\,yp+\gamma'\,xr) & \frac{2}{\sqrt{3}}(\gamma\,xp+\gamma'\,r) & -\frac{1}{\sqrt{3}}(\gamma\,p+\gamma'\,yr) \\
-\frac{1}{\sqrt{3}}(\gamma\,xp+\gamma'\,r) & -\frac{1}{\sqrt{3}}(\gamma\,p+\gamma'\,yr) &  \frac{2}{\sqrt{3}}(\gamma\,yp+\gamma'\,xr)
 \end{array}\right)}\,,
 \label{MR1}
  \end{eqnarray}
 where $p=1+2xy$, $r=y^2+2xy$, $\gamma=\gamma_2/\gamma_1$, $\gamma'=\gamma'_2/\gamma_1$, and the common factor $M$ can be replaced by the QCD axion decay constant $F_a$,
   \begin{eqnarray}
   M\equiv|\gamma_1Y^3_1\langle\chi\rangle|=\frac{|\gamma_1|}{2}F_a|N_CY^3_1|\,.
  \label{MR5}
  \end{eqnarray}
The terms with $\gamma_2$ in Eq.(\ref{MR1}) are derived by taking  the modular form $Y^{(6)}_{{\bf 3},1}$ safisfying Eq.(\ref{modu6}), whereas the terms with $\gamma'_2$ are derived by taking  $Y^{(6)}_{{\bf 3},2}$.
Eq.(\ref{MR1}) has three unknown complex parameters, $\gamma$ $\gamma'$, and $\gamma_1$, where the phase of $\gamma_1$ contributes as an overall factor after seesawing. Other variables such as $x, y, Y_1$  are determined from the analysis for the quark and charged-lepton sectors, and $\langle\chi\rangle$ is fixed from the seesaw formula Eq.(\ref{neut2}) whose scale is given  by PQ scale Eq.(\ref{sof2}).
The Dirac mass term in the Lagrangian\,(\ref{AxionLag2}) reads
  \begin{eqnarray}
m_{D}={\left(\begin{array}{ccc}
 \beta_{1}\,\Delta^{|g_e|}_\chi &  \beta_{2}\,y\Delta^{|g_\mu|}_\chi  &  \beta_{3}\,x\Delta^{|g_\tau|}_\chi  \\
 \beta_{1}\,y\Delta^{|g_e|}_\chi  & \beta_{2}\,x\Delta^{|g_\mu|}_\chi &  \beta_{3}\,\Delta^{|g_\tau|}_\chi \\
 \beta_{1}\,x\Delta^{|g_e|}_\chi & \beta_{2}\,\Delta^{|g_\mu|}_\chi & \beta_{3}\,y\Delta^{|g_\tau|}_\chi
 \end{array}\right)} {\left(\begin{array}{ccc}
e^{ig_e\frac{A_X}{u_\chi}} & 0 &  0 \\
 0 &  e^{ig_\mu\frac{A_X}{u_\chi}} & 0 \\
 0 & 0 & e^{ig_\tau\frac{A_X}{u_\chi}}
 \end{array}\right)}\,Y_1\,v_u\,.
 \label{Ynu1}
 \end{eqnarray}
The coefficients $\beta_i$ and $\gamma_i$ in the neutrino sector, like in the quark and charged-lepton sectors, are complex numbers corrected by higher-dimensional operators, resulting in an effective absolute value satisfying Eq.(\ref{AFN1}). Eq.(\ref{Ynu1}) contains three complex parameters ($\beta_1, \beta_2$, and $\beta_3$), where one of the phases can be removable as an overall factor after seesawing. As shwon before, the parameter $\Delta_\chi$ can be determined from quark and charged-lepton sectors. In addition, its $U(1)_X$ quantum number $g_\alpha$ can be determined from the numerical analysis for the neutrino sector with the help of the condition of  $U(1)_X$-mixed gravitational anomaly-free given in in Eq.(\ref{uyxf}).

 After integrating out the right-handed heavy Majorana neutrinos, the effective neutrino mass matrix ${\cal M}_{\nu}$ is given at leading order by
 \begin{eqnarray}
 {\cal M}_{\nu}\simeq -m^T_DM^{-1}_Rm_D=U^{\ast}_{\nu}{\rm diag.}(m_{\nu_1}, m_{\nu_2}, m_{\nu_3})U^{\dag}_{\nu}\,,
   \label{neut2}
 \end{eqnarray}
where $U_{\nu}$ is the rotation matrix diagonalizing ${\cal M}_{\nu}$ and $m_{\nu_i}$ ($i=1,2,3$) are the light neutrino masses. 
Then the PMNS mixing matrix becomes
  \begin{eqnarray}
 U_{\rm PMNS}=U_\nu .
   \label{pmns0}
 \end{eqnarray}
 The matrix $U_{\rm PMNS}$ is expressed in terms of three mixing angles, $\theta_{12}, \theta_{13}, \theta_{23}$, and a Dirac type \cp ~violaitng phase $\delta_{CP}$ and two additional \cp~ violating phases $\varphi_{1,2}$ if light neutrinos are Majorana particle as\,\cite{PDG}
 \begin{eqnarray}
  U_{\rm PMNS}=
  {\left(\begin{array}{ccc}
   c_{13}c_{12} & c_{13}s_{12} & s_{13}e^{-i\delta_{CP}} \\
   -c_{23}s_{12}-s_{23}c_{12}s_{13}e^{i\delta_{CP}} & c_{23}c_{12}-s_{23}s_{12}s_{13}e^{i\delta_{CP}} & s_{23}c_{13}  \\
   s_{23}s_{12}-c_{23}c_{12}s_{13}e^{i\delta_{CP}} & -s_{23}c_{12}-c_{23}s_{12}s_{13}e^{i\delta_{CP}} & c_{23}c_{13}
   \end{array}\right)}Q_{\nu}\,,
 \label{PMNS1}
 \end{eqnarray}
where $s_{ij}\equiv \sin\theta_{ij}$, $c_{ij}\equiv \cos\theta_{ij}$ and $Q_{\nu}={\rm Diag}(e^{-i\varphi_1/2}, e^{-i\varphi_2/2},1)$.
Then the neutrino masses are obtained by the transformation
 \begin{eqnarray}
  U^T_{\rm PMNS}\,{\cal M}_{\nu}\,U_{\rm PMNS}={\rm Diag}.(m_{\nu_1}, m_{\nu_2}, m_{\nu_3})\,.
 \label{nu_mas}
 \end{eqnarray}
Here $m_{\nu_i}$ ($i=1,2,3$) are the light neutrino masses.
The observed hierarchy $|\Delta m^{2}_{\rm Atm}|= |m^{2}_{\nu_3}-(m^{2}_{\nu_1}+m^{2}_{\nu_2})/2|\gg\Delta m^{2}_{\rm Sol}\equiv m^{2}_{\nu_2}-m^{2}_{\nu_1}>0$ and the requirement of a Mikheyev-Smirnov-Wolfenstein resonance\,\cite{Wolfenstein:1977ue} for solar neutrinos lead to two possible neutrino mass spectra: normal mass ordering (NO) $m^2_{\nu_1}<m^2_{\nu_2}<m^2_{\nu_3}$ and inverted mass ordering (IO) $m^2_{\nu_3}<m^2_{\nu_1}<m^2_{\nu_2}$. Nine physical observables can be derived from Eqs.(\ref{PMNS1}) and (\ref{nu_mas}): $\theta_{23}$, $\theta_{13}$, $\theta_{12}$, $\delta_{CP}$, $\varphi_1$, $\varphi_2$, $m_{\nu_1}$, $m_{\nu_2}$, and $m_{\nu_3}$. 
\begin{table}[h]
\caption{\label{exp_nu} The global fit of three-flavor oscillation parameters at the best-fit and $3\sigma$ level with Super-Kamiokande atmospheric data\,\cite{Esteban:2020cvm}. NO = normal neutrino mass ordering; IO = inverted mass ordering. And $\Delta m^{2}_{\rm Sol}\equiv m^{2}_{\nu_2}-m^{2}_{\nu_1}$, $\Delta m^{2}_{\rm Atm}\equiv m^{2}_{\nu_3}-m^{2}_{\nu_1}$ for NO, and  $\Delta m^{2}_{\rm Atm}\equiv m^{2}_{\nu_2}-m^{2}_{\nu_3}$ for IO.}
\begin{ruledtabular}
\begin{tabular}{cccccccccccc} &$\theta_{13}[^{\circ}]$&$\delta_{CP}[^{\circ}]$&$\theta_{12}[^{\circ}]$&$\theta_{23}[^{\circ}]$&$\Delta m^{2}_{\rm Sol}[10^{-5}{\rm eV}^{2}]$&$\Delta m^{2}_{\rm Atm}[10^{-3}{\rm eV}^{2}]$\\
\hline
$\begin{array}{ll}
\hbox{NO}\\
\hbox{IO}
\end{array}$&$\begin{array}{ll}
8.58^{+0.33}_{-0.35} \\
8.57^{+0.37}_{-0.34}
\end{array}$&$\begin{array}{ll}
232^{+118}_{-88} \\
276^{+68}_{-82}
\end{array}$&$33.41^{+2.33}_{-2.10}$&$\begin{array}{ll}
42.2^{+8.8}_{-2.5} \\
49.0^{+2.5}_{-9.1}
\end{array}$&$7.41^{+0.62}_{-0.59}$
 &$\begin{array}{ll}
2.507^{+0.083}_{-0.080} \\
2.486^{+0.084}_{-0.080}
\end{array}$ \\
\end{tabular}
\end{ruledtabular}
\end{table}
Recent global fits\,\cite{Esteban:2018azc, deSalas:2017kay, Capozzi:2018ubv} of neutrino oscillations have enabled a more precise determination of the mixing angles and mass squared differences, with large uncertainties remaining for $\theta_{23}$ and $\delta_{CP}$ at 3$\sigma$. The most recent analysis\,\cite{Esteban:2020cvm} lists global fit values and $3\sigma$ intervals for these parameters in Table-\ref{exp_nu}.
Furthermore, recent constraints on the rate of $0\nu\beta\beta$ decay have added to these findings. Specifically, the most tight upper bounds for the effective Majorana mass (${\cal M}_{\nu})_{ee}$, which is the modulus of the $ee$-entry of the effective neutrino mass matrix, are given by
  \begin{eqnarray}
 ({\cal M}_{\nu})_{ee}< 0.036-0.156\,{\rm eV}\,~ (^{136}\text{Xe-based experiment\,\cite{KamLAND-Zen:2022tow}})
 \label{nubb}
 \end{eqnarray}
at $90\%$ confidence level.
$0\nu\beta\beta$ decay is a low-energy probe of lepton number violation and its measurement could provide the strongest evidence for lepton number violation at high energy. Its discovery would suggest the Majorana nature of neutrinos and, consequently, the existence of heavy Majorana neutrinos via the seesaw mechanism\,\cite{Minkowski:1977sc}.
 
Transforming the neutrino fields by chiral rotations of Eq.(\ref{X-tr}) under $U(1)_X$ we obtain the flavored-QCD axion interactions to neutrinos
 \begin{eqnarray}
  -{\cal L}^{a\nu} &\simeq&\frac{a}{2u_{\chi}}\sum_{i\neq j}\overline{\nu_i}\big\{(m_{\nu_i}-m_{\nu_j}){\rm Im}[V]_{ij}-i\gamma_{5}(m_{\nu_i}+m_{\nu_j}){\rm Re}[V]_{ij}\big\}\nu_j
    +\frac{\partial a}{4u_{\chi}}\overline{N}\gamma_\mu\gamma_{5}N^c
  \label{neut1}
 \end{eqnarray}
where $m_{\nu_i}$ are real and positive, ${\cal Q}[V]_{ij}=(g_e-\frac{1}{2}){\cal Q}[U_{1i}U^\ast_{1j}]+(g_\mu-\frac{1}{2}){\cal Q}[U_{2i}U^\ast_{2j}]+(g_\tau-\frac{1}{2}){\cal Q}[U_{3i}U^\ast_{3j}]$ with ${\cal Q}={\rm Re}$ or ${\rm Im}$, and ${\rm Im}[V]_{ij}=-{\rm Im}[V]_{ji}$ with $U\equiv U_{\rm PMNS}$.
Since the light neutrino mass is less than 0.1 eV, the coupling between the flavored-QCD axion and light neutrinos is subject to a stringent constraint given by Eq.(\ref{cons_1}), which significantly suppresses the interaction. Therefore, we will not take it into consideration. Ref.\cite{Kharusi:2021jez} provides the latest experimental constraints on Majoron-neutrino coupling, which are below the range of $(0.4-0.9)\times10^{-5}$.

\begin{table}[h]
\caption{\label{reps_v} $U(1)_X$ charges linked to seesaw scale.}
\begin{tabular}{cccccc}
\ &$|g_e|$&$|g_\mu|$&$|g_\tau|$ & $\langle\chi\rangle$/GeV  \\
\hline
NO&$6$  & $4$ & $5$ & $10^{13}$  \\
 & $5$ & $3$&$4$ & $5\times10^{13}$\\
& $4$ & $2$ & $3$ & $10^{14}$\\
& $3$ & $1$ & $2$ & $5\times10^{14}$ \\
& $2$ & $0$ & $1$ & $10^{15}$ \\
\hline
IO & $2$ & $3$ & 3 & $5\times10^{13}$ &  \\
 & $1$ & 2 & 2 &  $10^{14}$\\ 
 & $0$ & $1$ & 1 & $5\times10^{14}$ &  \\ 
\end{tabular}
\end{table} 
Once the lepton $U(1)_X$ quantum numbers are fixed, the seesaw scale $M\sim v_\chi$ of Eq.(\ref{MR5}) comparable to the PQ scale of Eq.(\ref{sof2}) can be roughly determined using the seesaw formula Eq.(\ref{neut2}). By putting Eqs.(\ref{MR1}, \ref{Ynu1}) into the seesaw formula Eq.(\ref{neut2}), we obtain numerically a range of values for $\langle\chi\rangle$. 
For instance, see Table-\ref{reps_v}: it implies that for normal mass ordering, the maximum scale should be below $\sim10^{15}$ GeV, and for inverted mass ordering, the maximum scale should be below $\sim5\times10^{14}$ GeV. Refer to Table-\ref{q_n} and -\ref{q_m} for NO, and Table-\ref{l_n} for IO, with $F_a=2|\langle\chi\rangle/N_C|$.
By using the seesaw formula Eq.(\ref{neut2}), one can set the scale $\langle\chi\rangle$ along with the $U(1)_X$ quantum numbers $g_\alpha$ without loss of generality. Doing so results in an effective mass matrix with nine physical degrees of freedom
 \begin{eqnarray}
m_0\equiv\Big|\frac{\beta^2_1}{\gamma_1 Y_{\bf 1}}\Big|\frac{v^2_u}{\langle\chi\rangle}\,,~|\gamma|\,,~|\gamma'|\,,~|\tilde{\beta}_2|\,,~|\tilde{\beta}_3|\,,~\arg(\gamma)\,,~\arg(\gamma')\,,~\arg(\tilde{\beta}_2)\,,~\arg(\tilde{\beta}_3)\,,
 \label{dof_nu}
 \end{eqnarray}
in which $m_0$ is an overall factor of Eq.(\ref{neut2}), $\tilde{\beta}_2\equiv\beta_2/\beta_1, \tilde{\beta}_3\equiv\beta_3/\beta_1$, and $1-2\Delta^2_\chi\leq|\tilde{\beta}_{2(3)}|, |\gamma|, |\gamma'|\leq\frac{1}{1-2\Delta^2_\chi}$.
Out of the nine observables corresponding to Eq.(\ref{dof_nu}), the five measured quantities ($\theta_{12}$, $\theta_{23}$, $\theta_{13}$, $\Delta m^2_{\rm Sol}$, $\Delta m^2_{\rm Atm}$) can be used as constraints. The remaining four degrees of freedom correspond to four measurable quantities ($\delta_{CP}$, $\varphi_{1,2}$, and the $0\nu\beta\beta$-decay rate), which can be determined through measurements.

\section{Numerical analysis for quark, lepton, and a QCD axion}
\label{num1}
To simulate and match experimental results for quarks and leptons (Eqs.(\ref{ckmmixing}, \ref{qumas}) and Table-\ref{exp_nu}), we use linear algebra tools from Ref.\cite{Antusch:2005gp}. By analyzing experimental data for quarks and charged leptons, we determined the $U(1)_X$ quantum numbers listed in Tables-\ref{q_n} and -\ref{q_m} for normal neutrino mass ordering and Table-\ref{l_n} for inverted neutrino mass ordering. We also ensured the $U(1)_X$-mixed gravitational anomaly-free condition of Eq.(\ref{uyxf}) and consistency of the seesaw scale discussed above Eq.(\ref{dof_nu}) with the PQ breaking scale of Eq.(\ref{sof2}).

Notably, in our model, the flavored-QCD axion mass (and its associated PQ breaking or seesaw scale) is closely linked to the soft SUSY-breaking mass. Our analysis covers the PQ scale $\langle\chi\rangle$ from roughly $10^{13}$ GeV to $10^{15}$ GeV due to Table-\ref{reps_v}, corresponding to $m_{3/2}$ values of 1 TeV to $10^6$ TeV, by considering Eq.(\ref{sof2}) and Table-\ref{reps_v}. The given $U(1)_X$ quantum numbers can then be used to predict the branching ratio of $K^+\rightarrow \pi^++a$ (Eq.(\ref{Gkp})), as well as the axion coupling to photon (Eq.(\ref{gagg})) and electron (Eq.(\ref{gaTe})). See Table-\ref{q_n}, -{\ref{q_m}, and -{\ref{l_n} for more details. Our proposed model's predictions can be tested by current axion search experiments. KLASH\,\cite{Alesini:2019nzq} is sensitive to the mass range of $0.27-0.93\,\mu{\rm eV}$, while FLASH\,\cite{Gatti:2021cel} covers the mass range of $0.5-1.5\,\mu{\rm eV}$.
\begin{figure}[t]
\hspace*{-0.5cm}
\begin{minipage}[h]{8.3cm}
\includegraphics[width=8.3cm]{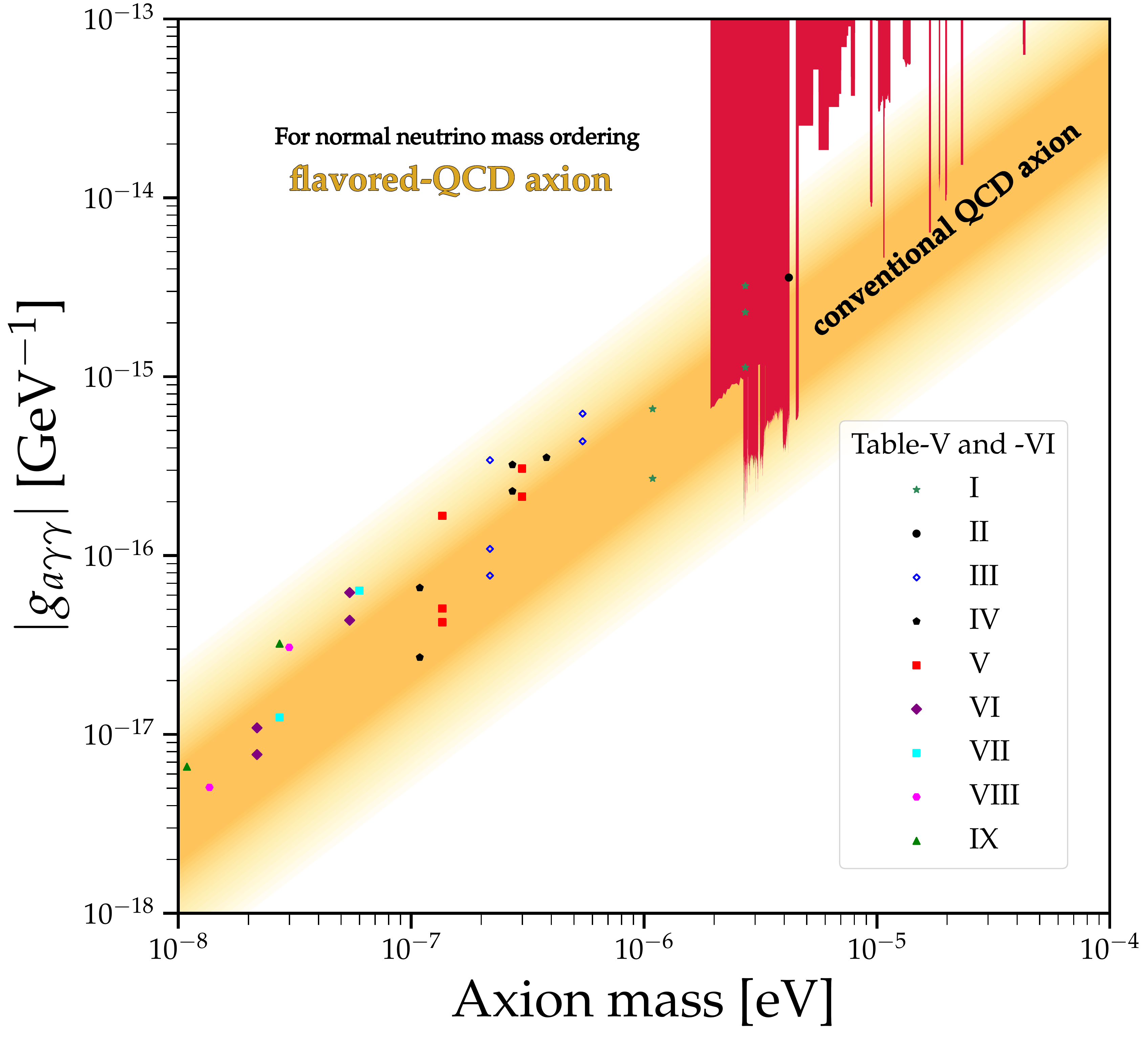}
\end{minipage}
\begin{minipage}[h]{8.3cm}
\includegraphics[width=8.3cm]{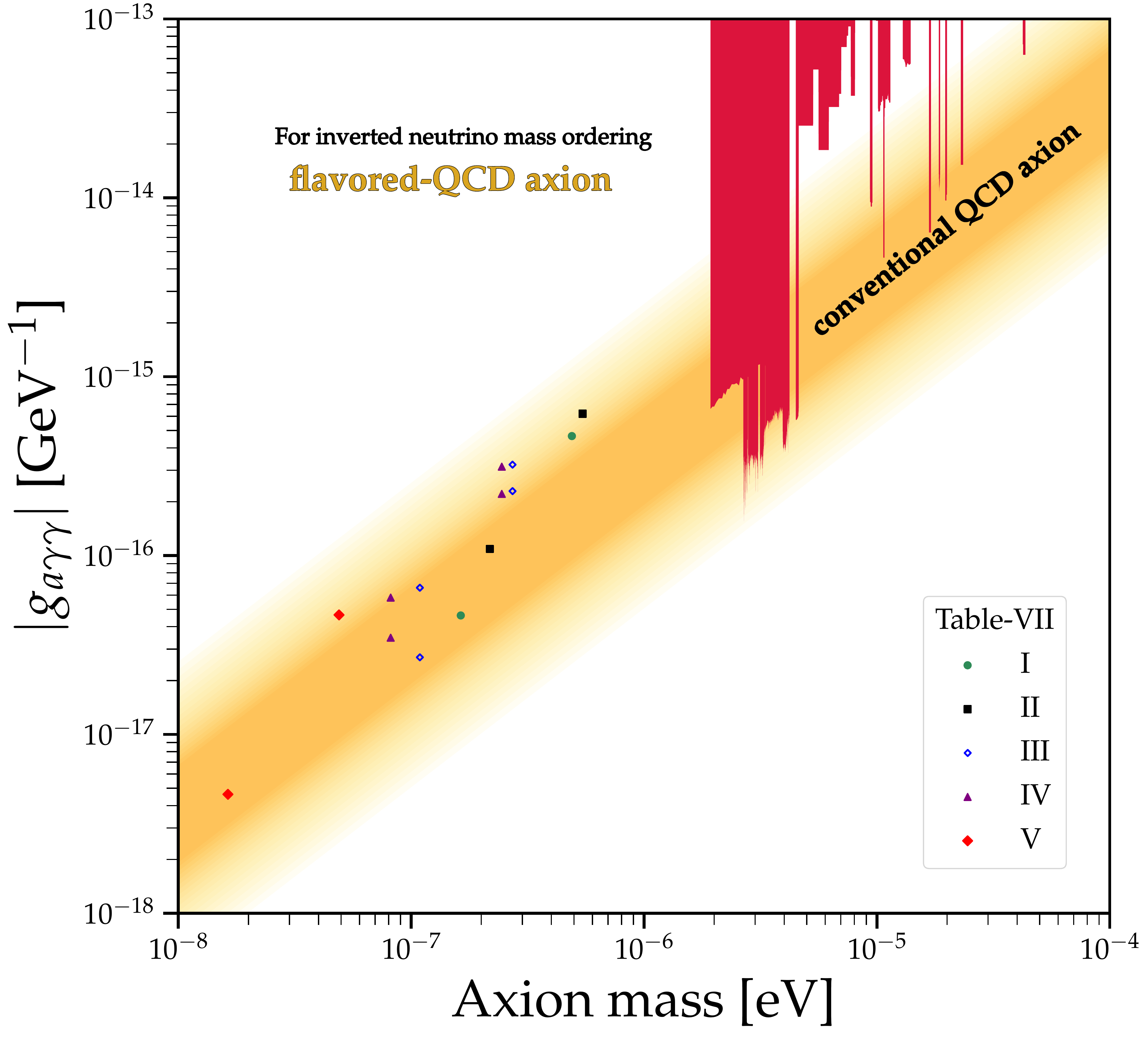}
\end{minipage}
\caption{\label{Fig1} Plots for axion-photon coupling $|g_{a\gamma\gamma}|$ as a function of the flavored-QCD axion mass $m_{a}$ for NO and IO. Orange shaded region and vertical red lines indicate the conventional QCD axion predictions and the exclusion region of various axion search experiments, respectively, see Ref.\cite{PDG}.}
\end{figure}
Fig.\ref{Fig1} illustrates plots of the axion-photon coupling $|g_{a\gamma\gamma}|$ as a function of the flavored-QCD axion mass $m_{a}$ for NO (left) and IO (right), respectively. Each plotted point corresponds to values listed in Tables-\ref{q_n}, -\ref{q_m}, and -\ref{l_n}, which are consistent with the experimental constraints described in Eqs.(\ref{cons_1}), (\ref{gagg_1}), and (\ref{alee0}). And Fig.\ref{Fig1} illustrates that certain data points in Table-\ref{q_n} (I-c, I-d, I-e) have been fully excluded by the ADMX experiment\,\cite{ADMX}, while another data point (II) in Table-\ref{q_n} has been marginally excluded by the same experiment. Fig.\ref{Fig2} shows plots for axion-electron coupling $|g_{aee}|$ as a function of the flavored-QCD axion mass $m_{a}$ for NO (left) and IO (right). 
\begin{figure}[t]
\hspace*{-0.5cm}
\begin{minipage}[h]{8.3cm}
\includegraphics[width=8.3cm]{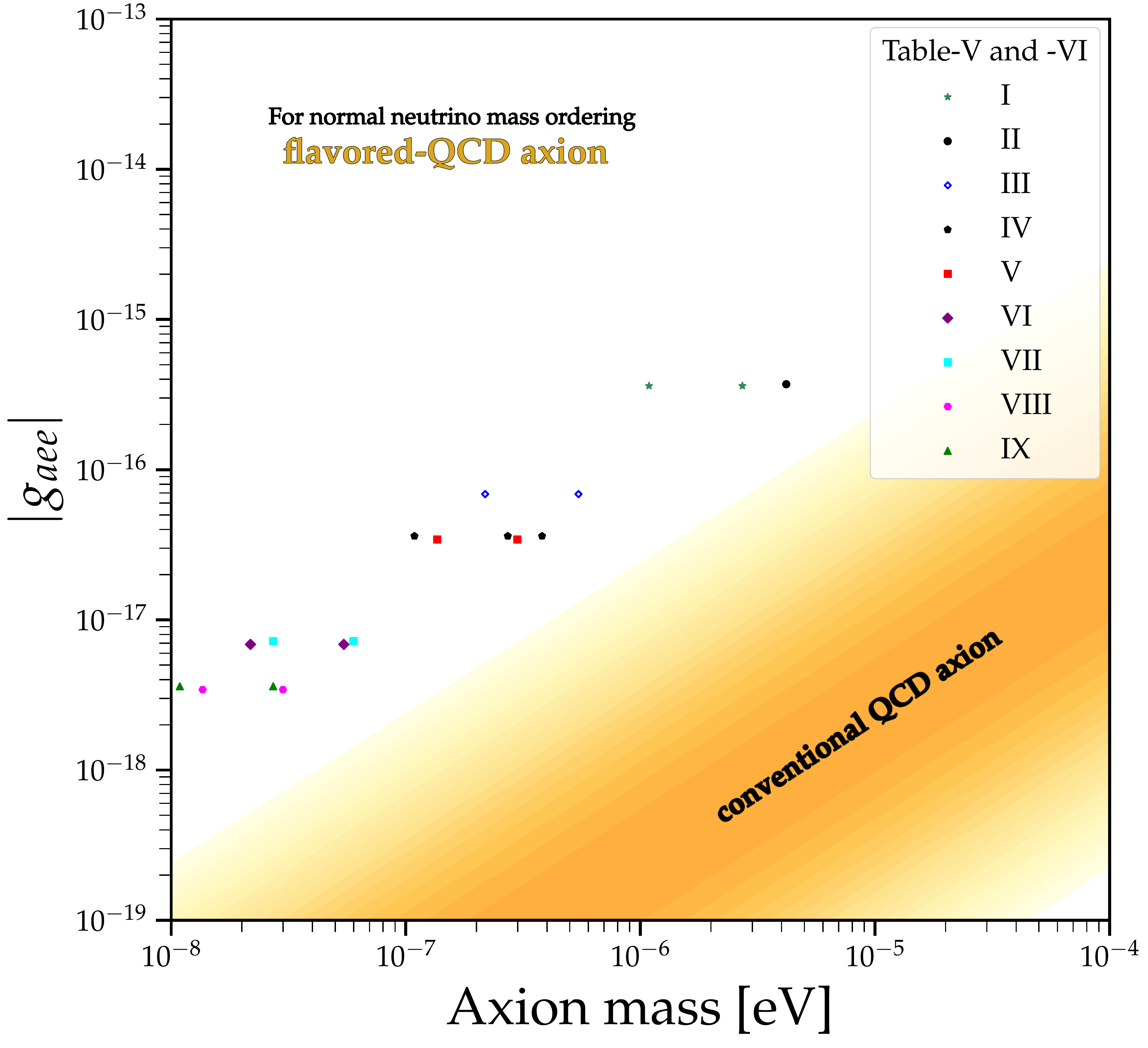}
\end{minipage}
\begin{minipage}[h]{8.3cm}
\includegraphics[width=8.3cm]{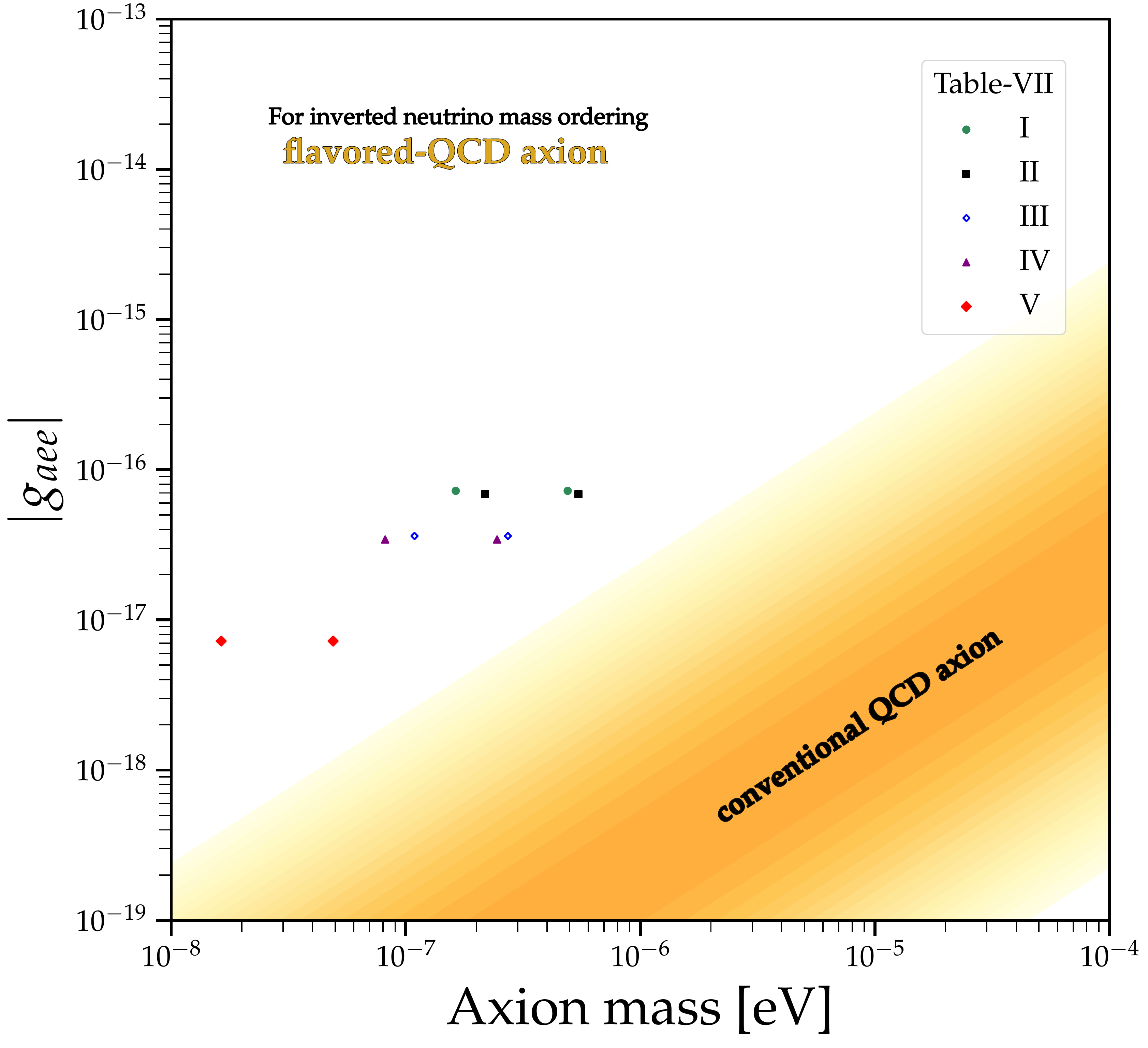}
\end{minipage}
\caption{\label{Fig2} Plots for axion-electron coupling $|g_{aee}|$ as a function of the flavored-QCD axion mass $m_{a}$ for NO (left) and IO (right). Orange shaded region indicates the conventional QCD axion predictions.}
\end{figure}

\begin{table}[h]
\caption{\label{q_n} 
The quark and lepton $U(1)_X$ quantum numbers that satisfy experimental results, including Eqs.(\ref{ckmmixing}, \ref{qumas}, \ref{cons_1}) and Table-\ref{exp_nu}, as well as the $U(1)$-mixed gravitational anomaly-free condition for normal neutrino mass ordering.
And QCD anomaly coefficient $N_C$, QCD axion decay constant $F_a=2|\langle\chi\rangle/N_C|$ (with PQ breakdown or seesaw scale $\langle\chi\rangle$), the ratio of QCD and QED anomaly coefficients $E/N_C$, axion-electron coupling $g_{ae}$, and for $z=0.47$ axion-photon coupling $g_{a\gamma\gamma}$, axion mass $m_a$, and branching ratio ${\rm Br}(K^+\rightarrow\pi^++a)\equiv{\rm Br}(K\pi a)$.}
\begin{ruledtabular}
\begin{tabular}{ccccccccccccccccccc}
$U(1)_X$ &$f_{u}$& $f_{c}$& $f_{d}$&$f_s$ & $f_b$&$f_e$ & $f_\mu$ & $f_\tau$ & $g_e$ & $g_\mu$ & $g_\tau$ & $N_C$ & $\frac{F_a}{\rm GeV}$& $\frac{E}{N_C}$ & $\frac{g_{ae}}{10^{-17}}$ & $\frac{|g_{a\gamma\gamma}|}{10^{-17}\,{\rm GeV}^{-1}}$& $\frac{m_a}{10^{-7}\,{\rm eV}}$& ${\rm Br}(K\pi a)$\\
\hline
 &$20$&$8$&$14$ &$11$&$5$&$20$&$9$ &$4$ &$6$ &$4$ &$5$ &&&&&&&$4.2^{+1.9}_{-1.7}\times10^{-15}$ \\
\hline
I-a&$\mp$&$\pm$&$\pm$ &$\mp$&$\pm$&$\pm$&$\mp$ &$\pm$ &$\pm$ &$\mp$ &$\mp$ &$\pm4$ &$5\times10^{12}$ & $-\frac{5}{6}$&$36.13$&$66.00$& $10.89$\\
\hline
I-b&$\mp$&$\pm$&$\pm$ &$\mp$&$\pm$&$\pm$&$\mp$ &$\mp$ &$\pm$ &$\pm$ &$\mp$ &$\pm4$ &$5\times10^{12}$ & $\frac{19}{6}$&$36.13$&$26.94$& $10.89$\\
\hline
I-c&$\mp$&$\pm$&$\mp$ &$\pm$&$\pm$&$\pm$&$\mp$ &$\pm$ &$\pm$ &$\pm$ &$\pm$ &$\pm10$ &$2\times10^{12}$ & $\frac{1}{15}$&$36.13$&$112.71$&$27.21$\\
\hline
I-d&$\mp$&$\pm$&$\mp$ &$\pm$&$\pm$&$\pm$&$\pm$ &$\mp$ &$\pm$ &$\pm$ &$\mp$ &$\pm10$ &$2\times10^{12}$ & $-\frac{29}{15}$&$36.13$&$228.88$&$27.21$\\
\hline
I-e&$\mp$&$\pm$&$\mp$ &$\pm$&$\pm$&$\pm$&$\pm$ &$\pm$ &$\pm$ &$\mp$ &$\mp$ &$\pm10$ &$2\times10^{12}$ & $-\frac{53}{15}$&$36.13$&$321.82$&$27.21$\\
\hline\hline
 &$21$&$8$&$14$ &$11$&$5$&$20$&$9$ &$4$ &$5$ &$3$ &$4$&&&&&&&$4.2^{+1.9}_{-1.7}\times10^{-15}$ \\
\hline
II&$\mp$&$\pm$&$\pm$ &$\mp$&$\mp$&$\pm$&$\pm$ &$\pm$ &$\pm$ &$\pm$ &$\pm$ &$\pm15$ &$1.3\times10^{12}$ & $-2$&$37.06$&$358.09$&$41.87$\\
\hline\hline
 &$20$&$8$&$14$ &$11$&$5$&$19$&$9$ &$4$ &$5$ &$3$ &$4$ &&&&&&&$1.7^{+0.8}_{-0.7}\times10^{-16}$\\
\hline
III-a&$\mp$&$\pm$&$\pm$ &$\mp$&$\pm$&$\pm$&$\mp$ &$\mp$ &$\pm$ &$\mp$ &$\pm$ &$\pm4$ &$2.5\times10^{13}$ & $\frac{11}{3}$&$6.87$&$7.71$&$2.18$\\
\hline
III-b&$\mp$&$\pm$&$\pm$ &$\mp$&$\pm$&$\pm$&$\pm$ &$\mp$ &$\mp$ &$\mp$ &$\mp$ &$\pm4$ &$2.5\times10^{13}$ & $-\frac{16}{3}$&$6.87$&$34.11$&$2.18$\\
\hline
III-c&$\mp$&$\pm$&$\pm$ &$\mp$&$\pm$&$\pm$&$\mp$ &$\pm$ &$\pm$ &$\mp$ &$\mp$ &$\pm4$ &$2.5\times10^{13}$ & $-\frac{1}{3}$&$6.87$&$10.88$&$2.18$\\
\hline
III-d&$\mp$&$\pm$&$\mp$ &$\pm$&$\pm$&$\pm$&$\pm$ &$\pm$ &$\pm$ &$\mp$ &$\mp$ &$\pm10$ &$10^{13}$ & $-\frac{10}{3}$&$6.87$&$62.04$&$5.44$\\
\hline
III-e&$\mp$&$\pm$&$\mp$ &$\pm$&$\pm$&$\pm$&$\pm$ &$\mp$ &$\pm$ &$\mp$ &$\pm$ &$\pm10$ &$10^{13}$ & $-\frac{26}{15}$&$6.87$&$43.45$&$5.44$\\
\hline\hline
 &$20$&$8$&$14$ &$11$&$5$&$20$&$9$ &$4$ &$4$ &$2$ &$3$&&&&&&&$4.2^{+1.9}_{-1.7}\times10^{-17}$ \\
\hline
IV-a&$\mp$&$\pm$&$\pm$ &$\mp$&$\pm$&$\pm$&$\mp$ &$\pm$ &$\mp$ &$\mp$ &$\pm$ &$\pm4$ &$5\times10^{13}$ & $-\frac{5}{6}$&$3.61$&$6.60$&$1.09$\\
\hline
IV-b&$\mp$&$\pm$&$\pm$ &$\mp$&$\pm$&$\pm$&$\mp$ &$\mp$ &$\pm$ &$\mp$ &$\pm$ &$\pm4$ &$5\times10^{13}$ & $\frac{19}{6}$&$3.61$&$2.69$&$1.09$\\
\hline
IV-c&$\mp$&$\pm$&$\mp$ &$\pm$&$\pm$&$\pm$&$\pm$ &$\mp$ &$\pm$ &$\mp$ &$\pm$ &$\pm10$ &$2\times10^{13}$ & $-\frac{29}{15}$&$3.61$&$22.89$&$2.72$\\
\hline
IV-d&$\mp$&$\pm$&$\mp$ &$\pm$&$\pm$&$\pm$&$\pm$ &$\pm$ &$\mp$ &$\mp$ &$\pm$ &$\pm10$ &$2\times10^{13}$ & $-\frac{53}{15}$&$3.61$&$32.18$&$2.72$\\
\hline
IV-e&$\mp$&$\pm$&$\pm$ &$\mp$&$\mp$&$\pm$&$\pm$ &$\pm$ &$\pm$ &$\pm$ &$\pm$ &$\pm14$ &$1.43\times10^{13}$ & $-\frac{7}{3}$&$3.61$&$35.30$&$3.81$\\
\end{tabular}
\end{ruledtabular}
\end{table}
\begin{table}[h]
\caption{\label{q_m} The same as in Table \ref{q_n}.}
\begin{ruledtabular}
\begin{tabular}{ccccccccccccccccccc}
$U(1)_X$ &$f_{u}$& $f_{c}$& $f_{d}$&$f_s$ & $f_b$&$f_e$ & $f_\mu$ & $f_\tau$ & $g_e$ & $g_\mu$ & $g_\tau$ & $N_C$ & $\frac{F_a}{\rm GeV}$& $\frac{E}{N_C}$ & $\frac{g_{ae}}{10^{-17}}$ & $\frac{|g_{a\gamma\gamma}|}{10^{-17}\,{\rm GeV}^{-1}}$& $\frac{m_a}{10^{-7}\,{\rm eV}}$& ${\rm Br}(K\pi a)$\\
\hline
 &$21$&$8$&$14$ &$11$&$5$&$19$&$9$ &$4$ &$4$ &$2$ &$3$&&&&&&&$4.2^{+1.9}_{-1.7}\times10^{-17}$ \\
\hline
V-a&$\mp$&$\pm$&$\pm$ &$\mp$&$\pm$&$\pm$&$\mp$ &$\pm$ &$\mp$ &$\pm$ &$\pm$ &$\pm5$ &$4\times10^{13}$ & $\frac{4}{15}$&$3.43$&$5.05$&$1.36$\\
\hline
V-b&$\mp$&$\pm$&$\pm$ &$\mp$&$\pm$&$\pm$&$\mp$ &$\mp$ &$\pm$ &$\pm$ &$\pm$ &$\pm5$ &$4\times10^{13}$ & $\frac{52}{15}$&$3.43$&$4.24$&$1.36$\\
\hline
V-c&$\mp$&$\pm$&$\pm$ &$\mp$&$\pm$&$\pm$&$\pm$ &$\mp$ &$\mp$ &$\mp$ &$\mp$ &$\pm5$ &$4\times10^{13}$& $-\frac{56}{15}$&$3.43$&$16.67$&$1.36$\\
\hline
V-d&$\mp$&$\pm$&$\mp$ &$\pm$&$\pm$&$\pm$&$\pm$ &$\pm$ &$\mp$ &$\pm$ &$\pm$ &$\pm11$ &$1.8\times10^{13}$ & $-\frac{92}{33}$&$3.43$&$30.64$&$2.99$\\
\hline
V-e&$\mp$&$\pm$&$\mp$ &$\pm$&$\pm$&$\pm$&$\pm$ &$\mp$ &$\pm$ &$\pm$ &$\pm$ &$\pm11$ &$1.8\times10^{13}$ & $-\frac{4}{3}$&$3.43$&$21.34$&$2.99$\\
\hline\hline
 &$20$&$8$&$14$ &$11$&$5$&$19$&$9$ &$4$ &$3$ &$1$ &$2$&&&&&&&$1.7^{+0.8}_{-0.7}\times10^{-18}$\\
\hline
VI-a&$\mp$&$\pm$&$\pm$ &$\mp$&$\pm$&$\pm$&$\mp$ &$\pm$ &$\mp$ &$\mp$ &$\pm$ &$\pm4$ & $2.5\times10^{14}$ & $-\frac{1}{3}$&$0.69$&$1.09$&$0.22$\\
\hline
VI-b&$\mp$&$\pm$&$\pm$ &$\mp$&$\pm$&$\pm$&$\mp$ &$\mp$ &$\pm$ &$\pm$ &$\pm$ &$\pm4$ & $2.5\times10^{14}$ & $\frac{11}{3}$&$0.69$&$0.77$&$0.22$\\
\hline
VI-c&$\mp$&$\pm$&$\mp$ &$\pm$&$\pm$&$\pm$&$\pm$ &$\pm$ &$\mp$ &$\mp$ &$\pm$ &$\pm10$ & $10^{14}$ & $-\frac{10}{3}$&$0.69$&$6.20$&$0.54$\\
\hline
VI-d&$\mp$&$\pm$&$\mp$ &$\pm$&$\pm$&$\pm$&$\pm$ &$\mp$ &$\pm$ &$\pm$ &$\pm$ &$\pm10$ & $10^{14}$ & $-\frac{26}{15}$&$0.69$&$4.35$&$0.54$\\
\hline\hline
 &$21$&$8$&$14$ &$11$&$5$&$20$&$9$ &$4$ &$3$ &$1$ &$2$&&&&&&&$1.7^{+0.8}_{-0.7}\times10^{-18}$\\
\hline
VII-a&$\mp$&$\pm$&$\pm$ &$\mp$&$\pm$&$\pm$&$\mp$ &$\pm$ &$\mp$ &$\pm$ &$\pm$ &$\pm5$ & $2\times10^{14}$ & $-\frac{2}{15}$&$0.72$&$1.24$&$0.27$\\
\hline
VII-b&$\mp$&$\pm$&$\mp$ &$\pm$&$\pm$&$\pm$&$\pm$ &$\pm$ &$\pm$ &$\mp$ &$\mp$ &$\pm11$ &$9\times10^{13}$  & $-\frac{98}{33}$&$0.72$&$6.36$&$0.60$\\
\hline\hline
 &$21$&$8$&$14$ &$11$&$5$&$19$&$9$ &$4$ &$2$ &$0$ &$1$ &&&&&&&$4.2^{+1.9}_{-1.7}\times10^{-19}$\\
\hline
VIII-a&$\mp$&$\pm$&$\pm$ &$\mp$&$\pm$&$\pm$&$\mp$ &$\pm$ &$\pm$ &$0$ &$\mp$ &$\pm5$ & $4\times10^{14}$ & $\frac{4}{15}$&$0.34$&$0.51$&$0.14$\\
\hline
VIII-b&$\mp$&$\pm$&$\mp$ &$\pm$&$\pm$&$\pm$&$\pm$ &$\pm$ &$\pm$ &$0$ &$\mp$ &$\pm11$ & $1.8\times10^{14}$ & $-\frac{92}{33}$&$0.34$&$3.06$&$0.30$\\
\hline\hline
 &$20$&$8$&$14$ &$11$&$5$&$20$&$9$ &$4$ &$2$ &$0$ &$1$&&&&&&&$4.2^{+1.9}_{-1.7}\times10^{-19}$ \\
\hline
IX-a&$\mp$&$\pm$&$\pm$ &$\mp$&$\pm$&$\pm$&$\mp$ &$\pm$ &$\mp$ &$0$ &$\mp$ &$\pm4$ & $5\times10^{14}$ & $-\frac{5}{6}$&$0.36$&$0.66$&$0.11$\\
\hline
IX-b&$\mp$&$\pm$&$\mp$ &$\pm$&$\pm$&$\pm$&$\pm$ &$\pm$ &$\mp$ &$0$ &$\mp$ &$\pm10$ &$2\times10^{14}$ & $-\frac{53}{15}$&$0.36$&$3.22$&$0.27$\\
\end{tabular}
\end{ruledtabular}
\end{table}

\begin{figure}[t]
\begin{minipage}[h]{8.0cm}
\includegraphics[width=8.0cm]{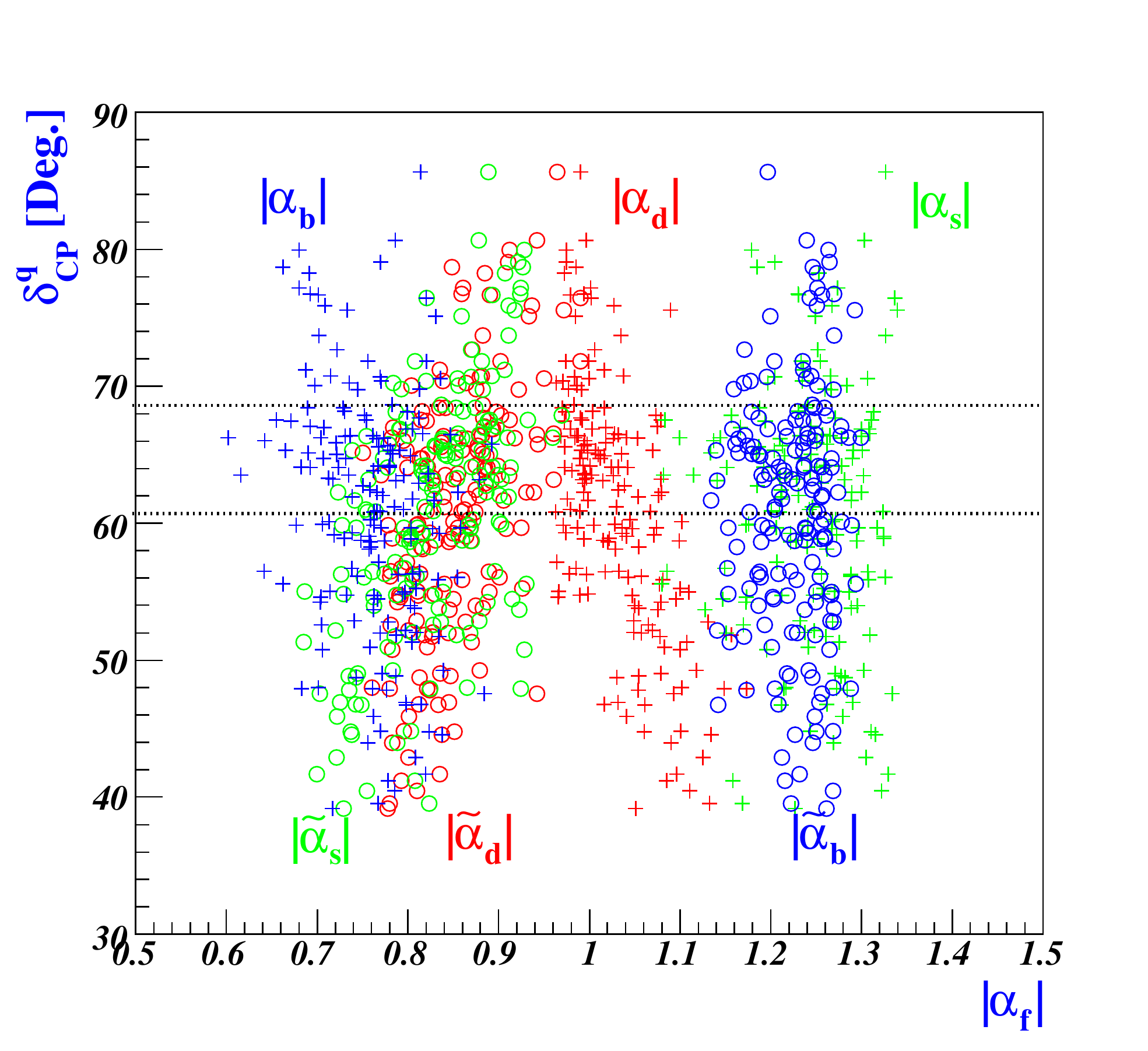}
\end{minipage}
\begin{minipage}[h]{8.0cm}
\includegraphics[width=8.0cm]{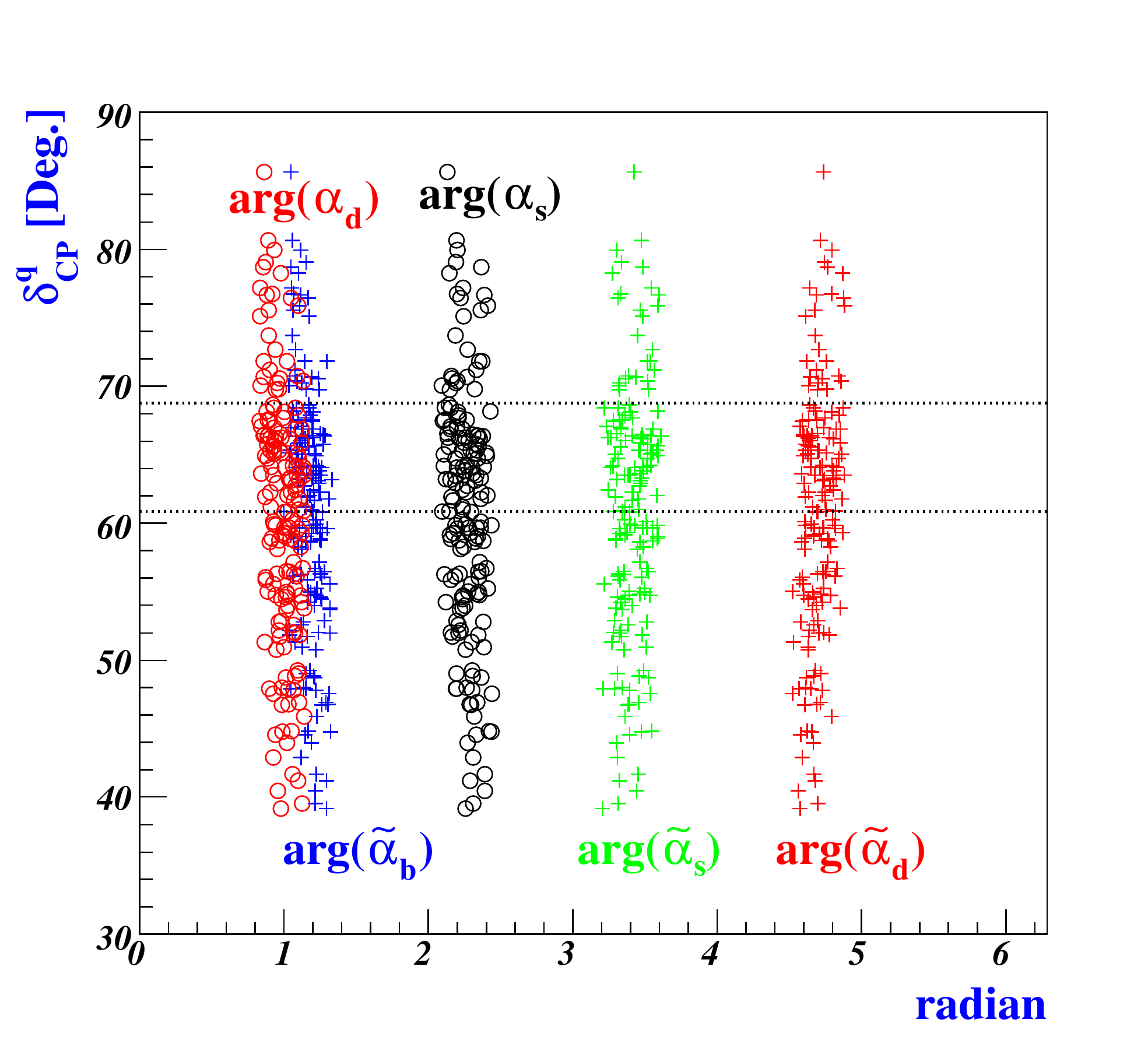}
\end{minipage}\\
\begin{minipage}[h]{8.0cm}
\includegraphics[width=8.0cm]{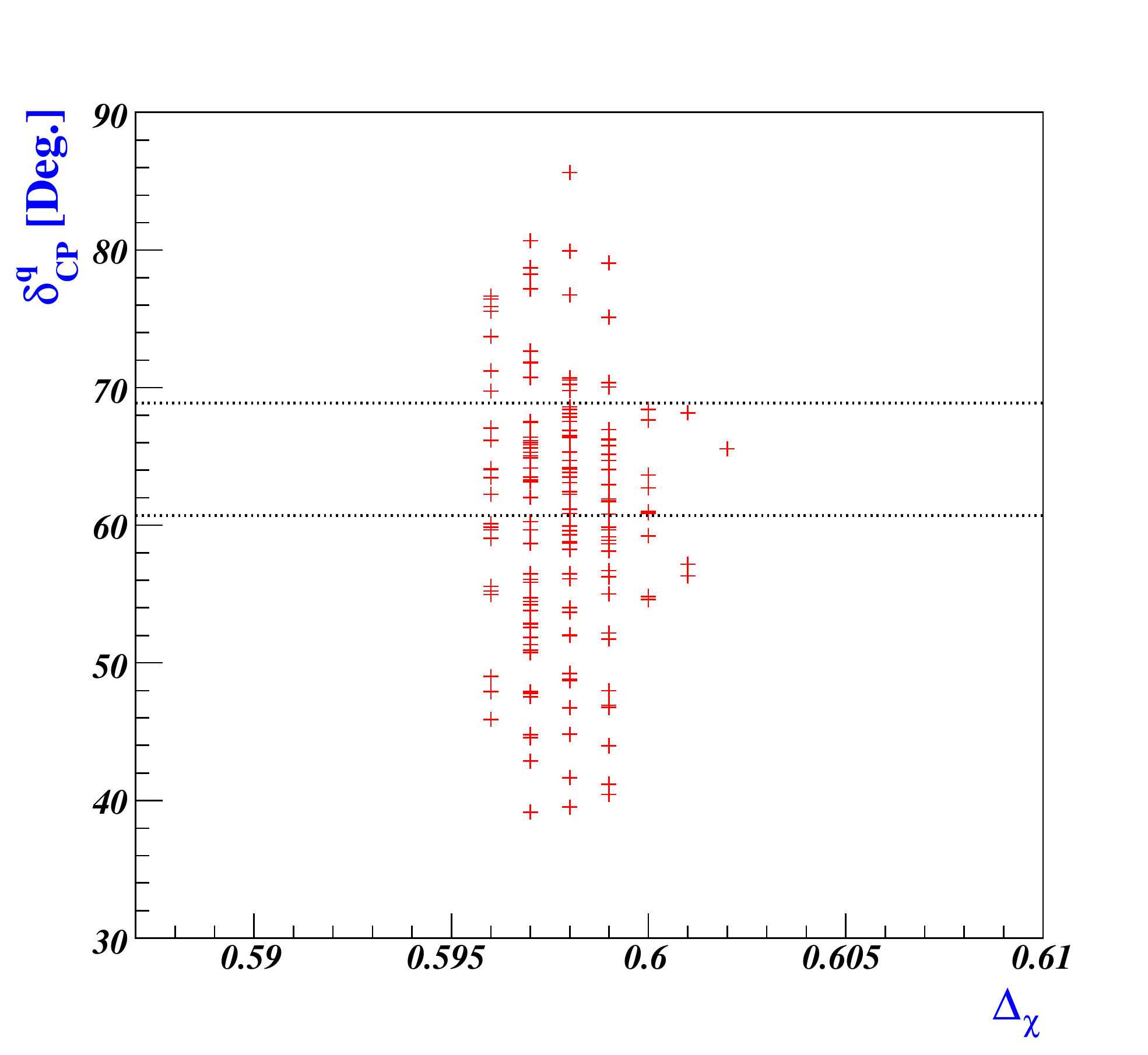}
\end{minipage}
\begin{minipage}[h]{8.0cm}
\includegraphics[width=8.0cm]{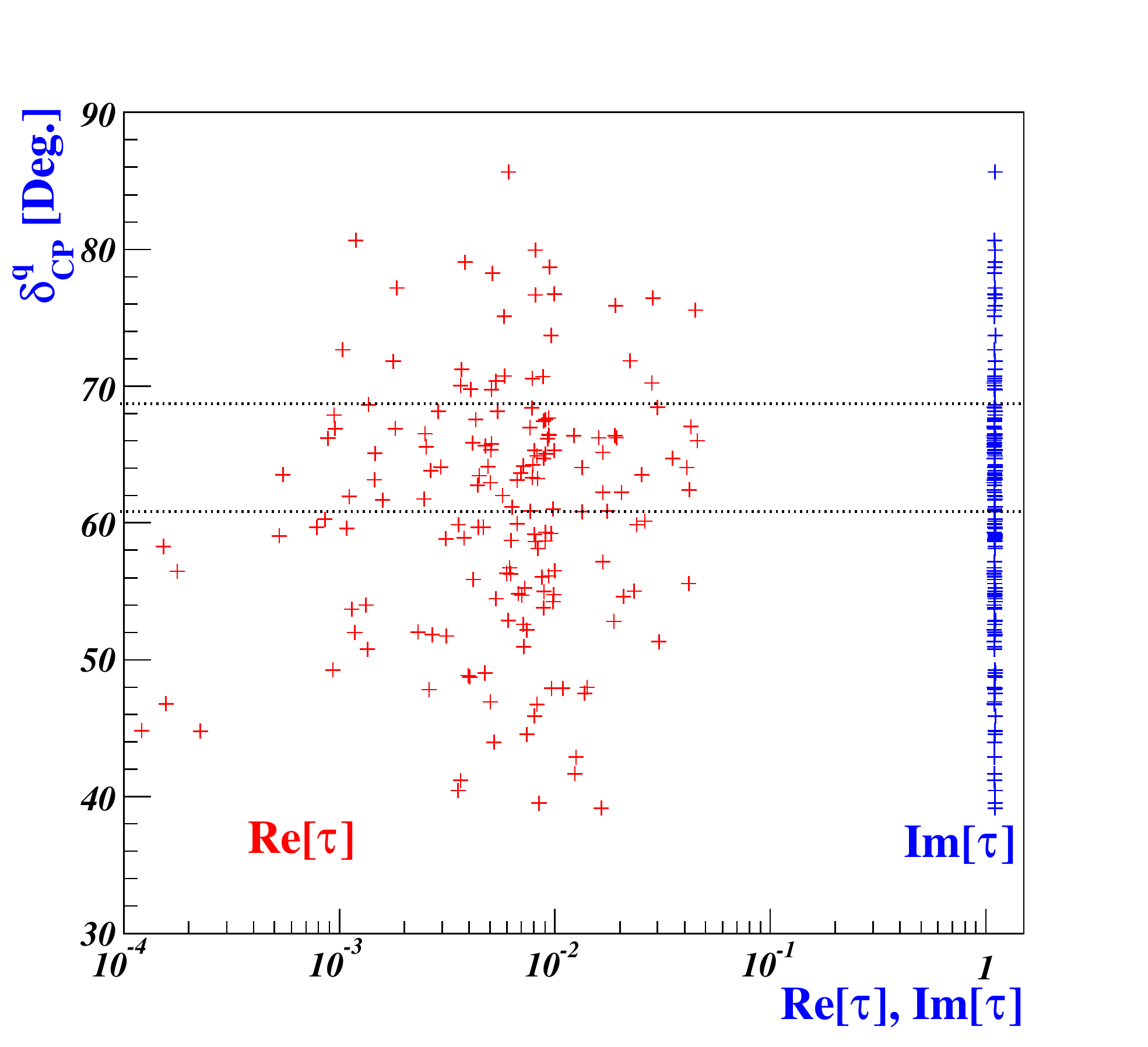}
\end{minipage}
\caption{\label{FigA3}  Model predictions for $\delta^q_{\rm CP}$ are shown, left upper, right upper, and left lower panel, as a function of the parameters that are constrained by other empirical results. The horizontal black-dotted lines indicate the $3\sigma$ experimental bound.}
\end{figure}

\subsection{Quark and charged-lepton}
The Yukawa matrices for charged fermions in the SM, as given in Eqs.(\ref{Ch2}, \ref{Ch1}, \ref{ChL1}), are taken at the scale of $U(1)_X$ symmetry breakdown. Hence their masses are subject to quantum corrections. Subsequently, these matrices are run down to $m_t$ and diagonalized. We assume that the Yukawa matrices at the scale of $U(1)_X$ breakdown are the same as those at the scale $m_t$, since the one-loop renormalization group running effect on observables for hierarchical mass spectra is expected to be negligible. The low-energy Yukawa couplings required for experimental values are obtained from the physical masses and mixing angles compiled by the PDG\,\cite{PDG} and CKMfitter\,\cite{ckm}.

We have thirteen physical observables in the quark and charged-lepton sector: $m_d, m_s, m_b$, $m_u, m_c, m_t, m_e, m_\mu, m_\tau$, and $\theta^q_{12}, \theta^q_{23}, \theta^q_{13}, \delta^q_{CP}$. These observables are used to determine thirteen effective model parameters: 21 parameters ($|\alpha_d|$, $|\alpha_s|$, $|\alpha_b|$, $|\tilde{\alpha}_d|$, $|\tilde{\alpha}_s|$, $|\tilde{\alpha}_b|$, $\alpha_e$, $\alpha_\mu$, $\alpha_\tau$, $\alpha_t$, $\alpha_c$, $\alpha_u$; $\arg(\alpha_d), \arg(\alpha_s), \arg(\tilde{\alpha}_d), \arg(\tilde{\alpha}_s), \arg(\tilde{\alpha}_b)$; $\Delta_\chi$, $\tan\beta$; ${\rm Re[\tau]}, {\rm Im}[\tau]$) among whihc 8 parameters are fixed by quantum numbers ($f_{e,\mu,\tau}, f_{d, s, b}, f_{u, c}$).
Using highly precise data as constraints for both quarks and charged leptons, with the exception of the quark Dirac CP phase, as described in Eqs.(\ref{ckmmixing}), (\ref{qumas}), and (\ref{clepmass}), we scanned all parameter ranges and determined that 
\begin{eqnarray}
&\Delta_\chi=[0.596, 0.602]\,,\qquad \tan\beta=[6.8, 7.3]\,,\nonumber\\
&\tau=(0.0001\sim0.046)+(1.0906\sim1.1086)i\,.
\label{para_sp}
\end{eqnarray}
 ${\rm Re}(\tau)$ contributes to the phase of the Yukawa coupling, while ${\rm Im}(\tau)$ influences the magnitude of Yukawa coupling, as demonstrated in Eqs.(\ref{Ch2}, \ref{Ch1}, \ref{ChL1}, \ref{MR5}, \ref{Ynu1}). When $\langle\tau\rangle=i$, resulting in real values of $x, y$ with $Y^{(2)}_{\bf 3}=Y_1(i)(1,1-\sqrt{3}, -2+\sqrt{3})$, it becomes apparent that the empirical results of quark masses and CKM mixing angles cannot be satisfied due to the overall factors in Eq.(\ref{Ch1}) being real. Therefore, a departure of $\tau$ from $i$ is necessary.
Fig.\ref{FigA3} shows how the quark Dirac CP phase $\delta^q_{CP}$ behaves based on certain constrained parameters. Our model predicts that $\delta^q_{CP}$ falls between $38^\circ$ and $87^\circ$, which aligns well with experimental data. The horizontal black-dotted lines in Fig.\ref{FigA3} represent the $3\sigma$ experimental bound for $\delta^{q}_{CP}$. Notably, the effective Yukawa coefficients satisfying the experimental data fall well within the bound specified in Eq.(\ref{AFN1}), as shown in the top left panel of Fig.\ref{FigA3}. This reflects that these coefficients have a natural size of unity, as stated in Eq.(\ref{AFN}).

We choose reference values, for example, that satisfy the experimental data :
\begin{eqnarray}
&\Delta_\chi=0.597\,,\qquad\tau=0.0074+1.0997i\,,\qquad\tan\beta=6.8
\label{delchi}
\end{eqnarray}
which result in effective Yukawa coefficients from Eq.(\ref{AFN1}) satisfying $0.45\lesssim|\alpha_i|\lesssim1.55$.
With the inputs
 \begin{eqnarray}
 &\arg(\alpha_d)=1.007,~\arg(\alpha_s)=2.232,~\arg(\tilde{\alpha}_d)=4.723,~\arg(\tilde{\alpha}_s)=3.388,~\arg(\tilde{\alpha}_b)=1.164,\nonumber\\
&\alpha_u=1.320~\text{for $|f_u|=21$}(\alpha_u=0.788~\text{for $|f_u|=20$}),~\alpha_c=0.950,~\alpha_t=1.006,\nonumber\\
&|\alpha_d|=1.039,~|\alpha_s|=1.218,~|\alpha_b|=0.790,~|\tilde{\alpha}_d|=0.896,~|\tilde{\alpha}_s|=0.822,~|\tilde{\alpha}_b|=1.158,
 \label{quarkvalue21}
 \end{eqnarray}
 we obtain the mixing angles and Dirac CP phase $\theta^q_{12}=12.980^{\circ}$, $\theta^q_{23}=2.320^{\circ}$, $\theta^q_{13}=0.218^{\circ}$, $\delta^q_{CP}=64.216^{\circ}$ compatible with the $3\sigma$ Global fit of CKMfitter\,\cite{ckm}, see Eq.(\ref{ckmmixing}); the quark masses $m_d=4.593$ MeV, $m_s=103.819$ MeV, $m_b=4.206$ GeV, $m_u=2.164$ MeV, $m_c=1.271$ GeV, and $m_t=173.1$ GeV compatible with the values in PDG\,\cite{PDG}, see Eq.(\ref{qumas}). 
Here, without loss of generality, the up-type quark masses $m_u$, $m_c$, and $m_t$ are a one-to-one correspondence with $\alpha_u$, $\alpha_c$, and $\alpha_t$, which have been taken real, and we have set $\arg(\alpha_b)=0$.

The masses of the charged leptons $m_e$, $m_\mu$, and $m_\tau$ are in a one-to-one correspondence with the real parameters $\alpha_e$, $\alpha_\mu$, and $\alpha_\tau$ from Eq.(\ref{cLep1}). Using the numerical results of Eq.(\ref{delchi}) from the quark sector, with the inputs
 \begin{eqnarray}
\alpha_e=1.268~\text{for}\, |f_e|=20 \,(\alpha_e=0.757~ \text{for}\, |f_e|=19), \alpha_\mu=0.900, \alpha_\tau=1.148\,,
 \label{clep01}
 \end{eqnarray}
 we obtain the charged lepton masses, which well agree with the empirical values of Eq.(\ref{clepmass}).

\begin{table}[h]
\caption{\label{l_n} 
The same as in Table \ref{q_n}, except for the inverted neutrino mass ordering.}
\begin{ruledtabular}
\begin{tabular}{ccccccccccccccccccc}
$U(1)_X$ &$f_{u}$& $f_{c}$& $f_{d}$&$f_s$ & $f_b$&$f_e$ & $f_\mu$ & $f_\tau$ & $g_e$ & $g_\mu$ & $g_\tau$ & $N_C$ & $\frac{F_a}{\rm GeV}$& $\frac{E}{N_C}$ & $\frac{g_{ae}}{10^{-17}}$ & $\frac{|g_{a\gamma\gamma}|}{10^{-17}\,{\rm GeV}^{-1}}$& $\frac{m_a}{10^{-7}\,{\rm eV}}$& ${\rm Br}(K\pi a)$\\
\hline
 &$19$&$8$&$14$ &$11$&$5$&$20$&$9$ &$4$ &$2$ &$3$ &$3$ &&&&&&&$1.7^{+0.8}_{-0.7}\times10^{-16}$\\
\hline
I-a&$\mp$&$\pm$&$\pm$ &$\mp$&$\pm$&$\pm$&$\mp$ &$\mp$ &$\pm$ & $\begin{array}{ll}
\mp \\
\pm
\end{array}$& $\begin{array}{ll}
\pm \\
\mp
\end{array}$ &$\pm3$ &$3.33\times10^{13}$ & $\frac{10}{3}$&$7.23$&$4.62$& $1.63$\\
\hline
I-b&$\mp$&$\pm$&$\mp$ &$\pm$&$\pm$&$\pm$&$\pm$ &$\mp$ &$\pm$ &$\begin{array}{ll}
\mp \\
\pm
\end{array}$ &$\begin{array}{ll}
\pm \\
\mp
\end{array}$ &$\pm9$ &$1.11\times10^{13}$ & $-\frac{22}{9}$&$7.23$&$46.54$& $4.90$\\
\hline\hline
 &$20$&$8$&$14$ &$11$&$5$&$19$&$9$ &$4$ &$2$ &$3$ &$3$&&&&&&&$1.7^{+0.8}_{-0.7}\times10^{-16}$ \\
\hline
II-a&$\mp$&$\pm$&$\pm$ &$\mp$&$\pm$&$\pm$&$\mp$ &$\pm$ &$\mp$ & $\begin{array}{ll}
\mp \\
\pm
\end{array}$& $\begin{array}{ll}
\pm \\
\mp
\end{array}$ &$\pm4$ &$2.5\times10^{13}$ & $-\frac{1}{3}$&$6.87$&$10.88$& $2.18$\\
\hline
II-b&$\mp$&$\pm$&$\mp$ &$\pm$&$\pm$&$\pm$&$\pm$ &$\pm$ &$\mp$ & $\begin{array}{ll}
\mp \\
\pm
\end{array}$& $\begin{array}{ll}
\pm \\
\mp
\end{array}$ &$\pm10$ &$10^{13}$ & $-\frac{10}{3}$&$6.87$&$62.04$& $5.44$\\
\hline\hline
 &$20$&$8$&$14$ &$11$&$5$&$20$&$9$ &$4$ &$1$ &$2$ &$2$ &&&&&&&$4.2^{+1.9}_{-1.7}\times10^{-17}$ \\
\hline
III-a&$\mp$&$\pm$&$\pm$ &$\mp$&$\pm$&$\pm$&$\mp$ &$\mp$ &$\pm$ &$\pm$ &$\pm$ &$\pm4$ &$5\times10^{13}$ & $\frac{19}{6}$&$3.61$&$6.60$&$1.09$\\
\hline
III-b&$\mp$&$\pm$&$\pm$ &$\mp$&$\pm$&$\pm$&$\mp$ &$\pm$ &$\pm$ &$\mp$ &$\mp$ &$\pm4$ &$5\times10^{13}$ & $-\frac{5}{6}$&$3.61$&$2.69$&$1.09$\\
\hline
III-c&$\mp$&$\pm$&$\mp$ &$\pm$&$\pm$&$\pm$&$\pm$ &$\mp$ &$\pm$ &$\pm$ &$\pm$ &$\pm10$ &$2\times10^{13}$ & $-\frac{29}{15}$&$3.61$&$22.89$&$2.72$\\
\hline
III-d&$\mp$&$\pm$&$\mp$ &$\pm$&$\pm$&$\pm$&$\pm$ &$\pm$ &$\pm$ &$\mp$ &$\mp$ &$\pm10$ &$2\times10^{13}$ & $-\frac{53}{15}$&$3.61$&$32.18$&$2.72$\\
\hline\hline
 &$19$&$8$&$14$ &$11$&$5$&$19$&$9$ &$4$ &$1$ &$2$ &$2$  &&&&&&&$4.2^{+1.9}_{-1.7}\times10^{-17}$\\
\hline
IV-a&$\mp$&$\pm$&$\pm$ &$\mp$&$\pm$&$\pm$&$\mp$ &$\mp$ &$\mp$ &$\pm$ &$\pm$ &$\pm3$ &$6.67\times10^{13}$ & $4$&$3.43$&$3.47$&$8.16$\\
\hline
IV-b&$\mp$&$\pm$&$\pm$ &$\mp$&$\pm$&$\pm$&$\mp$ &$\pm$ &$\mp$ &$\mp$ &$\mp$ &$\pm3$ &$6.67\times10^{13}$ & $-\frac{4}{3}$&$3.43$&$5.82$&$8.16$\\
\hline
IV-c&$\mp$&$\pm$&$\mp$ &$\pm$&$\pm$&$\pm$&$\pm$ &$\mp$ &$\mp$ &$\pm$ &$\pm$ &$\pm9$ &$2.22\times10^{13}$ & $-\frac{20}{9}$&$3.43$&$22.11$&$2.45$\\
\hline
IV-d&$\mp$&$\pm$&$\mp$ &$\pm$&$\pm$&$\pm$&$\pm$ &$\pm$ &$\mp$ &$\mp$ &$\mp$ &$\pm9$ &$2.22\times10^{13}$  & $-4$&$3.43$&$31.40$&$2.45$\\
\hline\hline
 &$19$&$8$&$14$ &$11$&$5$&$20$&$9$ &$4$ &$0$ &$1$ &$1$&&&&&&&$1.7^{+0.8}_{-0.7}\times10^{-18}$ \\
\hline
V-a&$\mp$&$\pm$&$\pm$ &$\mp$&$\pm$&$\pm$&$\mp$ &$\mp$ &$0$ &$\pm$ &$\pm$ &$\pm3$ &$3.33\times10^{14}$ & $\frac{10}{3}$&$0.72$&$0.46$&$0.16$\\
\hline
V-b&$\mp$&$\pm$&$\mp$ &$\pm$&$\pm$&$\pm$&$\pm$ &$\mp$ &$0$ &$\pm$ &$\pm$ &$\pm9$ &$1.11\times10^{14}$ & $-\frac{22}{9}$&$0.72$&$4.65$&$0.49$\\
\end{tabular}
\end{ruledtabular}
\end{table}

\subsection{Neutrino}
The seesaw mechanism in Eq.(\ref{neut2}) operates at the $U(1)_X$ symmetry breakdown scale, while its implications are measured by experiments below EW scale. Therefore, quantum corrections to neutrino masses and mixing angles can be crucial, especially for degenerate neutrino masses\,\cite{Antusch:2005gp}. 
However, based on our observation that the neutrino mass spectra exhibit hierarchy at the scale of $U(1)_X$ breakdown (as depicted in Fig.\ref{Fig4}), we can safely assume that the renormalization group running effect on observables can be ignored.
\begin{figure}[h]
\begin{minipage}[h]{8.0cm}
\includegraphics[width=8.0cm]{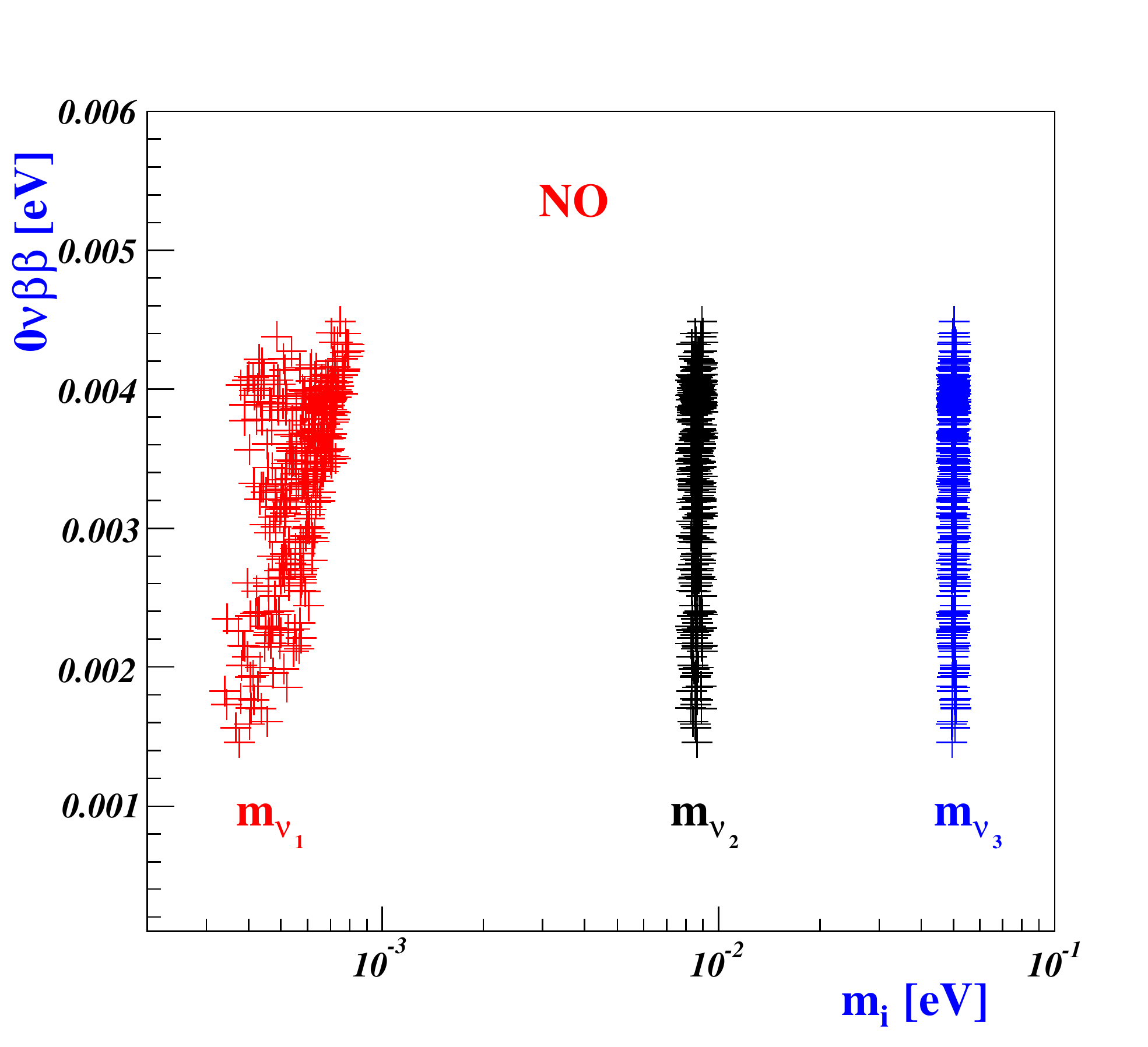}
\end{minipage}
\begin{minipage}[h]{8.0cm}
\includegraphics[width=8.0cm]{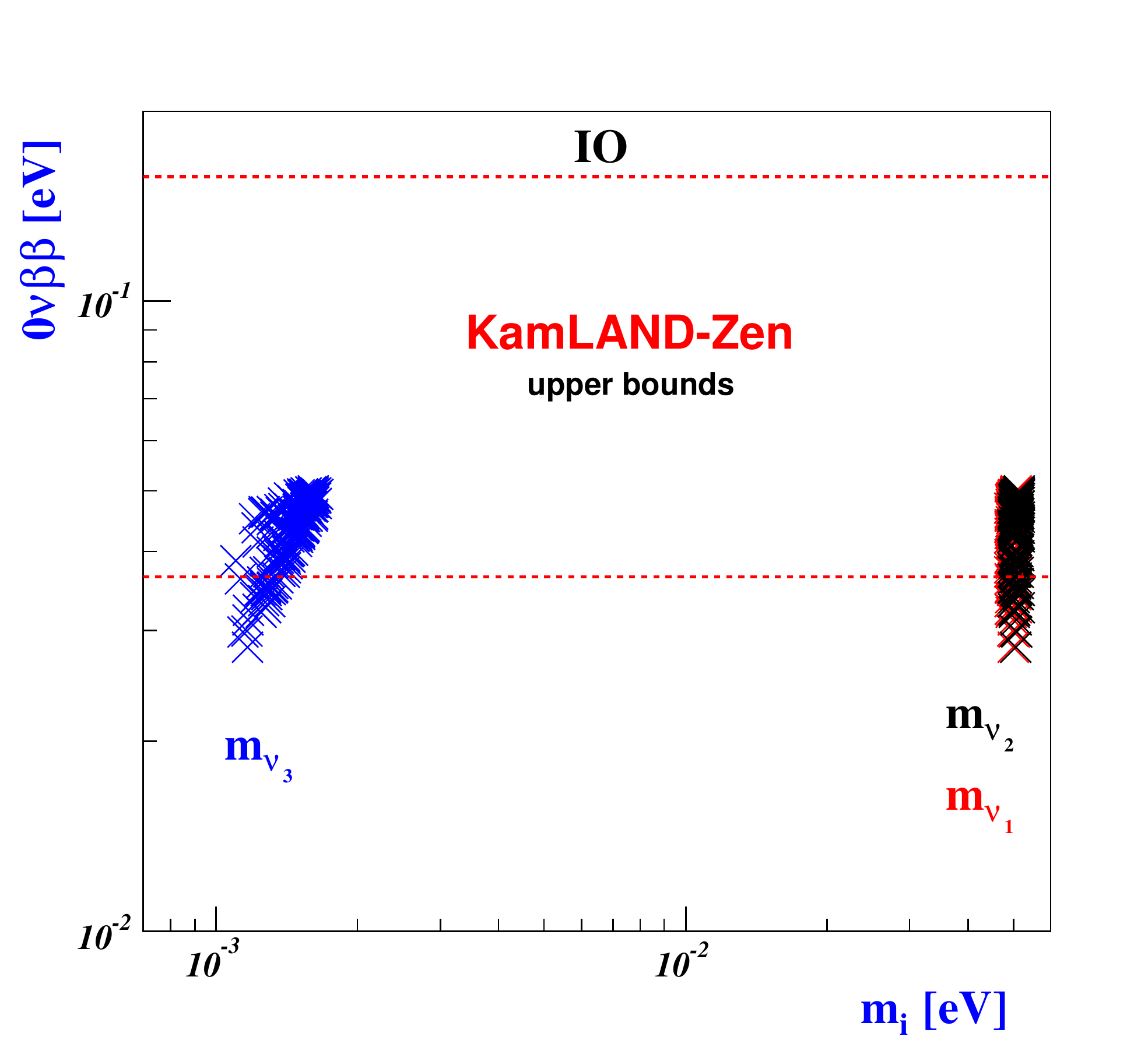}
\end{minipage}\\
\begin{minipage}[h]{8.0cm}
\includegraphics[width=8.0cm]{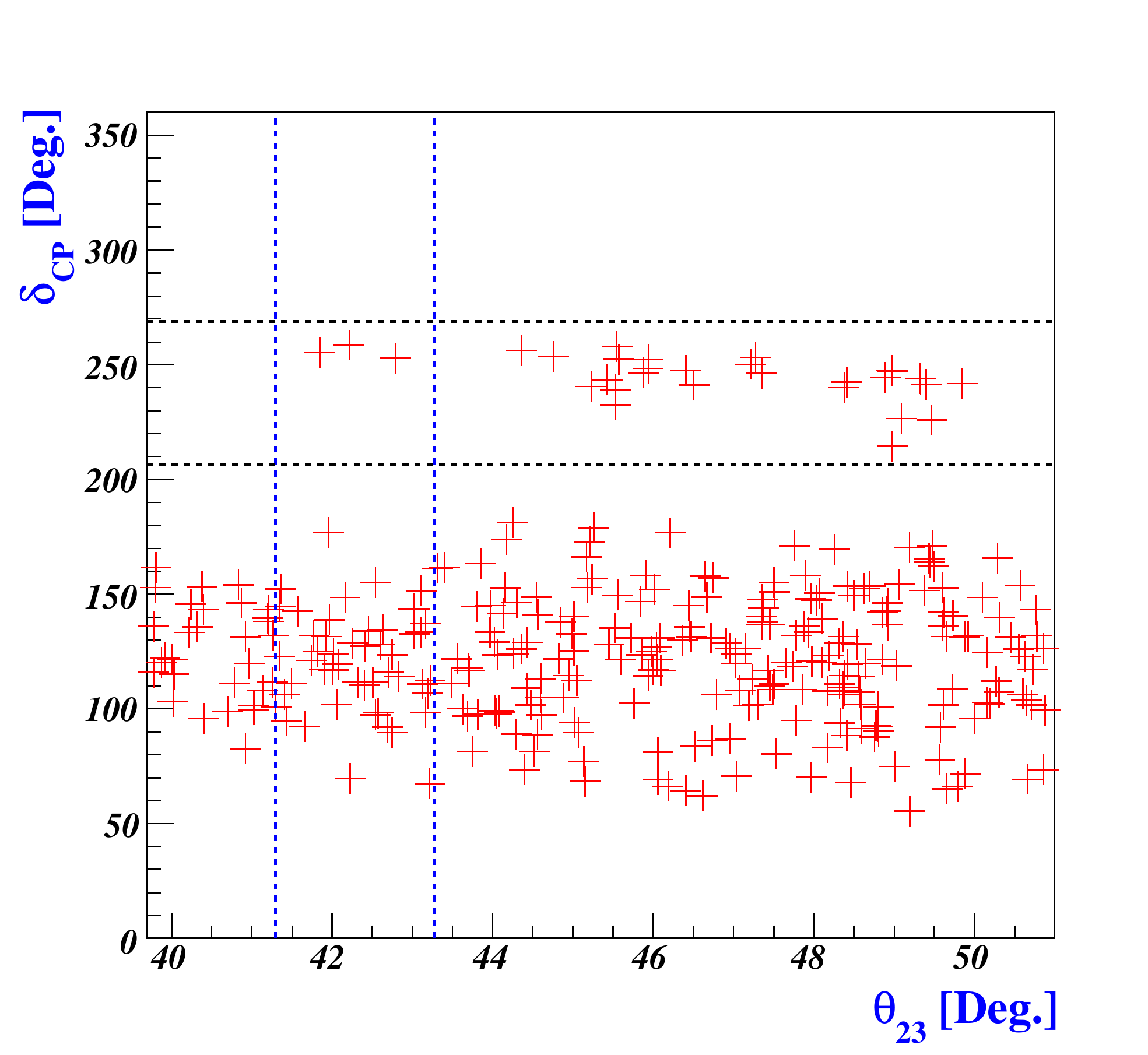}
\end{minipage}
\begin{minipage}[h]{8.0cm}
\includegraphics[width=8.0cm]{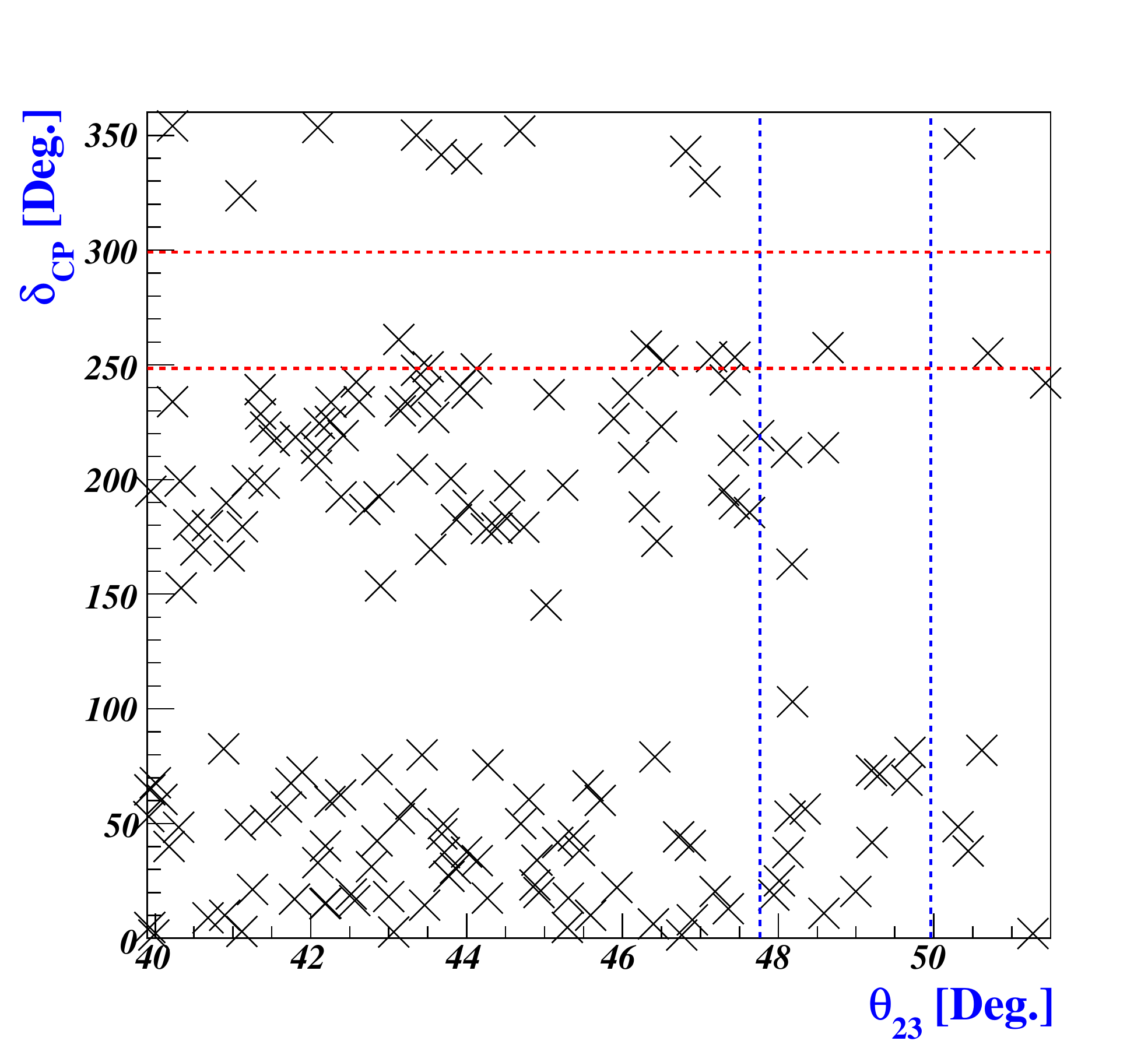}
\end{minipage}
\caption{\label{Fig4} Plots for $0\nu\beta\beta$-decay rate (upper panel) and leptonic Dirac CP phase $\delta_{CP}$ (lower panel) as a function of the neutrino masses $m_{\nu_i}$ and the atmospheric mixing angle $\theta_{23}$, respectively, for NO (left) and IO (right). Vertical and horizontal dashed lines represent the $1\sigma$ bounds for $\theta_{23}$ and $\delta_{CP}$, respectively, in Table-\ref{exp_nu}. Horizontal red-lines indicate the upper bound of KamLAND-Zen result of Eq.(\ref{nubb})\,\cite{KamLAND-Zen:2022tow}.}
\end{figure}

\begin{figure}[h]
\begin{minipage}[h]{8.0cm}
\includegraphics[width=8.0cm]{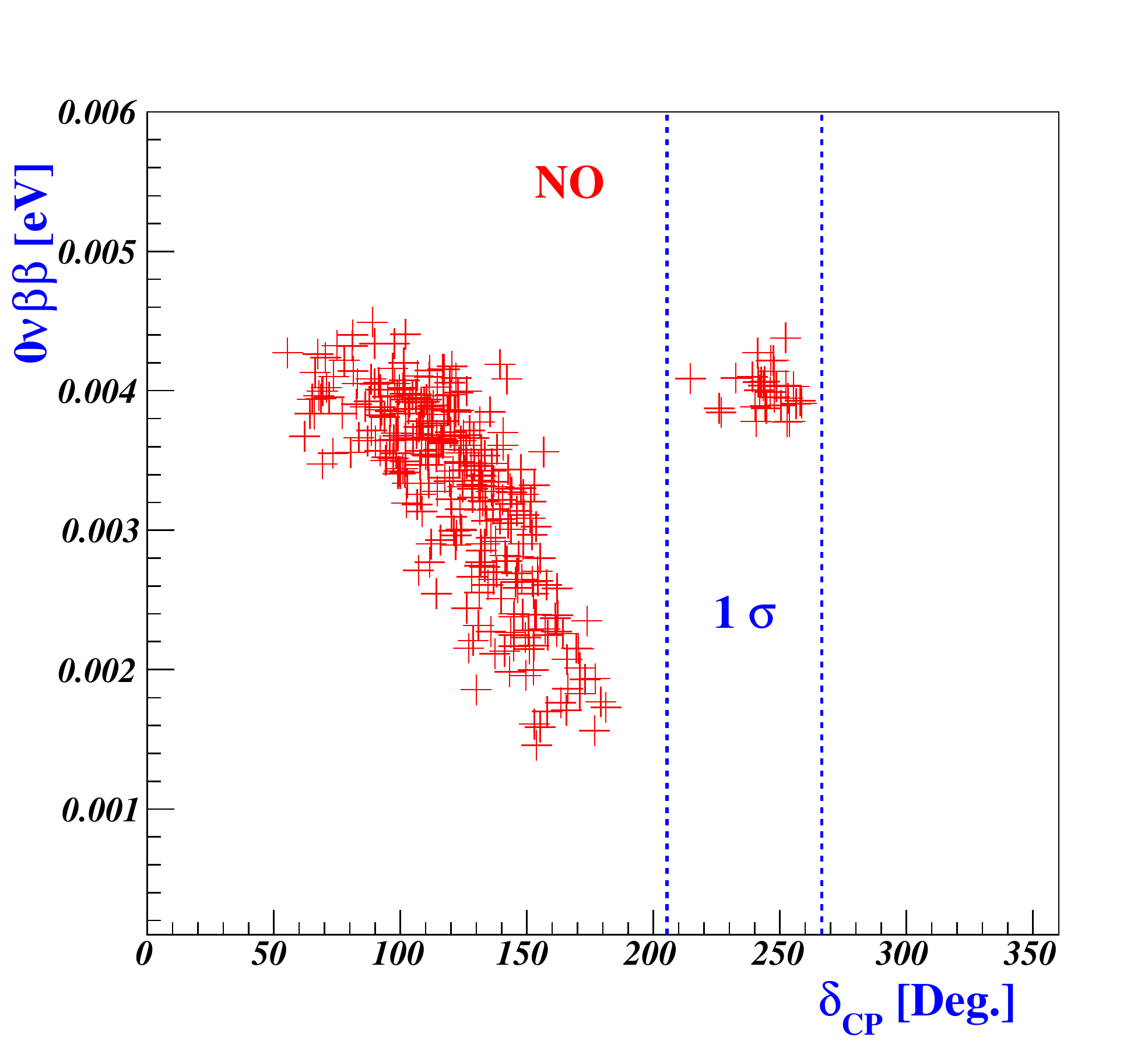}
\end{minipage}
\begin{minipage}[h]{8.0cm}
\includegraphics[width=8.0cm]{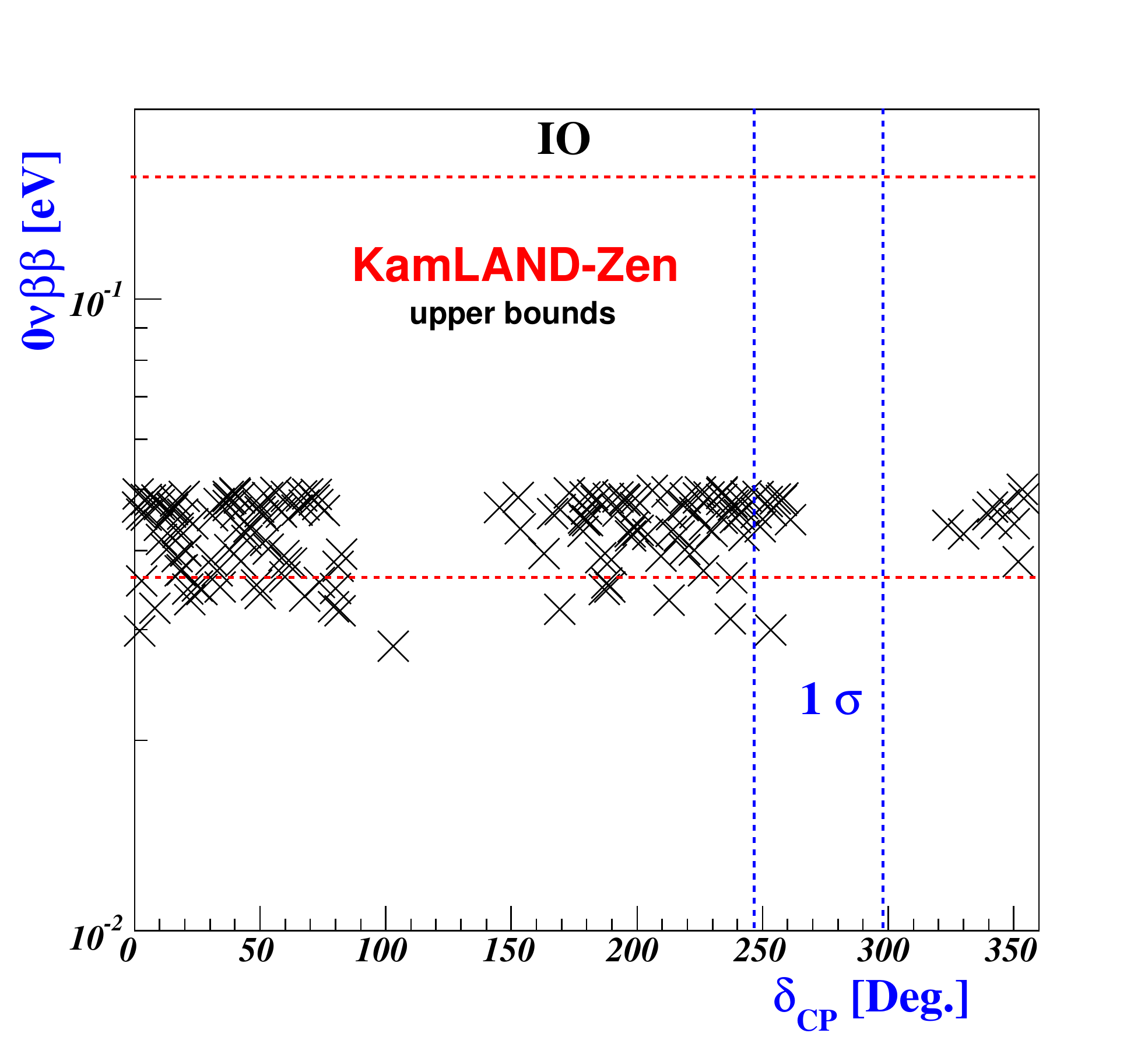}
\end{minipage}
\caption{\label{Fig5} Plots for $0\nu\beta\beta$-decay rate as a function of leptonic Dirac CP phase $\delta_{CP}$ for NO (left) and IO (right). Vertical dashed lines indicate the $1\sigma$ bound for $\delta_{CP}$ listed in Table-\ref{exp_nu}, while horizontal red-lines represent the upper bound of KamLAND-Zen result of Eq.(\ref{nubb})\,\cite{KamLAND-Zen:2022tow} for the $0\nu\beta\beta$-decay rate.}
\end{figure}
\begin{figure}[h]
\begin{minipage}[h]{8.0cm}
\includegraphics[width=8.0cm]{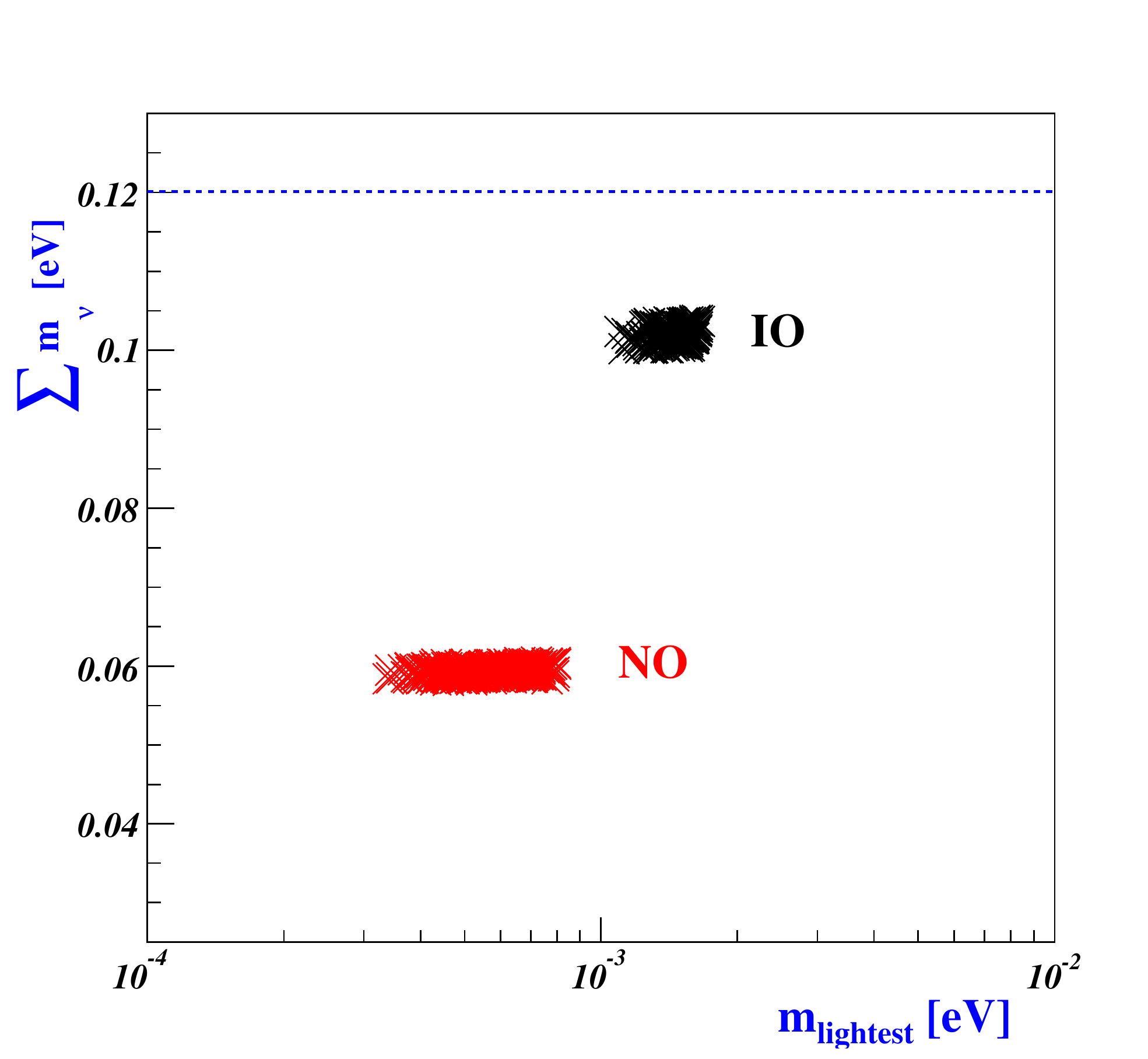}
\end{minipage}
\caption{\label{Fig6} Plot for $\sum m_{\nu}=m_{\nu_1}+m_{\nu_2}+m_{\nu_3}$ as a function of $m_{\rm lightest}$ for NO (red) and IO (black). }
\end{figure}
Neutrino oscillation experiments currently aim to make precise measurements of the Dirac CP-violating phase $\delta_{CP}$ and atmospheric mixing angle $\theta_{23}$. Using our model, we investigate which values of $\delta_{CP}$ and $\theta_{23}$ can predict the mass hierarchy of neutrinos (NO or IO) and identify observables that can be tested in current and next-generation experiments. 
To explore the parameter spaces, we scan the precision constraints $\{\theta_{13}$, $\theta_{23}$, $\theta_{12}$, $\Delta m^2_{\rm Sol}$, $\Delta m^2_{\rm Atm}\}$ at $3\sigma$ from Table-\ref{exp_nu}. Using the reference values from Eq.(\ref{delchi}) in the quark and charged-lepton sectors, we determine the input parameter spaces of Eq.(\ref{dof_nu}) for both NO and IO at the $U(1)_X$ breaking scale : for example, taking $\langle\chi\rangle=5\times10^{13}$ GeV (see, Table-\ref{q_m}, \ref{q_m}, and -\ref{l_n}) and for NO
\begin{eqnarray}
&\tilde{\beta}_2=[0.48, 1.10],\quad\tilde{\beta}_3=[0.61, 1.18]\,,\quad\arg(\tilde{\beta}_{2(3)})=[0, 2\pi]\nonumber\\
&m_0\langle\chi\rangle/v^2_u=[0.59, 1.84],\quad \gamma=[0.37, 0.84],\quad \gamma'=[0.37, 0.65]\nonumber\\
& \arg(\gamma)=[4.42, 5.55],\quad\arg(\gamma')=[1.30, 2.19];
 \label{neu01}
 \end{eqnarray} 
 where $\beta_1=[0.97, 1.47]$, $\beta_2=[0.61, 1.11]$, and $\beta_3=[0.69, 1.15]$;
 for IO
  \begin{eqnarray}
&\tilde{\beta}_2=[0.55, 0.87],\quad\tilde{\beta}_3=[0.56, 0.86]\,,\quad\arg(\tilde{\beta}_{2(3)})=[0, 2\pi]\nonumber\\
&m_0\langle\chi\rangle/v^2_u=[0.81, 1.39],\quad \gamma=[0.758, 1.154],\quad \gamma'=[0.350, 0.618]\nonumber\\
& \arg(\gamma)=[2.92, 4.19],\quad\arg(\gamma')=[0.1, 1.29]\,\&\,[5.66, 6.26]\,.
 \label{neu02}
 \end{eqnarray} 
 where $\beta_1=[0.98, 1.29]$, $\beta_2=[0.69, 0.95]$, and $\beta_3=[0.68, 0.93]$.
For these parameter regions, we investigate how $0\nu\beta\beta$-decay rate and Dirac CP phase can be determined for the normal and inverted mass ordering.
These predictions are represented by crosses and X-marks for NO and IO, respectively, in Fig.\ref{Fig5}.
Referring to the two-dimensional allowed regions at $3\sigma$ presented in Ref.\cite{Esteban:2020cvm}, we note that the most favored regions correspond to $\delta_{CP}\sim250^\circ$, whereas there are no favored regions with respect to $\theta_{23}$.} Ongoing experiments like DUNE\,\cite{DUNE:2018tke}, as well as proposed next-generation experiments such as Hyper-K\,\cite{Hyper-Kamiokande:2018ofw}, are poised to greatly reduce uncertainties in the values of $\theta_{23}$ and $\delta_{CP}$, providing a rigorous test for our proposed model. Furthermore, ongoing and future experiments on $0\nu\beta\beta$-decay like NEXT\,\cite{NEXT:2020amj}, SNO$+$\,\cite{SNO:2022trz}, KamLAND-Zen\,\cite{KamLAND-Zen:2022tow}, Theia\,\cite{Theia:2019non}, SuperNEMO\,\cite{Arnold:2004xq} may soon reach a sensitivity to exclude the inverted mass ordering of our model.
Cosmological and astrophysical measurements provide powerful constraints on the sum of neutrino masses. The upper bound on the sum of the three active neutrino masses can be summarized as $\sum m_{\nu}=m_{\nu_1}+m_{\nu_2}+m_{\nu_3}<0.120$ eV at $95\%$ CL for TT, TE, EE+lowE+lensing+BAO\,\cite{Planck:2018vyg}. 
Fig.\ref{Fig6} illustrates that the sum of neutrino masses lies in the range of $0.0582$ to $0.0605$ eV for the NO when $0.0003397 \lesssim m_{\rm lightest} = m_{\nu_1} \lesssim 0.0007945$ eV. Conversely, for the IO, the sum of neutrino masses falls within the range of $0.1003$ to $0.1038$ eV when $0.001125 \lesssim m_{\rm lightest} = m_{\nu_3} \lesssim 0.001647$ eV.

\section{conclusion}
We proposed a minimal extension of a modular invariant model that incorporates sterile neutrinos and a QCD axion (as a strong candidate of dark matter) into the SM to account for the mass and mixing hierarchies of quarks and leptons, as well as the strong CP problem. Our model, based on the 4D effective action, features the $G_{\rm SM}\times \Gamma_N\times U(1)_X$ symmetry.
To ensure the reliability of our model, we have examined the modular forms of the superpotential, corrected by K{\"a}hler transformation, under the $G_{\rm SM}\times \Gamma_N\times U(1)_X$ symmetry, while also considering the modular and $U(1)_X$ anomaly-free conditions. 
The model features a minimal set of fields that transform based on representations of $G_{\rm SM}\times \Gamma_N\times U(1)_X$, and includes modular forms of level $N$. These modular forms act as Yukawa couplings and transform under the modular group $\Gamma_N$. 
Our numerical analysis guarantees that, in the supersymmetric limit, all Yukawa coefficients in the superpotential are complex numbers with a unit absolute value, implying a democratic distribution.

We demonstrated, as an explicit example, a level 3 modular form-induced superpotential by introducing minimal supermultiplets.
The extension includes right-handed neutrinos ($N^c$) and SM gauge singlet scalar fields ($\chi$ and $\tilde{\chi}$) with zero modular weight and ($+$ and $-$) charge under $U(1)_X$. These scalar fields are crucial in generating the QCD axion, heavy neutrino mass, and fermion mass hierarchy.
 Modular invariance of both the superpotential and K{\"a}hler potential allows for K{\"a}hler transformation to correct modular form weight in the superpotential, enabling a $\tau$-independent superpotential for the scalar potential. The sterile neutrinos are introduced to satisfy the $U(1)_X$-mixed gravitational anomaly-free condition, explain small active neutrino masses via the seesaw mechanism, and provide a well-motivated PQ symmetry breaking scale. 
As the fields $\chi(\tilde{\chi})$ have modular weights of zero, any additional correction terms arising from higher weight modular forms are not permitted in the superpotentials. However, the combination $\chi\tilde{\chi}$ can trigger higher-order corrections that are permissible and do not modify the leading-order flavor structures.
 Taking into account both SUSY-breaking effects and supersymmetric next-leading-order Planck-suppressed terms, we have determined the low axion decay constant (or seesaw scale). This leads to an approximate range for the PQ scale $\langle\chi\rangle$ (equivalently, the seesaw scale) of $10^{13}$ GeV to $10^{15}$ GeV for $m_{3/2}$ values between 1 TeV and $10^6$ TeV, see Table-\ref{q_n}, -\ref{q_m} and -\ref{l_n}.
Interestingly enough, in our model, the PQ breaking scale (or axion mass) is closely linked to the seesaw scale and the soft SUSY breaking mass. Our model with $E/N_C$ could be tested by ongoing experiments such as  KLASH\,\cite{Alesini:2019nzq} and FLASH\,\cite{Gatti:2021cel}, see Fig.\ref{Fig1} and \ref{Fig2}, by considering the scale of $U(1)_X$ breakdown.

We explored numerical values of physical parameters that satisfy the highly precise data on the mass of quarks and charged-leptons, as well as the quark mixing angles, except for the quark Dirac CP phase. Our model predicts that the value of $\delta^q_{CP}$ falls within the range of $38^\circ$ to $87^\circ$, which is consistent with experimental data. Notably, the effective Yukawa coefficients satisfying the experimental data fall well within the bound specified in Eq.(\ref{AFN1}), as shown in the top left panel of Fig.\ref{FigA3}. This suggests that our assumption, as stated in Eq.(2), that the Yukawa coefficients have a natural size of unity is plausible.
Using precise neutrino oscillation data as constraints, we investigated how the $0\nu\beta\beta$-decay rate and Dirac CP phase could be determined for the normal and inverted mass ordering in the neutrino sector. Referring to the $3\sigma$ allowed regions in Ref.\cite{Esteban:2020cvm}, we note that the most favored regions for our proposed model are $\delta_{CP}\sim250^\circ$, with no favored regions with respect to $\theta_{23}$. Ongoing experiments, such as DUNE\,\cite{DUNE:2018tke} and proposed next-generation experiments, such as Hyper-K\,\cite{Hyper-Kamiokande:2018ofw}, are expected to greatly reduce uncertainties in the values of $\theta_{23}$ and $\delta_{CP}$, providing a rigorous test for our model. Additionally, ongoing and future experiments on $0\nu\beta\beta$-decay, such as NEXT\,\cite{NEXT:2020amj}, SNO$+$\,\cite{SNO:2022trz}, KamLAND-Zen\,\cite{KamLAND-Zen:2022tow}, Theia\,\cite{Theia:2019non}, and SuperNEMO\,\cite{Arnold:2004xq}, may soon have the sensitivity to exclude the inverted mass ordering in our model.

\acknowledgments{YHA was supported by the National Research Foundation of Korea(NRF) grant funded by the Korea government(MSIT) (No.2020R1A2C1010617). SKK was supported by the National Research Foundation of Korea(NRF) grant funded by the Korea government(MSIT)
 (No. 2019R1A2C1088953, No.2020K1A3A7A09080135, No.2023R1A2C1006091).

\appendix
\section{The group $A_4$}
\label{A4_i}
The group $A_4$ is the symmetry group of the tetrahedron, isomorphic to the finite group of the even permutations of four objects. The group $A_4$ has two generators, denoted $S$ and $T$, satisfying the relations $S^2=T^3=(ST)^3={\bf 1}$. In the three-dimensional complex representation, $S$ and $T$ are given by
 \begin{eqnarray}
 S=\frac{1}{3}{\left(\begin{array}{ccc}
 -1 & 2 & 2  \\
2 & -1 & 2   \\
 2 & 2  & -1
 \end{array}\right)}\,,\qquad  T={\left(\begin{array}{ccc}
 1 & 0 & 0  \\
0 & \omega & 0   \\
 0 & 0  & \omega^2
 \end{array}\right)}\,,
 \end{eqnarray}
where $\omega=e^{i2\pi/3}=-1/2+i\sqrt{3}/2$ is a complex cubic-root of unity. $A_4$ has four irreducible representations: three singlets ${\bf 1}, {\bf 1}'$, and ${\bf 1}''$ and one triplet ${\bf 3}$. An $A_4$ singlet ${\bf a}$ is invariant under the action of $S$ ($S{\bf a} = {\bf a}$), while the action of $T$ produces $T {\bf a} = {\bf a}$
for ${\bf 1}$, $T {\bf a} = \omega{\bf a}$ for ${\bf 1}'$, and $T{\bf a} = \omega^2{\bf a}$ for ${\bf 1}''$. Products of two $A_4$ representations decompose into irreducible representations according to the following multiplication rules: ${\mathbf3}\otimes{\mathbf3}={\mathbf3}_{s}\oplus{\mathbf3}_{a}\oplus{\mathbf1}\oplus{\mathbf1}'\oplus{\mathbf1}''$, ${\mathbf1}'\oplus{\mathbf1}'={\mathbf1}''$ and ${\mathbf1}''\oplus{\mathbf1}''={\mathbf1}'$.  Explicitly, if $(a_1, a_2, a_3)$ and $(b_1, b_2, b_3)$ denote two $A_4$ triplets, then we have Eq.(\ref{A4x}).

\section{The CKM mixing matrix}
\label{ckm_a}
The CKM mixing matrix is given in the Wolfenstein parametrization\,\cite{Wolfenstein:1983yz} by
 \begin{eqnarray}
 V_{\rm CKM}={\left(\begin{array}{ccc}
 1-\frac{1}{2}\lambda^2 & \lambda & A_d\lambda^3(\rho-i\eta)  \\
-\lambda & 1-\frac{1}{2}\lambda^2 & A_d\lambda^2   \\
 A_d\lambda^3(1-\rho-i\eta) & -A_d\lambda^2  & 1
 \end{array}\right)}+{\cal O}(\lambda^4)\,.
 \label{ckm0}
 \end{eqnarray}
where $\lambda=0.22500^{+0.00082}_{-0.00063}$, $A_d=0.813^{+0.026}_{-0.018}$, $\bar{\rho}=\rho/(1-\lambda^2/2)=0.157^{+0.036}_{-0.018}$, and $\bar{\eta}=\eta/(1-\lambda^2/2)=0.347^{+0.030}_{-0.020}$ with $3\sigma$ errors\,\cite{ckm}.

\newpage

\end{document}